\documentclass[12pt]{article}
\usepackage{amssymb, amsthm, amsbsy, amsmath, mathtools}
\usepackage{epsfig, graphicx, algorithm, algorithmic}
\usepackage{lipsum}
\usepackage{natbib, color, multirow, bm, indentfirst, url, layouts, enumerate}
\usepackage{subfigure}
\usepackage[margin=1in]{geometry}

\usepackage{mathtools}
\DeclareGraphicsExtensions{.pdf,.eps.gz,.eps,.epsi.gz,.epsi,.ps,.ps.gz}
\usepackage[colorlinks=true, allcolors=blue]{hyperref}%
\usepackage{centernot}

\newcommand\blfootnote[1]{%
  \begingroup
  \renewcommand\thefootnote{}\footnote{#1}%
  \addtocounter{footnote}{-1}%
  \endgroup
}

\long\def\symbolfootnote[#1]#2{\begingroup%
\def\thefootnote{\fnsymbol{footnote}}\footnote[#1]{#2}\endgroup}

\newcommand{\btheta}{{\bm\theta}}

\newcommand{\bLambda}{{\bm\Lambda}}

\newcommand{\bOmega}{{\bm\Omega}}

\def\a{{\bm a}}

\def\c{{\bm c}}

\def\e{{\bm e}}

\def\z{{\bm z}}

\def\A{{\bm A}}

\newtheorem{theorem}{Theorem}
\newtheorem{cor}{Corollary}%

\begin{document}
\title{\bf Fast Network Community Detection with Profile-Pseudo Likelihood Methods}
\vspace{0.5in}
\author{
Jiangzhou Wang$^{1,2}$, Jingfei Zhang$^{3}$, Binghui Liu$^{1}$, Ji Zhu$^{4}$, and Jianhua Guo$^{1}$\\
\fontsize{10}{10}\selectfont\itshape
$^{1}$\, School of Mathematics and Statistics \& KLAS, Northeast Normal University, Jilin, 130024, China. \\
\fontsize{10}{10}\selectfont\itshape
$^{2}$\, Department of Statistics and Data Science, Southern University of Science and Technology, Shenzhen, 518055, China. \\
\fontsize{10}{10}\selectfont\itshape
$^{3}$\,Department of Management Science, University of Miami, Coral Gables, FL, 33146, USA. \\
\fontsize{10}{10}\selectfont\itshape
$^{4}$\,Department of Statistics, University of Michigan, Ann Arbor, MI, 48109, USA.
}
\date{}
\maketitle

\blfootnote{The first three authors contributed equally to this work. For correspondence, please contact Jianhua Guo and Ji Zhu.}

\begin{abstract}%
The stochastic block model is one of the most studied network models for community detection, and fitting its likelihood function on large-scale networks is known to be challenging. One prominent work that overcomes this computational challenge is \citet{amini2013pseudo}, which proposed a fast pseudo-likelihood approach for fitting stochastic block models to large sparse networks. However, this approach does not have convergence guarantee, and may not be well suited for small and medium scale networks. In this article, we propose a novel likelihood based approach that decouples row and column labels in the likelihood function, enabling a fast alternating maximization. This new method is computationally efficient, performs well for both small and large scale networks, and has provable convergence guarantee. We show that our method provides strongly consistent estimates of communities in a stochastic block model. We further consider extensions of our proposed method to handle networks with degree heterogeneity and bipartite properties.
\end{abstract}

\vfill

\newpage
\baselineskip=24.5pt
\section{Introduction}
\label{sec:intro}

\noindent
One of the fundamental problems in network data analysis is community detection which aims to divide the nodes in a network into several communities such that nodes within the same community are densely connected, and nodes from different communities are relatively sparsely connected. Identifying such communities can provide important insights on the organization of a network. For example, in social networks, communities may correspond to groups of individuals with common interests \citep{moody2003structural}; in protein interaction networks, communities may correspond to proteins that are involved in the same cellular functions \citep{spirin2003protein}. There is a vast literature on network community detection contributed from different scientific communities, such as computer science, physics, social science and statistics. We refer to \citet{fortunato2010community, fortunato2016community,zhao2017survey} for comprehensive reviews on this topic.

In the statistics literature, the majority of community detection methods are model-based, which postulate and fit a probabilistic model that characterizes networks with community structures \citep{holland1983stochastic,airoldi2008mixed,karrer2011stochastic}. Within this family, the stochastic block model \citep[SBM;][]{holland1983stochastic} is perhaps the best studied and most commonly used. The SBM is a generative model, in which the nodes are divided into blocks, or communities, and the probability of an edge between two nodes only depends on which communities they belong to and is independent across edges once given the community assignment. \textcolor{black}{Several extensions of the SBM have been considered, notably the mixed membership model \citep{airoldi2008mixed}, which allows each node to be associated with multiple clusters, and the degree corrected stochastic block model \citep[DCSBM;][]{karrer2011stochastic}, which accommodates degree heterogeneity by including additional degree parameters. Due to the rapidly increasing interests, the statistical literature on community detection in SBMs is fast growing with great advances on algorithmic solutions \citep[][among others]{snijders1997estimation, nowicki2001estimation,daudin2008mixture,karrer2011stochastic,decelle2011asymptotic, amini2013pseudo,bickel2013asymptotic} and theoretical understandings of consistency and detection thresholds \citep[][among others]{bickel2009nonparametric,rohe2011spectral,zhao2012consistency,lei2015consistency, abbe2017community, gao2017achieving,gao2018community,su2019strong,abbe2020entrywise}.
}

\textcolor{black}{It is well known that fitting the block model (i.e., SBM and DCSBM) likelihood functions is a nontrivial task, and in principle optimizing over all possible community assignments is a NP-hard problem \citep{bickel2009nonparametric}. Many work have considered using spectral clustering for community detection in SBMs, which is computationally efficient and ensures weak consistency, that is, the proportion of misclassified nodes tends to zero as the network size increases, under certain regularity conditions \citep{rohe2011spectral,lei2015consistency,joseph2016impact}. As such, spectral clustering is often used to produce initializations for methods that aim to achieve strong consistency \citep{gao2017achieving}, that is, probability of the estimated label being equal to the true label converges to one as the network size grows, and methods that aim to directly maximize the nonconvex SBM and DCSBM likelihood functions \citep{amini2013pseudo,bickel2013asymptotic}.}

\begin{figure}[!t]
\centering
\includegraphics[trim=0 5mm 0 0, scale=0.65]{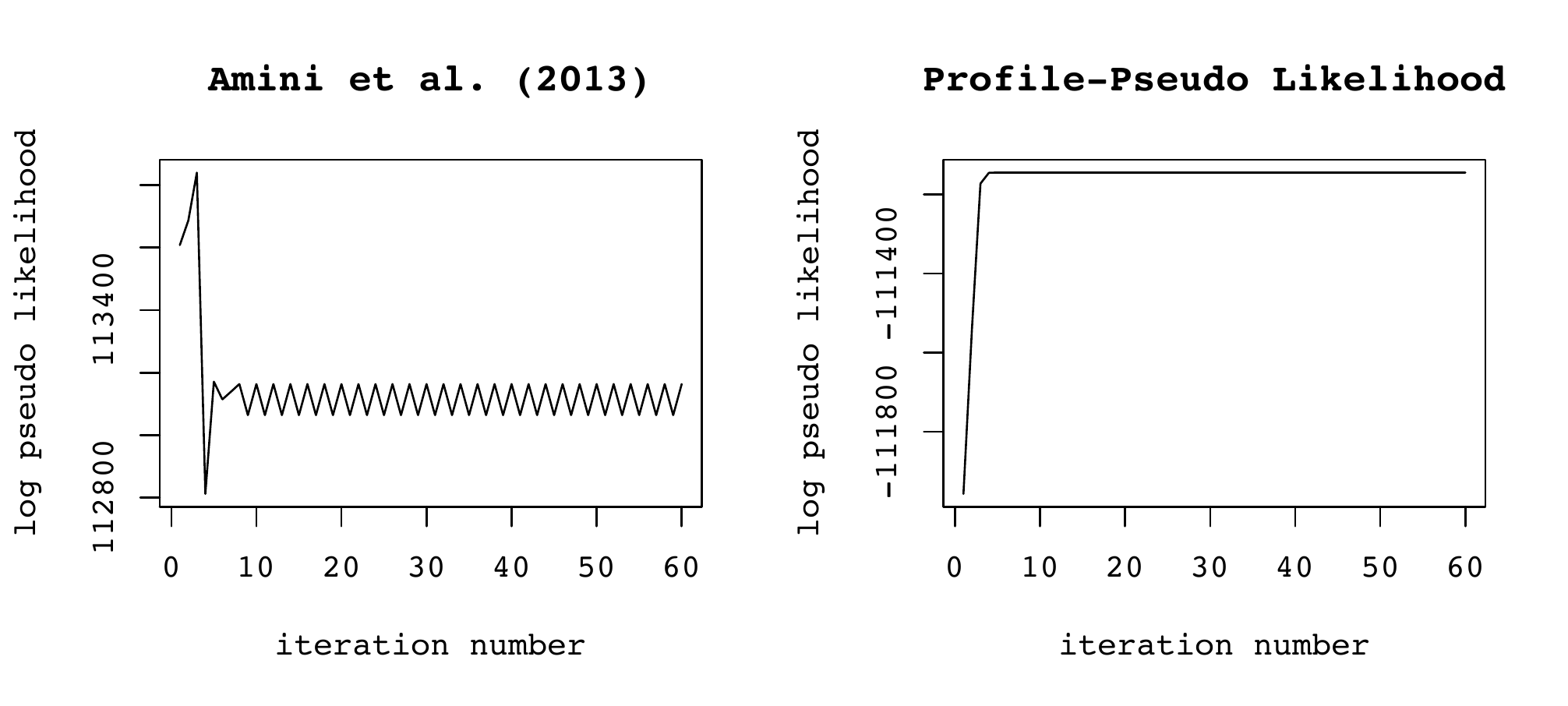}
\caption{An illustrative example comparing the pseudo likelihood method by \citet{amini2013pseudo} and the proposed profile-pseudo likelihood method. Details of the simulation setting are described in Section \ref{sim:sbm}.}
\label{fig1}
\end{figure}

{To overcome the computational challenge in fitting the SBM likelihood, \citet{amini2013pseudo} proposed a novel pseudo likelihood approach that approximates the row rums within blocks using Poisson random variables, and simplifies the likelihood function by lifting the symmetry constraint on the adjacency matrix.} This leads to a fast approximation to the block model likelihood, which subsequently enables efficient maximization that can easily handle up to millions of nodes. Additionally, it is shown that the maximum pseudo-likelihood estimator achieves (weak) community detection consistency, in the case of a sparse SBM with two communities. This pioneer work makes the SBM an attractive approach for network community detection, due to its computational scalability and theoretical properties such as the community detection consistency.
However, this method may have two drawbacks. First, in the examples that were presented in \citet{amini2013pseudo}, the authors found that empirically the pseudo-likelihood maximization algorithm converged fast. It is, however, not guaranteed that the algorithm will converge in general (see example in Figure \ref{fig1}). Convergence is a critical property as it guarantees that the final estimator exists, and is therefore important both computationally and statistically. Second, the pseudo likelihood approach may not be suitable for small and medium scale networks, as the Poisson approximation may have non negligible approximation errors in such cases.
\textcolor{black}{In the case of the DCSBM, cleverly employing the observation that the conditional distribution (on node degrees) of the Poisson variables is multinomial, \citet{amini2013pseudo} proposed a conditional pseudo likelihood approach that permits a fast estimation and adapts to both small and large scale networks. However, the algorithm still does not have convergence guarantees.}
%This, to some extent, may limit the application of the pseudo likelihood approach.

{Motivated by the pseudo likelihood approach, in this work, we propose a new SBM likelihood fitting method that decouples the membership labels of the rows and columns in the likelihood function, treating the row label as a vector of latent variables and the column label as a vector of unknown parameters.} Correspondingly, the likelihood can be maximized in an alternating fashion over the block model parameters and over the column label, where the maximization now involves a tractable sum over the distribution of latent row label.
Furthermore, we consider a profile-pseudo likelihood that adopts a hybrid framework of the profile likelihood and the pseudo likelihood, where the symmetry constraint on the adjacency matrix is also lifted. Our proposed method retains and improves on the computational efficiency of the pseudo likelihood method, performs well for both small and large scale networks and has provable convergence guarantee. We show that the community label (i.e., column label) estimated from our proposed method enjoys strong consistency, as long as the initial label has an overlap with the truth beyond that of random guessing. We further consider two extensions of the proposed method, including to the DCSBM and to the bipartite stochastic block model \citep[BiSBM;][]{larremore2014efficiently}.

Our work is closely related to a recent and growing literature on strong consistency (or exact recovery) pursuit in community detection \cite[see, for example,][]{abbe2015exact, lei2017generic, gao2017achieving,gao2018community}. The strong consistency property may be more desirable than weak consistency, as it enables establishing the asymptotic normality of the SBM plug-in estimators \citep{amini2013pseudo} and performing goodness of fit tests \citep{lei2016goodness,hu2019using}. To achieve strong consistency, the above methods usually consider a refinement step after obtaining the initial label, which is assumed to obey weak consistency. For example, in \cite{gao2017achieving}, a majority voting algorithm is applied to the clustering label obtained from spectral clustering. Similarly, our proposed profile-pseudo likelihood estimation can be viewed as a refinement on the initial label to achieve strong consistency. \textcolor{black}{Similar to other refinement algorithms, the scalability of our proposed method depends on the initialization step. While spectral clustering is used to produce initial solutions in our work, other initialization methods can be considered as well (see Section \ref{section:discuss}).
}
%One property of our proposal is that it only requires the initial label to have an overlap with the truth beyond that of random guessing, which is a weaker condition compared to weak consistency.

The rest of the paper is organized as follows.
Section \ref{section:model} introduces the profile-pseudo likelihood function and an efficient algorithm for its maximization. Moreover, we discuss the convergence guarantee of the algorithm.
Section \ref{section:theory} shows the strong consistency property of the community label estimated from the proposed algorithm.
Section \ref{section:extensions} considers two important extensions of the proposed method.
Section \ref{section:simulation} demonstrates the efficacy of the proposed method through comparative simulation studies.
Section \ref{section:examples} presents analyses of two real-world networks with communities. A discussion section concludes the paper.

\section{Profile-Pseudo Likelihood}
\label{section:model}

\noindent
Let $G(V,E)$ denote a network, where $V=\{1,2,\ldots,n\}$ is the set of $n$ nodes and $E$ is the set of edges between the nodes. The network $G(V,E)$ can be uniquely represented by the corresponding $n \times n$ adjacency matrix $\A$, where $A_{ij}=1$ if there is an edge $(i,j)\in E$ from node $i$ to node $j$ and $A_{ij}=0$ otherwise. In our work, we focus on unweighted and undirected networks, and thus $\A$ is a binary symmetric matrix. Under the stochastic block model, there are $K$ communities (or blocks) and each node belongs to only one of the communities. Let $\c=(c_{1}, c_{2}, \ldots, c_n) \in \{1, 2, \ldots, K\}^n$ denote the true community labels of the nodes, and assume that $c_i$'s are i.i.d. categorical variables with parameter vector $\bm\pi=(\pi_1,\ldots,\pi_K)$, where $\sum_k\pi_k=1$. Conditional on the community labels, the edge variables $A_{ij}$'s are independent Bernoulli variables with $\mathbb{E}(A_{ij}|\c)=P_{c_ic_j}$, where $\bm P\in[0,1]^{K\times K}$ is the symmetric edge-probability matrix with the $kl$-th entry $P_{kl}$ characterizing the probability of connection between nodes in communities $k$ and $l$. Let $\bOmega=(\bm \pi, \bm{P})$. Our objective is to estimate the unknown community labels $\c$ given the observed adjacency matrix $\A$.

Denote the rows of $\A$ as $\a_i=(A_{i1}, A_{i2}, \ldots, A_{in})$, $1\le i\le n$ and let $\e=(e_{1}, e_{2}, \ldots, e_{n})\in \{1, 2, \ldots, K\}^n$ denote the column labeling vector. Define the pseudo likelihood function as
\begin{equation}\label{apl1}
L_{\textrm{PL}}(\bOmega, \bm {e};  \{\a_i\})=\prod\limits_{i=1}^{n}\left\{\sum\limits_{l=1}^{K}\pi_{l}\prod\limits_{j=1}^{n}P_{le_{j}}^{A_{ij}}(1-P_{le_{j}})^{1-A_{ij}}\right\},
\end{equation}
with its logarithm as
\[
\ell_{\textrm{PL}}(\bOmega, \bm {e};  \{\a_i\})=\sum\limits_{i=1}^{n}\log\left\{\sum\limits_{l=1}^{K}\pi_{l}\prod\limits_{j=1}^{n}P_{le_{j}}^{A_{ij}}(1-P_{le_{j}})^{1-A_{ij}}\right\}.
\]
We make a few remarks on the objective function defined in \eqref{apl1}. First, in \eqref{apl1}, we treat the row labels as a vector of latent variables and the column labels $\e$ as a vector of unknown model parameters. That is, given $e_j$, each $A_{ij}$ is considered a mixture of $K$ Bernoulli random variables with mean $P_{le_j}$, $1\le l\le K$. This formulation decouples the row and column labels, and allows us to derive a tractable sum when optimizing for the column labels $\e$ and the block model parameter $\bOmega$.
Second, the objective function $L_{\textrm{PL}}(\bOmega, \bm {e};  \{\a_i\})$ is calculated while lifting the symmetry constraint on the adjacency matrix $\A$, or equivalently, ignoring the dependence among the rows $\a_i$'s. Hence, we refer to \eqref{apl1} as the pseudo likelihood function, which can be considered as an approximation to the SBM likelihood function.

We consider an iterative algorithm that alternates between updating $\e$ and updating $\bOmega$.
In each iteration, the estimation is carried out by first profiling out the nuisance parameter $\bOmega$ using $\max_{\bOmega}L_{\textrm{PL}}\left(\bOmega, \bm{e}; \{\a_i\}\right)$ given the current estimate of $\e$, and then maximizing the profile likelihood with respect to $\e$. This is referred to as the \textit{profile-pseudo likelihood} method. We show in Theorem \ref{thm1} the convergence guarantee of this efficient algorithm, and establish in Theorem \ref{thm2} the strong consistency of the estimated column labels $\e$.

The estimation procedure proceeds in detail as follows. First, given the current $\hat\e$ and treating the row labels as a vector of latent variables, $L_{\textrm{PL}}(\bOmega, \hat\e;  \{\a_i\})$ can be viewed as the likelihood of a mixture model with i.i.d. observations $\{\a_i\}$ and parameter $\bOmega$. Consequently, $L_{\textrm{PL}}(\bOmega,\hat\e;  \{\a_i\})$ can be maximized over $\bOmega$ using an expectation-maximization (EM) algorithm, where both the E-step and M-step updates have closed-form expressions. Next, given the estimated $\widehat\bOmega$, we update $\e$, treating $L_{\textrm{PL}}(\widehat\bOmega, \bm {e}; \{\a_i\})$ as the objective function. In this step, finding the maximizer of $L_{\textrm{PL}}(\widehat\bOmega, \bm {e}; \{\a_i\})$ with respect to $\e$ is a NP-hard problem since, in principle, it requires searching over all possible label assignments.  As an alternative, we propose a fast updating rule that leads to a non-decreasing objective function $L_{\textrm{PL}}(\widehat\bOmega, \bm {e}; \{\a_i\})$ (although not necessarily maximized), which ensures the desirable ascent property of the iterative algorithm. This algorithm is summarized in Algorithm \ref{algo1}.

In what follows, we discuss in details the profile-pseudo likelihood algorithm.
We refer to the iterations between updating $\e$ and $\bOmega$ as the outer iterations, and the iterations in the EM algorithm used to update $\bOmega$ as the inner iterations.
%In our proposed procedure, at the $(s+1)$-step of the (outer) iteration, given $\e^{(s)}$ from the previous step, we can write the pseudo likelihood as $L_{\textrm{PL}}(\bOmega, \e^{(s)};  \{\a_i\})=\sum_{\c}f(\{\a_i\},\c;\bOmega,\e^{(s)})$, where
%This formulation enables us to maximize $L_{\textrm{PL}}(\bOmega, \e^{(s)};  \{\a_i\})$ with respect to $\bOmega$ using an EM algorithm, treating $\c$ as the latent variable.
Specifically, in the $(t+1)$-th step of the EM (inner) iteration, given $\e^{(s)}$ and the parameter estimate from the previous EM update $\bOmega^{(s,t)}=(\bm\pi^{(s,t)},{\bm P}^{(s,t)})$, we {\color{black}let}
\begin{equation}\label{eqn:prob_update}
\tau_{ik}^{(s,t+1)}
%=f(c_i=k|\{\a_i\};\bOmega,\e^{(s)})
=\frac{{\pi}^{(s,t)}_{k}\prod\limits_{j=1}^{n}\left\{{{P}}^{(s,t)}_{k{e}^{(s)}_{j}}\right\}^{A_{ij}}\left\{1-{{P}}^{(s,t)}_{k{e}^{(s)}_{j}}\right\}^{1-A_{ij}}}
{\sum\limits_{l=1}^{K}{\pi}^{(s,t)}_{l}\prod\limits_{j=1}^{n}\left\{{{P}}^{(s,t)}_{l{e}^{(s)}_{j}}\right\}^{A_{ij}}\left\{1-{{P}}^{(s,t)}_{l{e}^{(s)}_{j}}\right\}^{1-A_{ij}}}
\end{equation}
for each $1\le i\le n$ and $1\le k\le K$, {\color{black} which calculates the conditional probability that the row label of node $i$ equals to $k$ at the $(t+1)$-th step of the EM iteration.}
Next, we define
\begin{eqnarray*}
& &Q(\bOmega|\bOmega^{(s,t)},\e^{(s)})=\mathbb{E}_{\z |\{\a_i\};\bm\Theta^{(s,t)},\e^{(s)}}\left\{\log f\left(\{\a_i\},\z;\bOmega,\e^{(s)}\right)\right\},
\end{eqnarray*}
where $\z$ denotes the latent row labels and
\begin{eqnarray*}
f(\{\a_i\},\z;\bOmega,\e^{(s)})=\prod\limits_{i=1}^{n}\left\{\pi_{z_{i}}\prod\limits_{j=1}^{n}P_{z_{i}e^{(s)}_{j}}^{A_{ij}}(1-P_{z_{i}e^{(s)}_{j}})^{1-A_{ij}}\right\}.
\end{eqnarray*}
In the M-step,  $\bOmega^{(s,t+1)}$ is updated by
\begin{eqnarray}
\bOmega^{(s,t+1)}=\arg\max_{\bOmega}Q(\bOmega|\bOmega^{(s,t)},\e^{(s)}),\nonumber
\end{eqnarray}
which has closed form solutions as follows
\begin{eqnarray}
\pi_{k}^{(s,t+1)}=\frac{1}{n}\sum\limits_{i=1}^{n}\tau_{ik}^{(s,t+1)}, \quad
P_{kl}^{(s,t+1)}=\frac{\sum\limits_{i=1}^{n}\sum\limits_{j=1}^{n}A_{ij}\tau_{ik}^{(s,t+1)}I(e^{(s)}_{j}=l)}{\sum\limits_{i=1}^{n}\sum\limits_{j=1}^{n}\tau_{ik}^{(s,t+1)}I(e^{(s)}_{j}=l)},
\label{eqn:theta_update}
\end{eqnarray}
for $1\le k,l\le K$.
Once the EM algorithm has converged, we let $\bOmega^{(s+1)}$ and $\left\{\tau_{il}^{(s+1)}\right\}$ take the values from the last EM update, respectively.
Next, given $\bOmega^{(s+1)}$, we propose to update $\e$ as follows:
\begin{eqnarray}
{e}^{(s+1)}_{j}=\arg\max_{k\in \{1,2,\ldots,K\}}\sum\limits_{i=1}^{n}\sum\limits_{l=1}^{K}\tau_{il}^{(s+1)}\left\{{A_{ij}}\log{P^{(s+1)}_{lk}}+(1-A_{ij})\log\left(1-P^{(s+1)}_{lk}\right) \right\}.
\label{proof2}
\end{eqnarray}
The update for $\e^{(s+1)}$ is obtained separately for each node, which can be carried out efficiently.
As we discussed earlier, this update is not guaranteed to maximize the pseudo likelihood function $L_{\textrm{PL}}(\bOmega^{(s+1)}, \bm {e}; \{\a_i\})$, which in fact is an intractable problem.
Nevertheless, it can be shown that the update in \eqref{proof2} leads to a non-negative increment in the pseudo likelihood. This gives the desirable ascent property, which we will formally state in the following theorem.
\begin{algorithm}[!t]
\caption{Profile-Pseudo Likelihood Maximization Algorithm.}
\begin{algorithmic}
\STATE \textbf{Step 1}: Initialize $\e^{(0)}$ using spectral clustering with permutations ({SCP}).
\STATE \textbf{Step 2}: Calculate $\bOmega^{(0)}=(\bm \pi^{(0)}, \bm P^{(0)})$. That is, for $1\le l,k\le K$,
\begin{eqnarray*}
&{\pi}^{(0)}_k=\frac{1}{n}\sum\limits_{i=1}^{n}I({e}^{(0)}_{i}=k),\quad
{{P}}^{(0)}_{kl}=\frac{\sum\limits_{i=1}^{n}\sum\limits_{j=1}^{n}A_{ij}I({e}^{(0)}_{i}=k)I({e}^{(0)}_{j}=l)}{\sum\limits_{i=1}^{n}\sum\limits_{j=1}^{n}I({e}^{(0)}_{i}=k)I({e}^{(0)}_{j}=l)}.
\end{eqnarray*}
\STATE \textbf{Step 3}: Initialize ${{\bOmega}}^{(0,0)}=({\bm \pi}^{(0,0)}, {{\bm P}}^{(0,0)})=({\bm \pi}^{(0)}, {{\bm P}}^{(0)})$.
\REPEAT
\REPEAT
\STATE \textbf{Step 4}: E-step: compute ${\tau}_{ik}^{(s,t+1)}$ using \eqref{eqn:prob_update} for $1\le k\le K$ and $1\le i\le n$.
\STATE \textbf{Step 5}: M-step: compute $\pi^{(s,t+1)}_k$ and $P^{(s,t+1)}_{kl}$ using \eqref{eqn:theta_update} for $1\le k,l\le K$.
\UNTIL{the EM algorithm converges.}
\STATE \textbf{Step 6}: Set $\bOmega^{(s+1)}$ and $\left\{\tau_{ik}^{(s+1)}\right\}$ to be the final EM update.
\STATE \textbf{Step 7}: Given ${{\bOmega}}^{(s+1)}$ and $\left\{\tau_{ik}^{(s+1)}\right\}$, update $e_j^{(s+1)}$, $1\le j\le n$, using \eqref{proof2}.
\UNTIL{the profile-pseudo likelihood converges.}
\end{algorithmic}\label{algo1}
\end{algorithm}

\begin{theorem}\label{thm1}
For a given initial labeling vector $\e^{(0)}$, Algorithm \ref{algo1} generates
 a sequence $\{\bOmega^{(s)}, \e^{(s)}\}$ such that
 \[
L_{\textrm{PL}}(\bOmega^{(s)}, \e^{(s)};  \{\a_i\})\leq L_{\textrm{PL}}(\bOmega^{(s+1)}, \e^{(s+1)};  \{\a_i\}).
\]
\end{theorem}
\noindent The proof of Theorem \ref{thm1} is provided in the supplemental material.
Theorem \ref{thm1} guarantees that the pseudo likelihood function is non-decreasing at each iteration in Algorithm \ref{algo1}.
Assuming that the parameter space for $\bOmega$ is compact, we arrive at the conclusion that $L_{\textrm{PL}}(\bOmega^{(s)}, \e^{(s)};  \{\a_i\})$ converges as the number of iterations $s$ increases. This is a desirable property that guarantees the stability of the proposed algorithm.
Since the pseudo likelihood function is not concave, Algorithm \ref{algo1} is not guaranteed to converge to the global optimum.
Whether it converges to a global or local solution depends on the initial value.
In practice, we find that the initialization procedure in Algorithm \ref{algo1} shows good performance, that is, we are able to achieve high clustering accuracy in our simulation studies.
To avoid local solutions in real data applications, we recommend considering multiple random initializations in addition to the initialization in Algorithm \ref{algo1}.

Finally, we summarize the differences between our proposal and the method in \citet{amini2013pseudo}.
Both our method and \citet{amini2013pseudo} consider algorithms that iterate through two parameter updating steps, namely, the step that updates the block model parameter $\bOmega$ using EM and the step that updates the membership label.
However, the likelihood function is treated very differently in these two methods.
As the row and column labels are enforced to be the same in \citet{amini2013pseudo}, a Poisson approximation is needed in the pseudo likelihood calculation.
%While the Poisson approximation enables a faster calculation in the case of a DCSBM, it may not work well for small and medium scale networks.
The label $\e$ in \citet{amini2013pseudo} is treated as an initial in the EM estimation, and its value is assigned heuristically in each iteration.
As such, the resulting procedure is not guaranteed to converge, as seen in Figure \ref{fig1}.
In comparison, our method decouples the row and column labels (i.e., $\z$ and $\e$), and does not require a Poisson approximation in the pseudo likelihood calculation.
When updating the column labels $\e$, we use $L_{\textrm{PL}}(\widehat\bOmega, \bm {e}; \{\a_i\})$ as the objective function that guides our updating routine. The proposed node-wise update enjoys the ascent property, which subsequently guarantees the convergence of the algorithm (see Theorem \ref{thm1}).
We also remark that due to the differences in our problem formulation, our theoretical analysis is nontrivial and new technical tools are needed. %While \citet{amini2013pseudo} only establishes weak consistency, our analysis establishes strong consistency of the estimated community label.

\section{Consistency Results}\label{section:theory}
\noindent
In this section, we investigate the strong consistency of the estimator obtained from one outer loop iteration (i.e., updating the column labels $\e$) of Algorithm \ref{algo1}, denoted as $\hat\c\{\e^{(0)}\}$, where $\e^{(0)}$ is an initial of Algorithm \ref{algo1}.
We first consider strong consistency in the case of SBMs with two balanced communities, and then extend our strong consistency result to SBMs with $K$ communities.

We first present the consistency result for directed SBMs with two communities, fitted to directed networks, and then modify the result to handle the more challenging case of undirected SBMs, fitted to undirected networks.
To separate the cases of directed and undirected SBMs, we adopt different notations for the corresponding adjacency matrices and edge-probability matrices.
First, for a directed SBM, we denote the adjacency matrix as $\widetilde\A$ and assume that its entries $\widetilde{A}_{ij}$'s are mutually independent given $\c$, that is,
\begin{eqnarray}
\textrm{(directed) } \widetilde{A}_{ij}|\c\sim \textrm{Bernoulli}(\widetilde{P}_{c_{i}c_{j}}), \,\,\textrm{for}\,\, 1\le i,j\le n.
\label{dirMM}
\end{eqnarray}
For an undirected SBM, we denote the adjacency matrix as $\A$ and assume its entries ${A}_{ij}$'s, $i\leq j$, are mutually independent given $\c$, that is,
\begin{eqnarray}
\textrm{(undirected) } A_{ij}|\c\sim \textrm{Bernoulli}(P_{c_{i}c_{j}})\textrm{ and }A_{ij}=A_{ji},\textrm{ for}\,\,1\le i\leq j\le n.
\label{undirMM}
\end{eqnarray}
Furthermore, we assume that the edge-probability matrix of the directed SBM has the form
\begin{eqnarray}
\widetilde{\bm P}=\frac{1}{m}\left(
  \begin{array}{cc}
    a & b\\
    b & a\\
  \end{array}
\right),
\label{dirP}
\end{eqnarray}
while that of the undirected SBM has the form
\begin{eqnarray}
\bm P=\frac{2}{m}\left(
    \begin{array}{cc}
     a & b\\
     b & a\\
    \end{array}
    \right)
    -\frac{1}{m^{2}}\left(
    \begin{array}{cc}
     a^{2} & b^{2}\\
     b^{2} & a^{2}\\
    \end{array}
    \right).
\label{undirP}
\end{eqnarray}
Such a coupling between the directed and undirected models makes it possible to extend the consistency result of the directed SBM to the undirected case.

Given an initial labeling vector $\e^{(0)}$, estimates $\hat a$, $\hat b$ and $(\hat\pi_1,\hat\pi_2)$,
the estimator $\hat\c\{\e^{(0)}\}$ can be written as
\begin{eqnarray}
\hat{c}_{j}\{\e^{(0)}\}=\arg\max_{k\in \{1,2\}}\sum\limits_{i=1}^{n}\sum\limits_{l=1}^{2}\hat{\tau}_{il}\left\{ A_{ij}\log(\widehat{P}_{lk})+(1-A_{ij})\log(1-\widehat{P}_{lk} \big)\right\},
\label{cjhat}
\end{eqnarray}
where $\hat{\tau}_{il}$ is defined as in \eqref{eqn:prob_update}, $\bm {\widehat{P}}$ is defined as in \eqref{dirP} for directed SBMs and as in \eqref{undirP} for undirected SBMs, with $a$ and $b$ replaced by $\hat{a}$ and $\hat{b}$, respectively.
Here the estimates $\hat a$, $\hat b$ and $(\hat\pi_1,\hat\pi_2)$ are outputs from the inner loop (i.e., EM) iterations, and are in effect initials for the outer loop calculation.
Consistency of the inner loop (i.e., EM) outputs $\hat a$, $\hat b$ and $(\hat\pi_1,\hat\pi_2)$ can be established using the result in \citet{amini2013pseudo}.
%The same proof strategy can be used to establish consistency of the inner loop outputs in Algorithm \ref{algo1}.
In our theoretical analysis, we focus our efforts on establishing strong consistency of the column labels $\e$ estimated in the outer loop, given that the outer loop initials satisfy $(\hat{a}, \hat{b}) \in \mathcal{P}_{a,b}^{\delta}$ in \eqref{Pab} and $\hat \pi_{1}=\hat \pi_{2}=1/2$.

For SBMs with two balanced communities, we make the following assumption:
\begin{enumerate}[({{\color{blue}A}})]
\item Assume that each community contains $m=n/2$ nodes and $\hat \pi_{1}=\hat \pi_{2}=1/2$.
\end{enumerate}
The assumption that $\hat \pi_{1}=\hat \pi_{2}=1/2$ is reasonable as the inner loop outputs $(\hat \pi_{1},\hat \pi_{2})$ are consistent estimators of $(\pi_1,\pi_2)=(1/2,1/2)$, as shown in \citet{amini2013pseudo}. Without loss of generality, let $c_{i}=1$ for $i \in \{1, \ldots, m\}$, and $c_{i}=2$ for $i \in \{m+1, \ldots, n\}$.
Assume that $\e^{(0)}\in \{1, 2\}^n$ assigns equal numbers of nodes to the two communities, i.e., the initial labeling vector is balanced.
Let $\e^{(0)}$ match with the truth on $\gamma m$ labels in each of the two communities for some $\gamma \in (0, 1)$.
We assume $\gamma m$ to be an integer.
Next, let $\mathcal{E}^{\gamma}$ denote the set that collects all such initial labeling vectors, i.e.,
\begin{eqnarray}
\mathcal{E}^{\gamma}=\left\{{\e^{(0)}}\in \{1, 2\}^{n} : \sum\limits_{i=1}^{m}I(e^{(0)}_{i}=1)=\gamma m,\sum\limits_{i=m+1}^{n}I(e^{(0)}_{i}=2)=\gamma m\right\}.\nonumber
\end{eqnarray}
Note that $\gamma=1/2$ corresponds to ``no correlation" between $\e^{(0)}$ and $\c$, whereas $\gamma=0$ and $\gamma=1$ both correspond to perfect correlation.
In our analysis, we do not require knowing the value of $\gamma$, or knowing which labels are matched.
In Theorem \ref{thm2}, we show that the amount of overlap $\gamma$ can be any value, as long as $\gamma\neq 1/2$.
Our goal is to establish strong consistency for $\hat{\c}\{\e^{(0)}\}$. %uniformly over $\e^{(0)}\in\mathcal{E}^{\gamma}$. \textcolor{blue}{(Is the term ``uniformly'' appropriate?)}
For a constant $\delta>1$, we define $\mathcal{P}_{a,b}^{\delta}$ as follows:
\begin{equation}\label{Pab}
\mathcal{P}_{a,b}^{\delta}=\left\{(\hat{a}, \hat{b}): \frac{\hat{a}}{\hat{b}}I(a>b)+\frac{\hat{b}}{\hat{a}}I(a<b)\geq\delta\right\}.
\end{equation}
The set $\mathcal{P}_{a,b}^{\delta}$ specifies that $(\hat a,\hat b)$ has the same ordering as $(a,b)$, and the relative difference between the estimates $\hat a$ and $\hat b$ is lower bounded. Our next theorem considers the collection of estimates $(\hat a, \hat b)$ in $\mathcal{P}_{a,b}^{\delta}$.

\begin{theorem}\label{thm2}
Assume (A) holds, $\delta>1$, $\gamma \in (0, 1)\backslash\{\frac{1}{2}\}$ and \textcolor{black}{$\frac{(a-b)^{2}}{(a+b)}\ge C\log n$ for a sufficiently large constant $C>0$}. For a directed SBM in \eqref{dirMM} with the edge-probabilities given by \eqref{dirP} with $a\neq b$, we have that for any $\epsilon >0$, there exists $N>0$ such that for all $n\geq N$, the following holds
\begin{eqnarray}
\textcolor{black}{\mathbb{P}\Bigg\{\bigcap_{(\hat{a}, \hat{b}) \in \mathcal{P}_{a,b}^{\delta}}\hat{\c}\{\e^{(0)}\}=\c\Bigg\}}
\geq 1-\left\{ne^{-\frac{(a-b)^{2}-4(a-b)\epsilon +4\epsilon^{2}}{4(a+b)}}+n(n+2)e^{-\frac{(2\gamma-1)^{2}(a-b)^2}{8(a+b)}}\right\},\nonumber
\end{eqnarray}
for any $\e^{(0)} \in \mathcal{E}^{\gamma}$, where $\hat{\c}\{\e^{(0)}\}=\c$ means that they belong to the same equivalent class of label permutations.
\end{theorem}
\noindent
The proof of Theorem \ref{thm2} is provided in the supplemental material.
It can be seen from Theorem \ref{thm2} that the one-step estimate $\hat{\c}\{\e^{(0)}\}$ for a directed SBM is a strongly consistent estimate of $\c$ for any ${\e^{(0)}}\in\mathcal{E}^{\gamma}$.
%\textcolor{red}{In order for Theorem \ref{thm2} to hold uniformly for all $\e^{(0)}$, similar to that in \citet{amini2013pseudo}, the initial community label $\e^{(0)}$ needs to be independent of the data under our current proof strategy. This is due to a technical argument used in establishing strong consistency; see more discussion in Section \ref{section:discuss}.}
%The condition $\frac{(a-b)^{2}}{(a+b)\,\mathrm{log}\,n} \rightarrow \infty$ holds, if $(a,b)$ is set to be $(a'n^{\alpha},b'n^{\alpha})$, for some $\alpha\in (0,1)$, where $a'$ and $b'$ are two fixed constants.
Note that weak consistency was established in \cite{amini2013pseudo} under the assumption that $\frac{(a-b)^{2}}{(a+b)} \rightarrow \infty$.
\textcolor{black}{In comparison, our result requires $\frac{(a-b)^{2}}{(a+b)}\ge C\log n$ to establish strong consistency.
In existing literature on strong consistency, the condition $\frac{\lambda_{n}}{\mathrm{log}n}\rightarrow \infty$ is often commonly imposed \citep{bickel2009nonparametric, zhao2012consistency}, where $\lambda_n$ denotes the average network degree.
Specifically, under the SBM setting considered in \citet{bickel2009nonparametric} and \citet{zhao2012consistency}, we have that $a-b\asymp \lambda_n$ and $a+b\asymp \lambda_n$, where $\asymp$ denotes that the two quantities on both sides are of the same order. In this case, $\frac{\lambda_{n}}{\mathrm{log}n}\rightarrow \infty$ implies $\frac{(a-b)^{2}}{(a+b)}\ge C\log n$ for any constant $C>0$.
}

\textcolor{black}{
Theorem \ref{thm2} guarantees strong consistency for any $\e^{(0)}\in \mathcal{E}^{\gamma}$. In comparison, the weak consistency in \cite{amini2013pseudo} holds uniformly for all $\e^{(0)}\in \mathcal{E}^{\gamma}$, even if it is derived from the data. Indeed, $\e^{(0)}$ is usually derived from data using initialization procedures such as the spectral clustering.
For the strong consistency result to apply, one may consider a data splitting strategy following the method in \citet{li2020network}.
Specifically, we may sample a proportion of the node pairs to produce an initial value $\e^{(0)}$ and estimate $\hat\c(\e^{(0)})$ using the rest of the node pairs. In this case, $\e^{(0)}$ is independent of the data used for community detection and the result in Theorem 2 can be used to ensure strong consistency of $\hat\c(\e^{(0)})$. In our numerical studies, for simplicity we did not use data splitting, while the simulation results show that the proposed method still performs well.  We also note that Theorem \ref{thm2} can be adapted to hold uniformly for all $\e^{(0)}\in \mathcal{E}^{\gamma}$, if stronger conditions are placed on $\gamma$ and $a,b$.
Specifically, if the misclassification ratio of $\e^{(0)}$ is, for example, $O(1/(a+b))$ and the condition on $a,b$ is strengthen to $(a-b) \gtrsim \sqrt{n \log n}$ (i.e., average degree is at least of order $\sqrt{n \log n}$), then strong consistency in Theorem \ref{thm2} holds uniformly for all such $\e^{(0)}$, even if it is derived from the data. This can be shown by combining the union bound argument and a Stirling approximation that gives $\log \left(\begin{array}{c}n \\ n_{\gamma}\end{array}\right)\le n \log (en/n_{\gamma})$, where $n_{\gamma}$ is the number of misclassified nodes.
The misclassification ratio of $O(1/(a+b))$ imposed above is known to hold with high probability for spectral clustering (see, for example, Corollary 3.2 in \cite{lei2015consistency}).
}

Next, we consider the case of undirected SBMs.
Let {$a_{\gamma}=\big[(1-\gamma)a+\gamma b\big]I(\gamma>\frac{1}{2})+\big[\gamma a+(1-\gamma)b\big]I(\gamma<\frac{1}{2})$}.
We have the following result on the strong consistency of $\hat\c\{\e^{(0)}\}$.

\begin{theorem}\label{thm3}
Assume (A) holds, $\delta>1$, $\gamma \in (0, 1)\backslash\{\frac{1}{2}\}$ and \textcolor{black}{$\frac{(a-b)^{2}}{(a+b)}\ge C\log n$ for a sufficiently large constant $C>0$}.
For an undirected SBM in \eqref{undirMM} with the edge-probabilities given by \eqref{undirP} with $2(1+\epsilon)a_{\gamma}\leq \epsilon |(1-2\gamma)(a-b)|$ for some $\epsilon\in(0, 1)$, there exist  $\rho\in(0,1)$ and $N>0$, such that for all $n\geq N$, the following holds
\begin{eqnarray*}
&&\textcolor{black}{\mathbb{P}\Bigg\{\bigcap_{(\hat{a}, \hat{b}) \in \mathcal{P}_{a,b}^{\delta}}\hat{\c}\{\e^{(0)}\}=\c\Bigg\}}\\
&\geq& 1-\left[3ne^{-\frac{(\frac{1-\rho}{4})^{2}(a-b)^{2}}{4(a+b)}}+n(n+2)\left\{
e^{-\frac{(\frac{1-\epsilon}{2})^{2}(2\gamma-1)^{2}(a-b)^{2}}{4(a+b)}}+2e^{-\frac{\epsilon^{2}/2}{1+\epsilon/2}a_{\gamma}}\right\}\right],
\end{eqnarray*}
for any $\e^{(0)} \in \mathcal{E}^{\gamma}$, where $\hat{\c}\{\e^{(0)}\}=\c$ means that they belong to the same equivalent class of label permutations.
\end{theorem}
\noindent
The proof of Theorem \ref{thm3} is provided in the supplemental material.
It can be seen that the one-step estimate $\hat{\c}\{\e^{(0)}\}$ for an undirected SBM is a strongly consistent estimate of $\c$, for any $\e^{(0)} \in \mathcal{E}^{\gamma}$.
Given $\epsilon$ and $\gamma$, the condition $2(1+\epsilon)a_{\gamma}\leq \epsilon |(1-2\gamma)(a-b)|$ places an upper bound on $b/a$.
For example, for $\epsilon=\frac{1}{3}$ and $\gamma<\frac{1}{10}$, the above condition is satisfied if $b/a\leq (1-10\gamma)/(9-10\gamma)$.

\textcolor{black}{Strong consistency can be more desirable than weak consistency, as it enables normal distribution based inference and goodness of fit tests (see numerical studies in Section \ref{section:inferencesim}). For example, consider a SBM with $K=2$, $\bm \pi=(\pi_{1}, \pi_{2})$ and true community labels $\c=(c_{1}, c_{2}, \ldots, c_{n})$.
Suppose we can construct a label vector $\hat{\c}^{(\text{w})}$ such that $\{\hat{c}_{i}^{(\text{w})}\}_{i=1}^{n}$ are independent with $\mathbb{P}(\hat{c}_{i}^{(\text{w})}\neq c_{i})=2p_{n}$ for $c_i=1$ and $\mathbb{P}(\hat{c}_{i}^{(\text{w})}\neq c_{i})=p_{n}$ for $c_i=2$, where $p_{n}=1/\log n$.
Then it can be shown that $\hat{\c}^{(\text{w})}$ is weakly consistent, with a misclassification ratio of $O_p(1/\log n)$, but not strongly consistent to $\bm c$. Let $\hat\pi^{\text{w}}_1=\sum\limits_{i=1}^{n}I(\hat{c}_{i}^{(\text{w})}=1)/n$. It holds that $\sqrt{n}\left\{\hat\pi^{\text{w}}_1-\left(\pi_{1}+\frac{1-3\pi_{1}}{\log n}\right)\right\}\stackrel{d}{\longrightarrow} N\left\{0, \pi_{1}(1-\pi_{1})\right\}$ (See the proof in the supplemental material). Thus the bias term of $\hat\pi^{\text{w}}_1$ is $O(1/\log{n})$, which can be non negligible for inference.  On the other hand, for a strongly consistent estimator $\hat{\bm c}^{(\text{s})}=\left(\hat{c}_{1}^{(\text{s})}, \hat{c}_{2}^{(\text{s})}, \ldots, \hat{c}_{n}^{(\text{s})}\right)$, letting $\hat\pi^{\text{s}}_1=\sum\limits_{i=1}^{n}I(\hat{c}_{i}^{(\text{s})}=1)/n$, it holds that
$\sqrt{n}\left\{\hat\pi^{\text{s}}_1-\pi_{1}\right\}\stackrel{d}{\longrightarrow} N\left\{0, \pi_{1}(1-\pi_{1})\right\}$.}

Next, we consider the more general case of directed and undirected SBMs with $K$ communities. Similar to Assumption (A), we make the following assumption:

\begin{enumerate}[({{\color{blue}B}})]
\item Assume that each community contains $m=n/K$ nodes and $\hat\pi_{k}=1/K$.
\end{enumerate}
Let the edge-probability matrix of the directed SBM be
\begin{eqnarray}
\widetilde{P}_{kl}=\frac{a}{m}1(k=l)+\frac{b}{m}1(k\neq l),
\label{dirPK}
\end{eqnarray}
and that of the undirected SBM be
\begin{eqnarray}
P_{kl}=\left(\frac{2a}{m}-\frac{a^{2}}{m^{2}}\right)1(k=l)+\left(\frac{2b}{m}-\frac{b^{2}}{m^{2}}\right)1(k\neq l),
\label{undirPK}
\end{eqnarray}
for $k, l=1, \ldots, K$.
Without loss of generality, let $c_{i}=k$ for $i \in \left\{(k-1)m+1, \ldots, km\right\}$ for $k=1, \ldots, K$.
Let $\mathcal{E}^{\gamma}$ denote the set that collects all initial labeling vectors such that
\begin{eqnarray}
\mathcal{E}^{\gamma}=\left\{{\e^{(0)}}\in \{1, \dots, K\}^{n} : \sum\limits_{i=(k-1)m+1}^{km}I(e^{(0)}_{i}=k)=\gamma_{k} m,\sum\limits_{i=1}^{n}I(e^{(0)}_{i}=k)=m, \,\,\, k=1, \dots, K\right\},\nonumber
\end{eqnarray}
where $\gamma=(\gamma_{1}, \ldots, \gamma_{K})$.% We assume that $\hat\pi_{k}=1/K$, $k=1,\ldots,K$.
Corollaries \ref{cor1} and \ref{cor2} establish the strong consistency of profile pseudo likelihood estimators for directed and undirected SBMs, respectively.

\begin{cor}\label{cor1}
Assume (B) holds, $\delta>1$, $\min\left\{\gamma_{1}, \gamma_{2}, \ldots, \gamma_{K}\right\} \in (\frac{1}{2}, 1)$
%\textcolor{blue}{(Double check if the lower bound is still 1/2?)}
and \textcolor{black}{$\frac{(a-b)^{2}}{(a+b)}\ge C\log n$ for a sufficiently large constant $C>0$}. For a directed SBM in \eqref{dirMM} with the edge-probabilities given by \eqref{dirPK} with $a\neq b$, we have that for each $\epsilon >0$, there exists $N>0$ such that for all $n\geq N$, the following holds
\begin{eqnarray}
&&\textcolor{black}{\mathbb{P}\Bigg\{\bigcap_{(\hat{a}, \hat{b}) \in \mathcal{P}_{a,b}^{\delta}}\hat{\c}\{\e^{(0)}\}=\c\Bigg\}}\\
&\geq& 1-\left\{(K-1)ne^{-\frac{(a-b)^{2}-4(a-b)\epsilon +4\epsilon^{2}}{4(a+b)}}+\frac{(10K-8)n^{2}}{K}\sum\limits_{k=1}^{K}\sum\limits_{l=1}^{K}e^{-\frac{(\gamma_{k}+\gamma_{l}-1)^{2}(a-b)^2}{8(a+b)}}\right\},\nonumber
\end{eqnarray}
for any $\e^{(0)} \in \mathcal{E}^{\gamma}$, where $\hat{\c}\{\e^{(0)}\}=\c$ means that they belong to the same equivalent class of label permutations.
\end{cor}
\begin{cor}\label{cor2}
Assume (B) holds, $\delta>1$, $\min\left\{\gamma_{1}, \gamma_{2}, \ldots, \gamma_{K}\right\} \in (\frac{1}{2}, 1)$ and \textcolor{black}{$\frac{(a-b)^{2}}{(a+b)}\ge C\log n$ for a sufficiently large constant $C>0$}. For an undirected SBM in \eqref{undirMM} with the edge-probabilities given by \eqref{undirPK} with $2(1+\epsilon)a_{\gamma_{k}}\leq \epsilon (\gamma_{k}+\gamma_{l}-1)(a-b)$ for all $1\le k, l\le K$ and some $\epsilon\in(0, 1)$, where $a_{\gamma_{k}}=(1-\gamma_{k})a+\gamma_{k}b$, there exist  $\rho\in(0,1)$ and $N>0$, such that for all $n\geq N$, the following holds
\begin{eqnarray*}
&&\textcolor{black}{\mathbb{P}\Bigg\{\bigcap_{(\hat{a}, \hat{b}) \in \mathcal{P}_{a,b}^{\delta}}\hat{\c}\{\e^{(0)}\}=\c\Bigg\}}\\
&\geq& 1-\left[3(K-1)ne^{-\frac{(\frac{1-\rho}{4})^{2}(a-b)^{2}}{2(a+b)}}+\frac{(10K-8)n^{2}}{K}\sum\limits_{k=1}^{K}\sum\limits_{l=1}^{K}\left\{
e^{-\frac{(\frac{1-\epsilon}{2})^{2}(\gamma_{k}+\gamma_{l}-1)^{2}(a-b)^{2}}{6(a+b)}}+2e^{-\frac{3\epsilon^{2}a_{\gamma_{k}}}{8(4+\epsilon)}}\right\}\right],
\end{eqnarray*}
for any $\e^{(0)} \in \varepsilon_{n}^{\gamma}$, where $\hat{\c}\{\e^{(0)}\}=\c$ means that they belong to the same equivalent class of label permutations.
\end{cor}
\noindent
The proofs of Corollaries \ref{cor1} and \ref{cor2} follow very similar steps as in the proofs of Theorems \ref{thm2} and \ref{thm3}, respectively.
We omit presenting the details.

\section{Extensions}\label{section:extensions}

\noindent
In this section, we study two useful extensions of the proposed method. First, we consider the case of fitting the degree corrected stochastic block model using the proposed profile-pseudo likelihood method.
Second, we consider the case of fitting the bipartite stochastic block model using the proposed profile-pseudo likelihood method (see Section \ref{section:bisbm} in the supplemental material).

It has often been observed that real-world networks exhibit high degree heterogeneity, with a few nodes having a large number of connections and the majority of the rest having a small number of connections. The stochastic block model, however, cannot accommodate such degree heterogeneity. To incorporate the degree heterogeneity in community detection, \citet{karrer2011stochastic} proposed the degree-corrected SBM. Specifically, conditional on the label vector $\c$, it is assumed that the edge variables $A_{ij}$ for all $i\leq j$ are mutually independent Poisson variables with
\[
\mathbb{E}[A_{ij}|\c]=\theta_{i}\theta_{j}\lambda_{c_ic_j},
\]
where $\bm \Lambda=[\lambda_{kl}]$ is a $K\times K$ symmetric matrix and $\btheta=(\theta_{1}, \theta_{2}, \ldots, \theta_{n})$ is a degree parameter vector, with the additional constraint $\sum\limits_{i=1}^{n}\theta_{i}/n=1$ that ensures identifiability \citep{zhao2012consistency}.

\begin{algorithm}[!t]
\caption{DCSBM Profile-Pseudo Likelihood Maximization Algorithm.}
\begin{algorithmic}
\STATE \textbf{Step 1}: Initialize $\e^{(0)}$ using spectral clustering with permutations (SCP).
\STATE \textbf{Step 2}: Calculate $\bOmega^{(0)}=(\bm \pi^{(0)}, \bLambda^{(0)}, \btheta^{(0)})$. That is, for $1\le l,k\le K$, $1\le i\le n$,
\begin{eqnarray*}
&{\pi}^{(0)}_k=\frac{1}{n}\sum\limits_{i=1}^{n}I({e}^{(0)}_{i}=k),\quad
{\theta}^{(0)}_i\propto d_i,\quad
\lambda^{(0)}_{kl}=\frac{\sum\limits_{i=1}^{n}\sum\limits_{j=1}^{n}A_{ij}I({e}^{(0)}_{i}=k)I({e}^{(0)}_{j}=l)}{\sum\limits_{i=1}^{n}\sum\limits_{j=1}^{n}I({e}^{(0)}_{i}=k)I({e}^{(0)}_{j}=l){\theta}^{(0)}_i{\theta}^{(0)}_j}.
\end{eqnarray*}
\STATE \textbf{Step 3}: Initialize ${{\bOmega}}^{(0,0)}=({\bm \pi}^{(0,0)}, \bLambda^{(0,0)},\btheta^{(0,0)})=({\bm \pi}^{(0)}, \bLambda^{(0)},\btheta^{(0)})$.
\REPEAT
\REPEAT
\STATE \textbf{Step 4}: E-step: compute $\tau_{ik}^{(s,t+1)}$ using \eqref{eqn:tau2} for $1\le k\le K$ and $1\le i\le n$.
\textcolor{black}{\STATE \textbf{Step 5}: CM-step: compute $\bm\pi^{(s,t+1)}$, $\bLambda^{(s,t+1)}$, $\btheta^{(s,t+1)}$.
For $1\le k,l\le K$, set
\[
\pi^{(s,t+1)}_k=\sum\limits_{i=1}^{n}\tau^{(s,t+1)}_{ik}/n, \quad \lambda^{(s,t+1)}_{kl}=\frac{\sum\limits_{i=1}^{n}\sum\limits_{j=1}^{n}\tau^{(s,t+1)}_{ik}I(e^{(s)}_{j}=l)A_{ij}}
{\sum\limits_{i=1}^{n}\sum\limits_{j=1}^{n}\tau^{(s,t+1)}_{ik}I(e^{(s)}_{j}=l)\theta^{(s,t)}_{i}\theta^{(s,t)}_{j}},
\]
Letting $g_{ij}^{(s,t+1)}=\sum\limits_{k,l=1}^{K}\tau^{(s,t+1)}_{ik}I(e^{(s)}_{j}=l)\lambda_{kl}^{(s,t+1)}$, for $1\le i\le n$, set
\[
\theta_{i}^{(s,t+1)}=\left(-h_{i}^{(s,t+1)}+\sqrt{{h_{i}^{(s,t+1)}}^{2}+8d_{i}g_{ii}^{(s,t+1)}}\right)\Big/{4g_{ii}^{(s,t+1)}},
\]
where $h_{i}^{(s,t+1)}=\sum\limits_{j=1}^{i-1}\theta_{j}^{(s,t+1)}g_{ij}^{(s,t+1)}+\sum\limits_{j=i+1}^{n}\theta_{j}^{(s,t)}g_{ij}^{(s,t+1)}$.
%{\color{red} Note that the above $\bm\pi^{(s,t+1)}$, $\bLambda^{(s,t+1)}$, $\btheta^{(s,t+1)}$ are obtained by maximizing the objective function with a sequence of alternating optimization.}
}
\UNTIL{the ECM algorithm converges.}
\STATE \textbf{Step 6}: Set $\bOmega^{(s+1)}$ to be the final ECM update.
\STATE \textbf{Step 7}: Given ${{\bOmega}}^{(s+1)}$, update $e_j^{(s+1)}$, $1\le j\le n$, using
\begin{eqnarray*}
&e^{(s+1)}_{j}=\arg\max_{k\in \{1,2,\ldots,K\}}\sum\limits_{i=1}^{n}\sum\limits_{l=1}^{K}\left\{-\theta^{(s+1)}_{i}
\theta^{(s+1)}_{j}\lambda^{(s+1)}_{lk}+A_{ij}\log(\lambda^{(s+1)}_{lk})\right\}\tau^{(s+1)}_{il}.
\end{eqnarray*}
\UNTIL{the profile-pseudo likelihood converges.}
\end{algorithmic}\label{algo2}
\end{algorithm}

Define $\bOmega=(\bm \pi, \bm \Lambda,\btheta)$. To fit the DCSBM to an observed adjacency matrix $\A$, we define the following log pseudo likelihood function
\[
\ell^{\textrm{DC}}_{\textrm{PL}}(\bOmega, \bm{e};  \{\a_i\})=\sum\limits_{i=1}^{n}\log\left\{\sum\limits_{l=1}^{K}\pi_{l}\prod\limits_{j=1}^{n}e^{-\theta_{i}\theta_{j}\lambda_{le_{j}}}(\theta_{i}\theta_{j}\lambda_{le_{j}})^{A_{ij}}\right\}.
\]
Let $d_i=\sum_{j=1}^nA_{ij}$, $1\le i\le n$. A profile-pseudo likelihood algorithm that maximizes $\ell^{\textrm{DC}}_{\textrm{PL}}(\bOmega, \bm{e};  \{\a_i\})$ is described in Algorithm \ref{algo2}.
{\color{black} At step 4, we update the conditional probabilities for the row labels by}
\begin{equation}\label{eqn:tau2}
{\color{black} \tau_{ik}^{(s,t+1)}
=\frac{\prod\limits_{j=1}^{n}\pi^{(s,t)}_{k}e^{-\theta^{(s,t)}_{i}\theta^{(s,t)}_{j}\lambda^{(s,t)}_{k e^{(s)}_{j}}}\left\{\theta^{(s,t)}_{i}\theta^{(s,t)}_{j}\lambda^{(s,t)}_{k e^{(s)}_{j}}\right\}^{A_{ij}}}
{\sum\limits_{l=1}^{K}\prod\limits_{j=1}^{n}\pi^{(s,t)}_{l}e^{-\theta^{(s)}_{i}\theta^{(s,t)}_{j}\lambda^{(s,t)}_{l e^{(s)}_{j}}}\left\{\theta^{(s,t)}_{i}\theta^{(s,t)}_{j}\lambda^{(s,t)}_{l e^{(s)}_{j}}\right\}^{A_{ij}}}.}
\end{equation}
{\color{black} At step 5, we update the parameters by sequentially solving the following optimization problems:
\begin{eqnarray*}
&&(\bm \pi^{(s,t+1)}, \bm \Lambda^{(s,t+1)})=\arg\max_{(\bm \pi, \bm \Lambda)}Q(\bm \pi, \bm \Lambda, \bm \theta^{(s,t)}|\bOmega^{(s,t)},\e^{(s)}),\\
&&\theta_{i}^{(s,t+1)}=\arg\max_{\theta_{i}}Q(\bm \pi^{(s,t+1)}, \bm \Lambda^{(s,t+1)}, \theta_{1}^{(s,t+1)},\ldots, \theta_{i-1}^{(s,t+1)}, \theta_{i}, \theta_{i+1}^{(s,t)}, \ldots, \theta_{n}^{(s,t)}|\bOmega^{(s,t)},\e^{(s)}).
\end{eqnarray*}
Here, the objective function $Q(\bOmega|\bOmega^{(s,t)},\e^{(s)})$ is defined as
\begin{eqnarray*}
& &Q(\bOmega|\bOmega^{(s,t)},\e^{(s)})=\mathbb{E}_{\z |\{\a_i\};\bm\Theta^{(s,t)},\e^{(s)}}\left\{\log f\left(\{\a_i\},\z;\bOmega,\e^{(s)}\right)\right\},
\end{eqnarray*}
where $\z=(z_{1},\cdots,z_{n})^\top$ denotes the row label vector and
\begin{eqnarray*}
f(\{\a_i\},\z;\bOmega,\e^{(s)})=\prod\limits_{i=1}^{n}\left[\pi_{z_{i}}\prod\limits_{j=1}^{n}e^{-\theta_{i}\theta_{j}\lambda_{z_{i}e_{j}^{(s)}}}\frac{\left\{\theta_{i}\theta_{j}\lambda_{z_{i}e_{j}^{(s)}}\right\}^{A_{ij}}}{A_{ij}!}\right].
\end{eqnarray*}
}The inner loop of Algorithm \ref{algo2}, {\color{black}i.e., steps 4 and 5,} is different from that in Algorithm \ref{algo1}, as it considers a conditional EM (ECM) update.
Specifically, the objective function {\color{black}$Q(\bOmega|\bOmega^{(s,t)},\e^{(s)})$} in the M-step,  {\color{black}i.e., step 5,} which solves for block parameters $\lambda_{kl}$'s and degree parameters $\theta_i$'s, is nonconvex and does not have closed form solutions. Hence, directly optimizing it using numerical techniques can be computationally costly and is not ensured to find the global optimum. The ECM algorithm replaces the challenging optimization problem in the M-step with a sequence of alternating updates, each of which has a closed-form solution.
It is easy to implement and enjoys the desirable ascent property \citep{meng1993maximum}. Consequently, Algorithm 2 has convergence guarantees, which improves over \cite{amini2013pseudo}.

\textcolor{black}{We also note that in our profile-pseudo likelihood approach, while the conditional distribution (on node degrees) of the Poisson variables is multinomial, the multinomial coefficient (i.e., the $\frac{d_i!}{b_{i1}!b_{i2}!\cdots b_{iK}!}$ factorial term) in the density function involves the column labels (in $b_{ik}$'s). As such, optimizing for the column labels in the outer loop becomes highly challenging. In Algorithm 2, we work with the pseudo likelihood without conditioning on node degrees and it requires estimating the degree parameters in the M-step.  This is different from that in \cite{amini2013pseudo}.}
%In the case of DCSBM, while the pseudo likelihood approach in \citet{amini2013pseudo} may not have convergence guarantee, it is computationally attractive as it avoids estimating the node-specific degree parameters and adapts to both small and large scale networks.

\section{Simulation Studies}\label{section:simulation}
\noindent
In this section, we carry out simulation studies to investigate the finite sample performance of our proposed profile-pseudo likelihood method (referred to as \texttt{PPL}), and to compare with existing solutions including the spectral clustering with permutations (referred to as \texttt{SCP}) and the pseudo likelihood method (referred to as \texttt{PL}) proposed in \cite{amini2013pseudo}. {Both \texttt{SCP} and \texttt{PL} are implemented using the code provided by \cite{amini2013pseudo}.}
\textcolor{black}{We also compare with the strongly consistent majority voting method proposed in \cite{gao2017achieving} (see Section \ref{sec:add} in the supplemental material).}

We consider two evaluation criteria. The first one is the normalized mutual information (NMI), which measures the distance between the true labeling vector and an estimated labeling vector. The NMI takes values between 0 and 1, and a larger value implies a higher accuracy.
The second one is the CPU running time, which measures the computational cost.
\textcolor{black}{Note the reported running time does not include the initialization step (see Section \ref{sec:add} in the supplemental material and discussions in Section \ref{section:discuss}).}
All methods are implemented in Matlab and run on a single processor of an Intel(R) Core(TM) i7-4790 CPU 3.60 GHz PC.

\subsection{SBM}
\label{sim:sbm}
\noindent
In this section, we simulate networks from SBMs. Three different settings are considered.
In Setting 1, we evaluate the convergence of \texttt{PPL} and \texttt{PL}; in Setting 2, we compare the performance of \texttt{PPL}, \texttt{SCP} and \texttt{PL} when the networks are small and dense; in Setting 3, we compare the three methods when the networks are large and sparse.

\begin{figure}[!t]
\centering
\includegraphics[trim=0 5mm 0 0, scale=0.45]{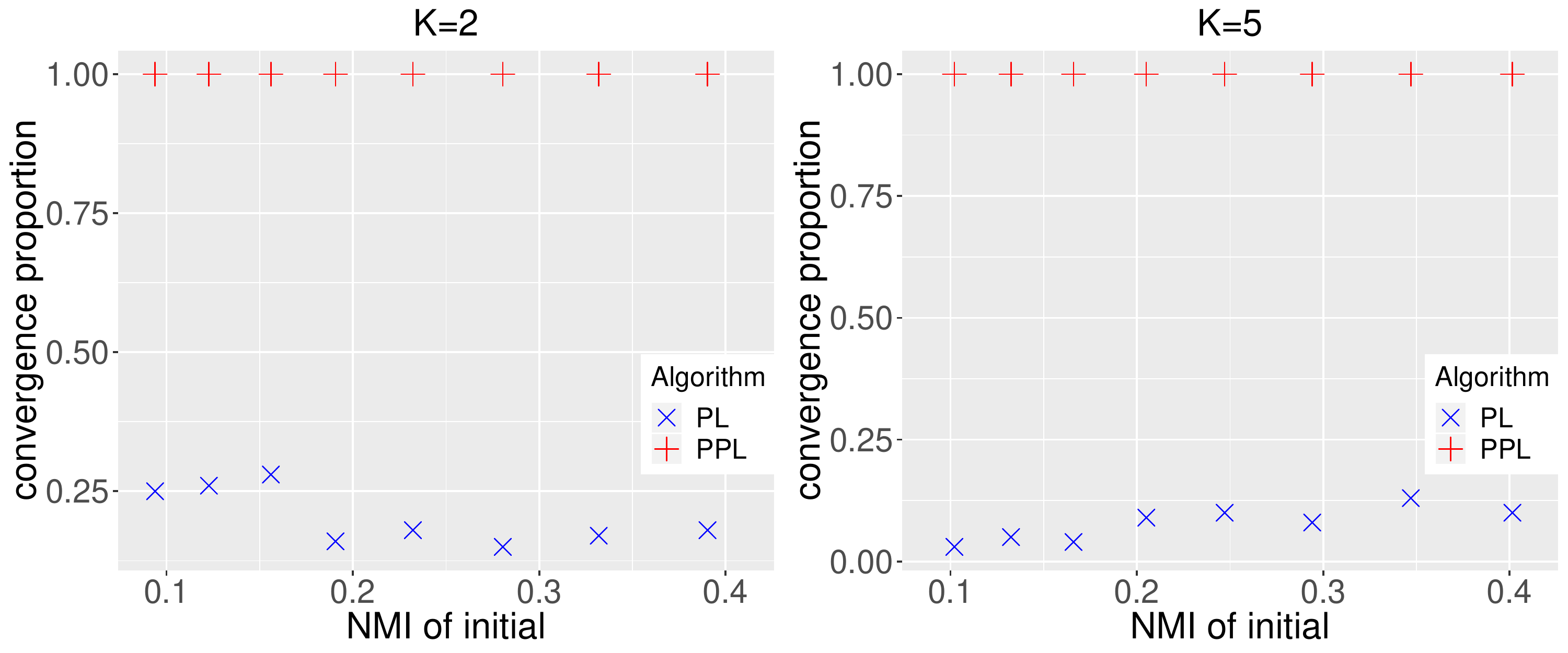}
\caption{Proportion of convergence of \texttt{PPL} and \texttt{PL} with initial labels of varying NMI. }
\label{fig2}
\end{figure}

\bigskip\noindent
\textbf{Setting 1}: In this simulation, we evaluate the convergence performance of \texttt{PPL} and \texttt{PL} with varying initial labeling vectors.
We simulate from SBMs with $n=500$ nodes that are divided into $K$ equal sized communities, and the within/between community connecting probabilities are $P_{kl}=p_1+p_2\times1(k=l), k,l=1,\ldots, K$.
We consider $(K, p_1, p_2)=(2, 0.13, 0.07)$, and $(K, p_1, p_2)=(5, 0.10, 0.13)$.
Both the \texttt{PPL} and \texttt{PL} algorithms are considered to have converged if the change of the latest update (relative to the previous one) is less than $10^{-6}$ or if the number of outer iterations exceeds 60.
We let the NMI of the initial labeling vector vary from 0.1 to 0.5.
All simulations are repeated 100 times.
The proportion of convergence for \texttt{PPL} and \texttt{PL} are presented in Figure \ref{fig2}.
It is seen that the \texttt{PL} does not have a satisfactory convergence performance.
One example (in the case of $K=2$) of the convergence of \texttt{PPL} and non-convergence of \texttt{PL} is shown in Figure \ref{fig1}, where it is observed that the \texttt{PL} algorithm did not converge, and the final estimate has a smaller log pseudo likelihood when compared to the initial value.

\begin{figure}[!t]
\centering
\includegraphics[trim=0 5mm 0 0, scale=0.35]{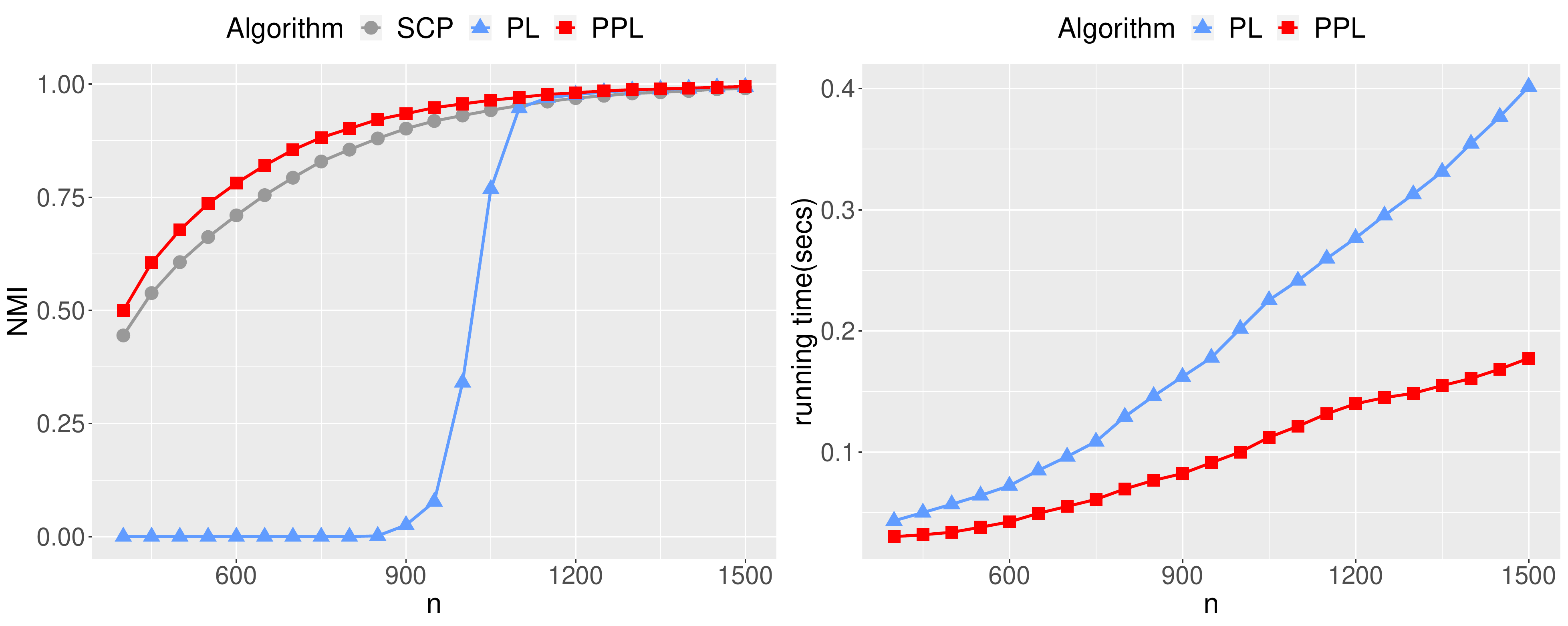}
\caption{NMI and computing time of \texttt{PPL} and \texttt{PL} with varying network size $n$.}
\label{fig3}
\end{figure}

\bigskip\noindent
\textbf{Setting 2}: In this simulation, we compare the performance of \texttt{SCP}, \texttt{PL}, and \texttt{PPL} on small scale and dense networks. The \texttt{PL} method is not expected to perform well in this setting due to the relatively large Poisson approximation error. We acknowledge that many networks in real applications are large and/or sparse, and we note that here we use simulated examples to investigate a limitation of the \texttt{PL} method.
We simulate from SBMs with $n$ nodes that are divided into $K=2$ equal sized communities, and the within/between community connecting probabilities are $P_{kl}=p_1+p_2\times1(k=l), k,l=1,\ldots,K$.
We consider $(p_1, p_2)=(0.84, 0.06)$.
Both \texttt{PPL} and \texttt{PL} are initialized by \texttt{SCP}.
Figure \ref{fig3} reports the NMI from the three methods based on 100 replications.
It is seen that \texttt{PPL} outperforms the \texttt{PL} both in terms of community detection accuracy (when $n<1000$) and computational efficiency.
The unsatisfactory performance of the \texttt{PL} method when $n<1000$ is due to the errors from approximating binomial random variables with Poisson random variables. This approximation is not expected to work well when $p_1$ (or $p_2$) is large and when $n$ is small \citep{hodges1960poisson}.
{Also note that the \texttt{PL} method may perform worse than the initial labels, as its iterations do not enjoy the ascent property.}
It can also be seen that as $n$ increases, the performance of \texttt{PL} improves notably.

\begin{figure}[!t]
\centering
\includegraphics[scale=0.55]{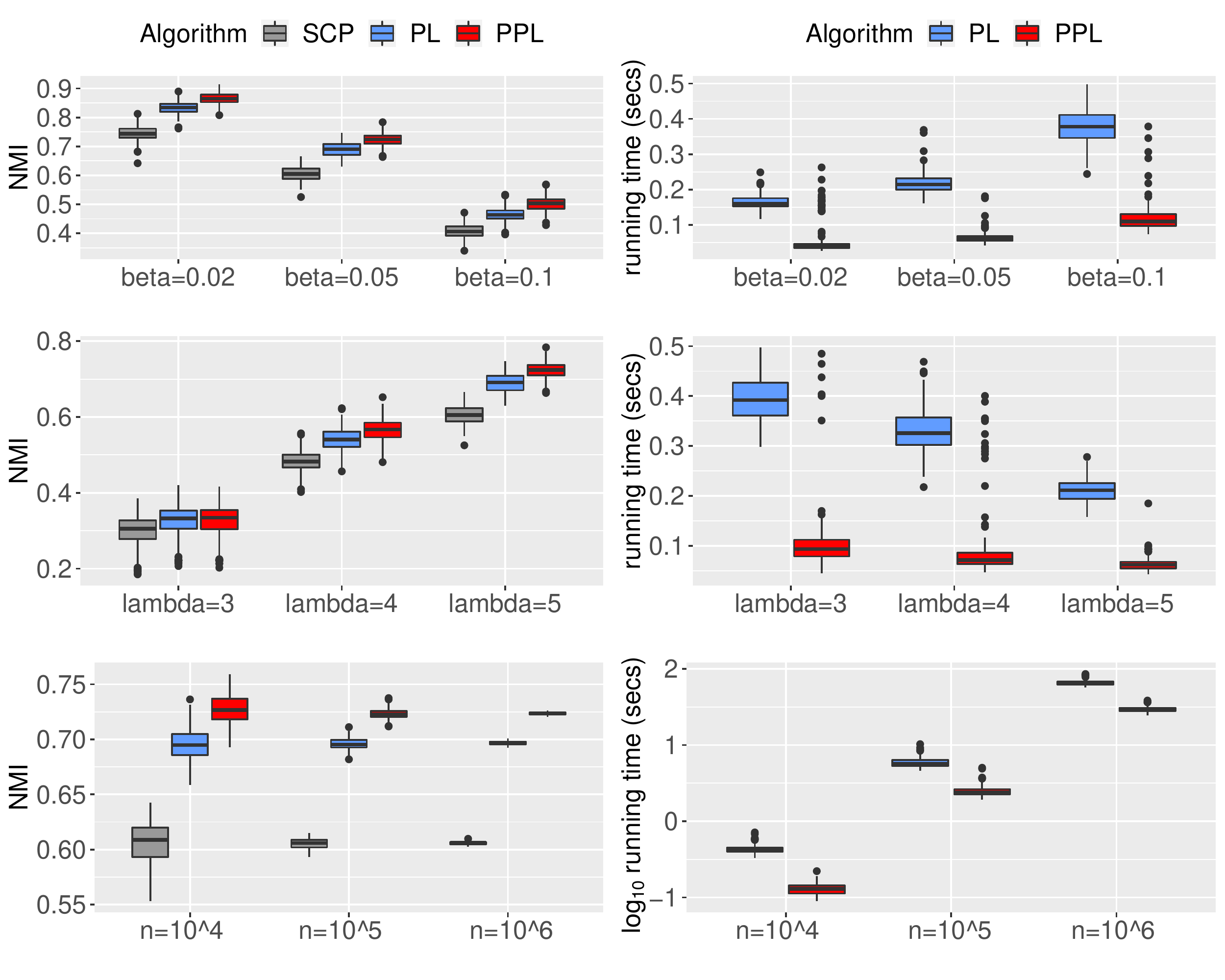}
\caption{Comparisons of the NMI and computing time from \texttt{SCP}, \texttt{PL} and \texttt{PPL} under different settings. The three rows correspond to the following three scenarios respectively: 1) varying $\beta$ while setting $\lambda=5$ and $n=4000$, 2) varying $\lambda$ while setting $\beta=0.05$, and $n=4000$, and 3) varying $n$ while setting $\lambda=5$ and $\beta=0.05$.}
\label{fig4}
\end{figure}

\bigskip\noindent
\textbf{Setting 3}: In this simulation, we compare the performance of \texttt{SCP}, \texttt{PL}, and \texttt{PPL} on large-scale and sparse networks.
We consider similar simulation settings as in \cite{amini2013pseudo}.
As in \cite{decelle2011asymptotic}, the edge-probability matrix $\bm{P}$ is controlled by the following two parameters: the ``out-in-ratio" $\beta$, varying from $0$ to $0.2$, and the weight vector $\omega$, determining the relative degrees within communities. We set $\omega=(1,1,1)$. Once $\beta=0$, $\bm{P}^*$ is set to be a diagonal matrix $\textrm{diag}(\omega)$, while otherwise we set the diagonal elements of $\bm{P}^*$ to be $\beta^{-1}\omega$ and set all the off-diagonal ones to $1$. Then, the overall expected network degree is set to be $\lambda$, which varies from 3 to 5. Finally, we re-scale $\bm{P}^*$ to obtain this expected degree, giving the resulting $\bm{P}$ as follows:
\begin{eqnarray}
\bm{P}=\frac{\lambda}{(n-1)(\bm\pi^{T}\bm{P}^*\bm\pi)}\bm{P}^*,
\label{P}
\end{eqnarray}
which generates sparse networks, since $P_{kl}=O(1/n)$.
In this simulation study, both \texttt{PL} and \texttt{PPL} are initialized by \texttt{SCP}.
We let $K=3$ and $\bm \pi=\left(0.2, 0.3, 0.5\right)$. We consider three scenarios: 1) varying $\beta$ while setting $\lambda=5$ and $n=4000$, 2) varying $\lambda$ while setting $\beta=0.05$, and $n=4000$, and 3) varying $n$ while setting $\lambda=5$ and $\beta=0.05$. Figure \ref{fig4} reports the NMI from the three methods and the computing time from \texttt{PPL} and \texttt{PL}, based on 100 replications.
\textcolor{black}{We note the reported running times for \texttt{PPL} and \texttt{PL} do not include the initialization step. For comparison, when $\lambda=5$, $\beta=0.05$ and $n=10^6$, the SCP initialization step takes less than 100 seconds (see Section \ref{sec:add} in the supplemental material).}
It is seen that \texttt{PPL} outperforms both \texttt{SCP} and \texttt{PL} in terms of community detection accuracy. Moreover, \texttt{PPL} consistently outperforms \texttt{PL} in terms of computational efficiency.

\subsection{Goodness of fit test and normality of plug-in estimators}\label{section:inferencesim}
\begin{figure}[!t]
\centering
\includegraphics[scale=0.375, trim=0 1cm 0 0]{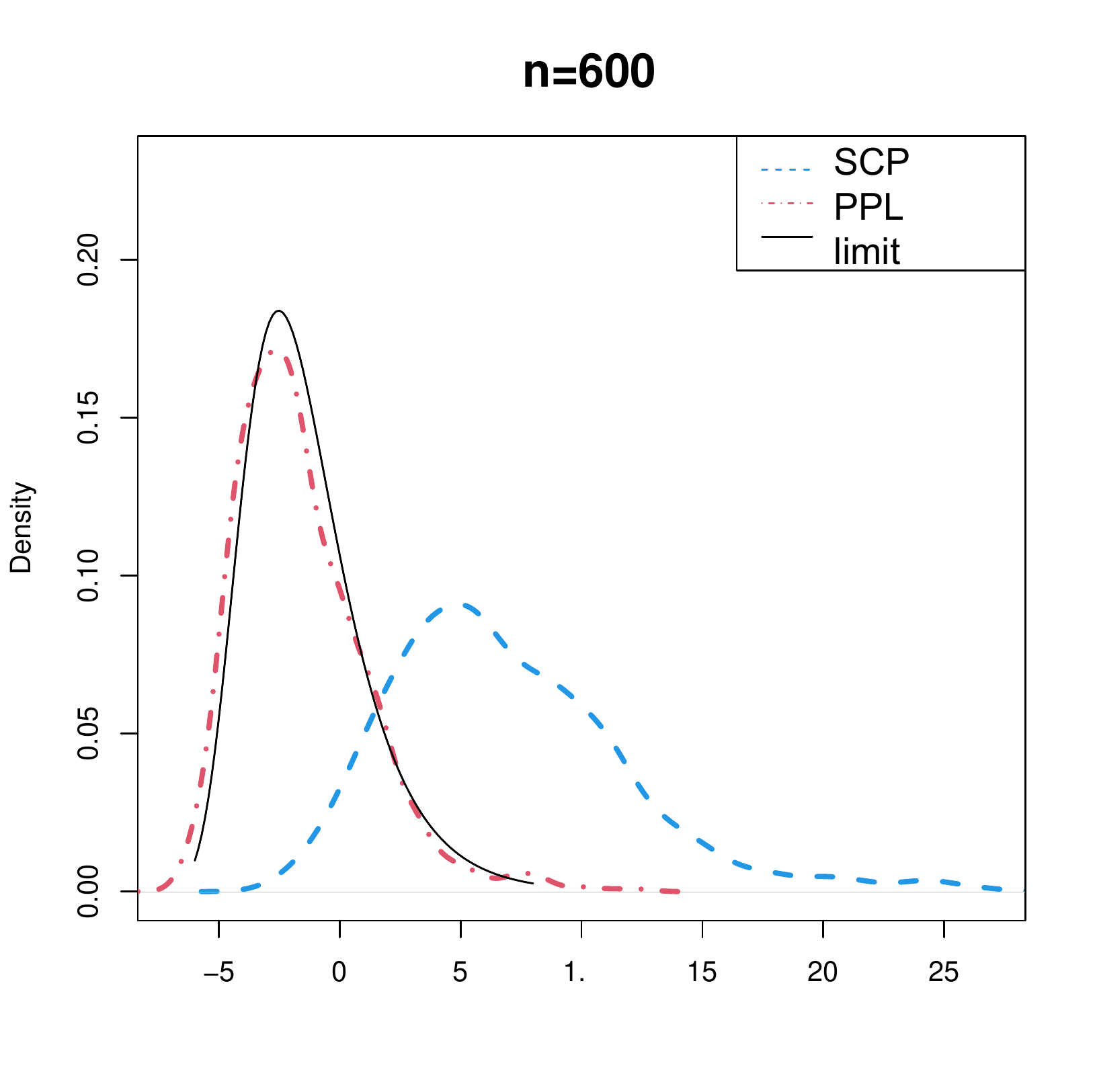}
\includegraphics[scale=0.375, trim=0 1cm 0 0]{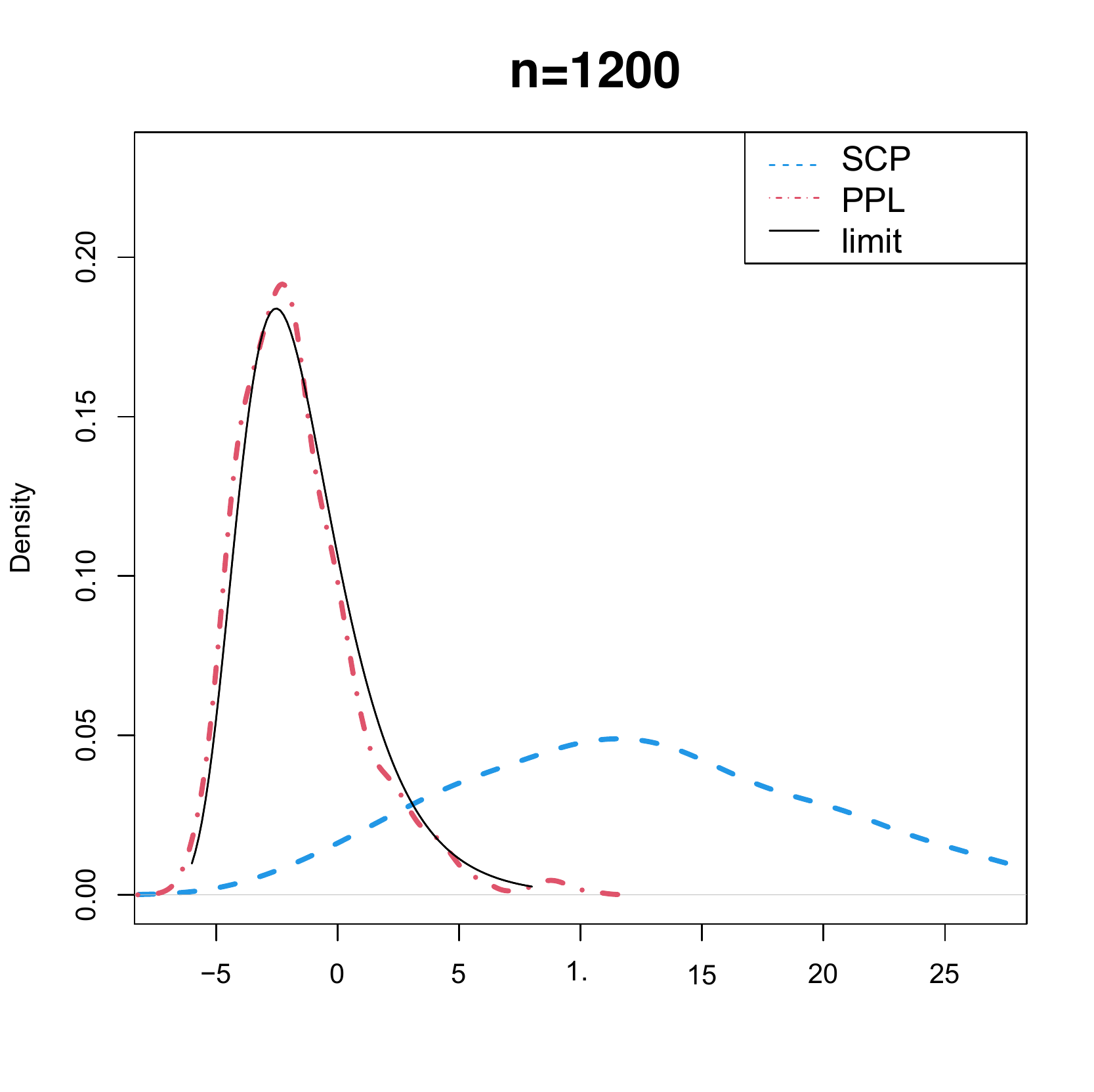}
\caption{\textcolor{black}{Null densities of the test statistic with $n = 600$ (left plot) and $n=1200$ (right plot). The blue dashed lines, red dash-dotted lines and black solid lines show the densities under \texttt{SCP}, \texttt{PPL} and the theoretical limit, respectively.}}
\label{fig:gumbel}
\end{figure}

\noindent
\textcolor{black}{To evaluate goodness of fit, we consider the maximum entry-wise deviation based testing procedure in \citet{hu2019using}. The authors showed that the distribution of the test statistic, denoted by $T_n$ and calculated with a strongly consistent community label, converges to a Gumbel distribution. In this simulation study, we consider a SBM with $K=3$, $\bm\pi=(0.2,0.3,0.5)$, and $P_{kl}=0.12+0.08\times I(k=l)$, and investigate the distribution of $T_n$ calculated using estimates from \texttt{PPL} and \texttt{SCP} respectively. The results over 1000 replications are shown in Figure \ref{fig:gumbel}. It is seen that
the sample null distribution of $T_n$ calculated with \texttt{PPL} is very close to the limiting distribution while that calculated with \texttt{SCP} deviates from the limit considerably.
This is due to that $T_n$ in \citet{hu2019using} is calculated based on maximum entry-wise deviation and as such, the misclassified nodes in \texttt{SCP}, albeit not many, may much inflate the test statistic.
With the refinement of \texttt{PPL}, the test statistic is seen to have a sample null distribution close to the theoretical limit, ensuring a well-controlled test size.}

\begin{figure}[!t]
\centering
\includegraphics[scale=0.415]{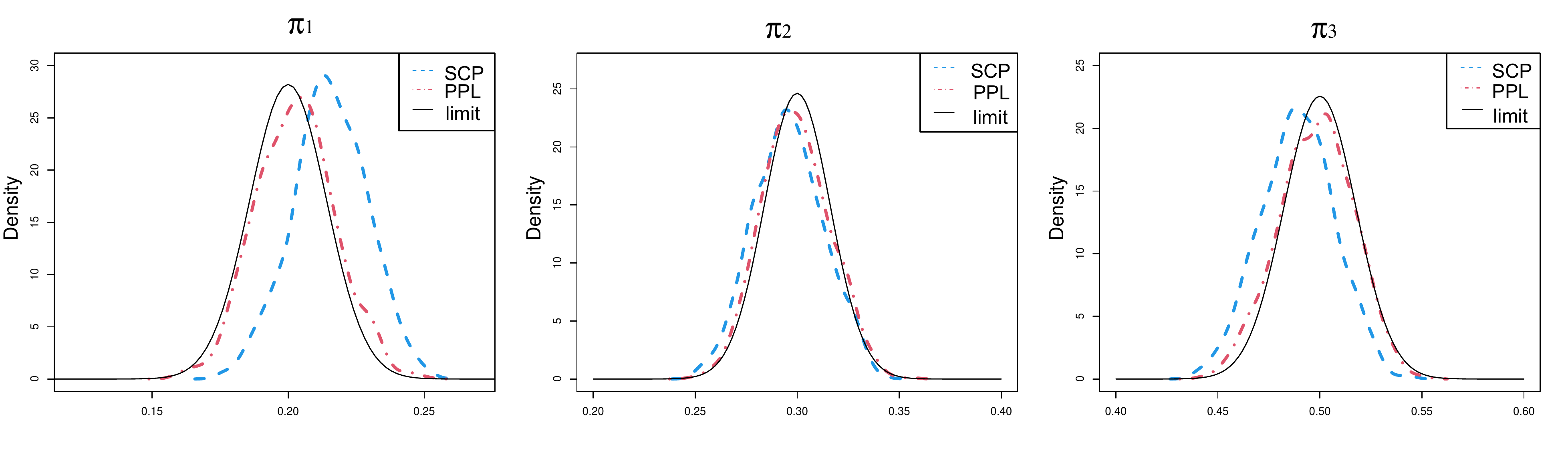}
\caption{\textcolor{black}{Empirical distributions of $\hat\pi_1$, $\hat\pi_2$ and $\hat\pi_3$. The blue dashed lines, red dash-dotted lines and black solid lines show the densities under \texttt{SCP}, \texttt{PPL} and the theoretical limit, respectively.}}
\label{fig:normal}
\end{figure}

\textcolor{black}{To examine normality of plug-in estimators, we consider a SBM with $K=3$, $\bm\pi=(0.2,0.3,0.5)$, $P_{kl}=0.12+0.08\times I(k=l)$ and $n=800$. We consider the empirical distribution of $\hat\pi_1$, $\hat\pi_2$ and $\hat\pi_3$ calculated using labels produced by \texttt{PPL} and \texttt{SCP}, respectively. The results over 1000 replications are shown in Figure \ref{fig:normal}.
It is seen that the empirical distributions calculated with \texttt{PPL} are very close to the limiting distributions while those calculated with \texttt{SCP} deviate, especially for $\hat\pi_1$ and $\hat\pi_3$, from the theoretical limits.
}

\subsection{DCSBM}\label{sec:sim5.2}
\noindent
In this section, we evaluate the performance of the profile-pseudo likelihood method under the DCSBM, referred to as \texttt{DC-PPL}.
{We fix $K=3$, $n=1200$, $\bm \pi=\left(0.2, 0.3, 0.5\right)$ and let $\bm P=10^{-2}\times\left[J_{K,K}+\text{diag}(2, 3, 4)\right]$}, where $J_{K,K}$ is a $K$ by $K$ matrix where every element is equal to one. The degree parameters $\{\theta_{i}\}_{i=1}^{n}$ are generated from \citep{zhao2012consistency}, i.e.,
$$
\mathbb{P}\left(\theta_{i}=m x\right)=\mathbb{P}\left(\theta_{i}=x\right)=1/2\quad\mathrm{with} \,\, x=\frac{2}{m+1},
$$
which ensures that $E(\theta_{i})=1$. We consider $m=2, 4, 6$.
Given $\c$ and $\btheta$, the edge variables $A_{ij}$'s are independently generated from a Bernoulli distribution with parameters $\theta_{i}\theta_{j}P_{c_{i}c_{j}}$, $1\le i\leq j\le n$.

\begin{figure}[!t]
\centering
\includegraphics[trim=0 10mm 0 0,height=0.24\textwidth, width=0.9\textwidth]{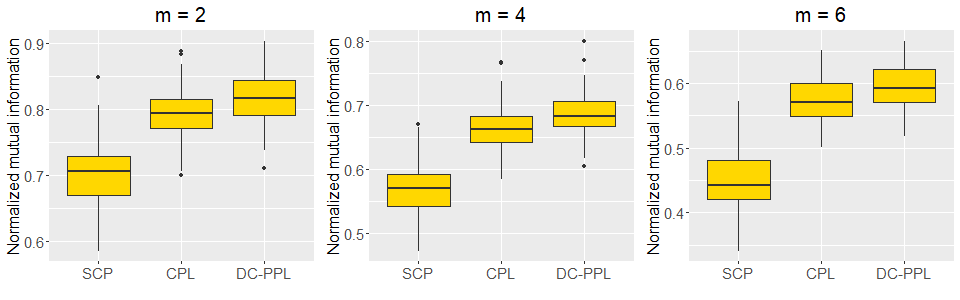}
\caption{Comparison of \texttt{SCP}, \texttt{CPL}, \texttt{DC-PPL} under DCSBM with varying $m$.}
\label{sim4.1}
\end{figure}

We compare \texttt{DC-PPL} with \texttt{SCP} as well as \texttt{CPL}, an extension of \texttt{PL} proposed for networks with degree heterogeneity in \cite{amini2013pseudo}. The results are summarized in Figure \ref{sim4.1}, based on 100 replications. We can see both \texttt{DC-PPL} and \texttt{CPL} outperform \texttt{SCP}, and \texttt{DC-PPL} performs better than \texttt{CPL} in terms community detection accuracy.

\section{Real-world Data Examples}\label{section:examples}
\subsection{Political blogs data}
\noindent
In this subsection, we apply our proposed method to the network of political blogs collected by \cite{Adamic2005The}. The nodes in this network are blogs on US politics and the edges are hyper-links between these blogs with directions removed. This data set was collected right after the 2004 presidential election and demonstrates strong divisions. In \cite{Adamic2005The}, all the blogs were manually labeled as liberal or conservative, and we take these labels as the ground truth. As in \citet{zhao2012consistency}, we focus on the largest connected component of the original network, which contains 1,222 nodes, 16,714 edges and has the average degree of approximately 27.

\begin{figure}[!t]
\centering
\subfigure[True]{{}%[{\bf $60\%$ quantile}]{\label{fig:subfig:a}
\includegraphics[width=0.305 \linewidth]{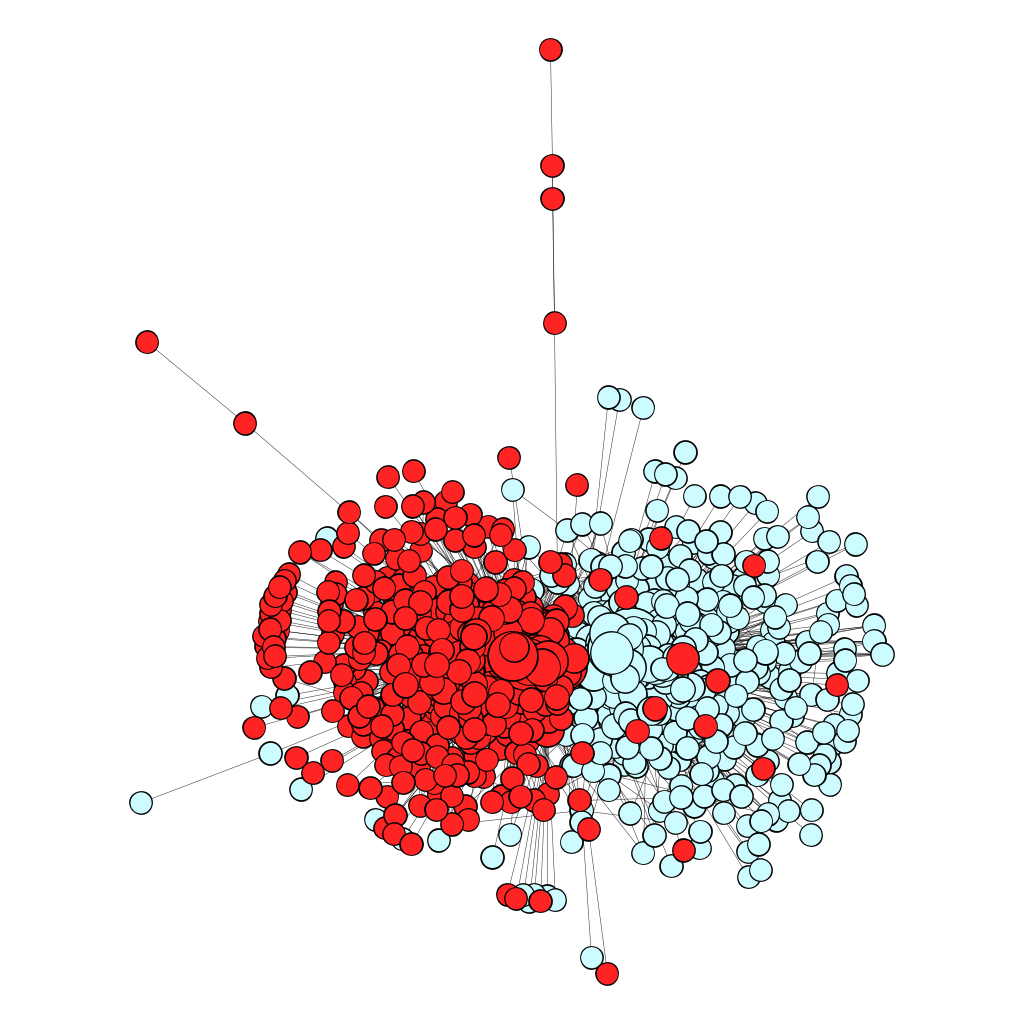}}
\hspace{0.01\linewidth}
\subfigure[PL]{{}%[{\bf $60\%$ quantile}]{\label{fig:subfig:a}
\includegraphics[width=0.305 \linewidth]{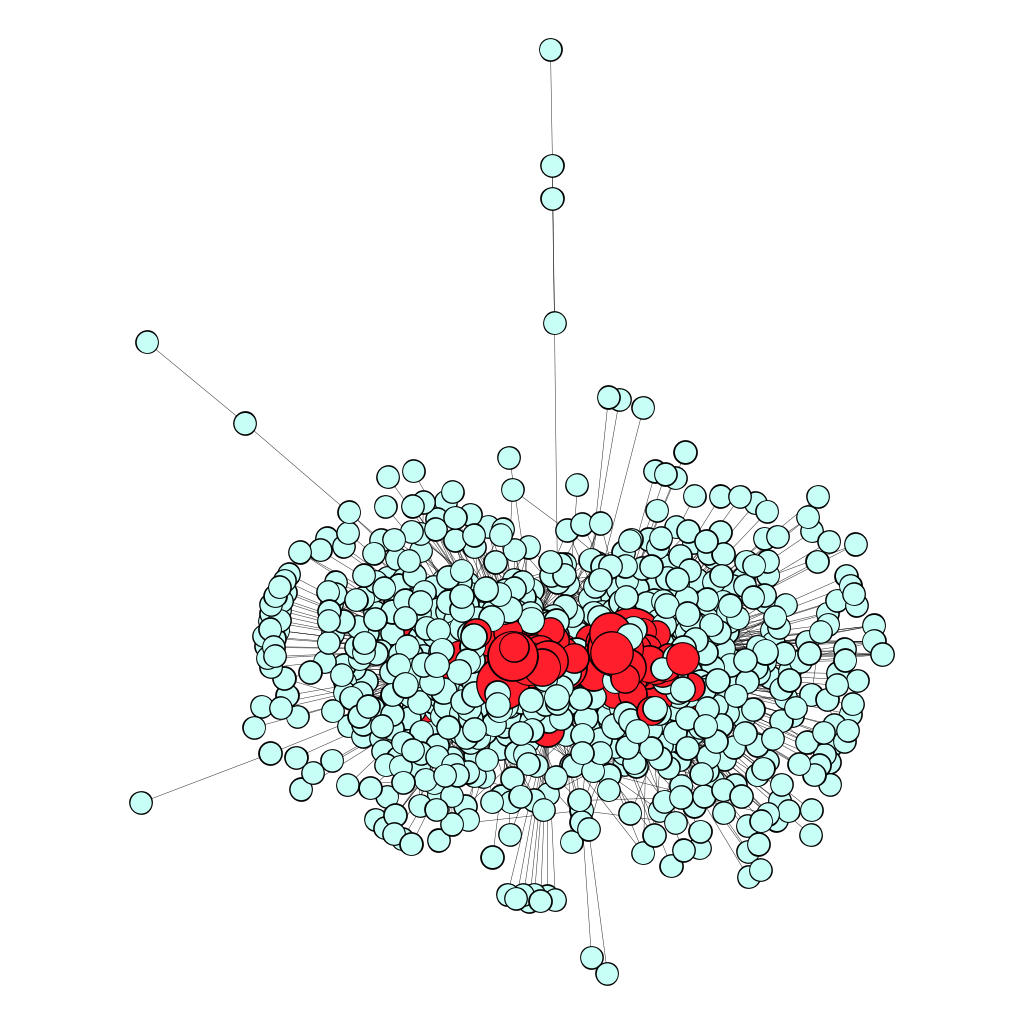}}
\hspace{0.01\linewidth}
\subfigure[PPL]{{}
\includegraphics[width=0.305 \linewidth]{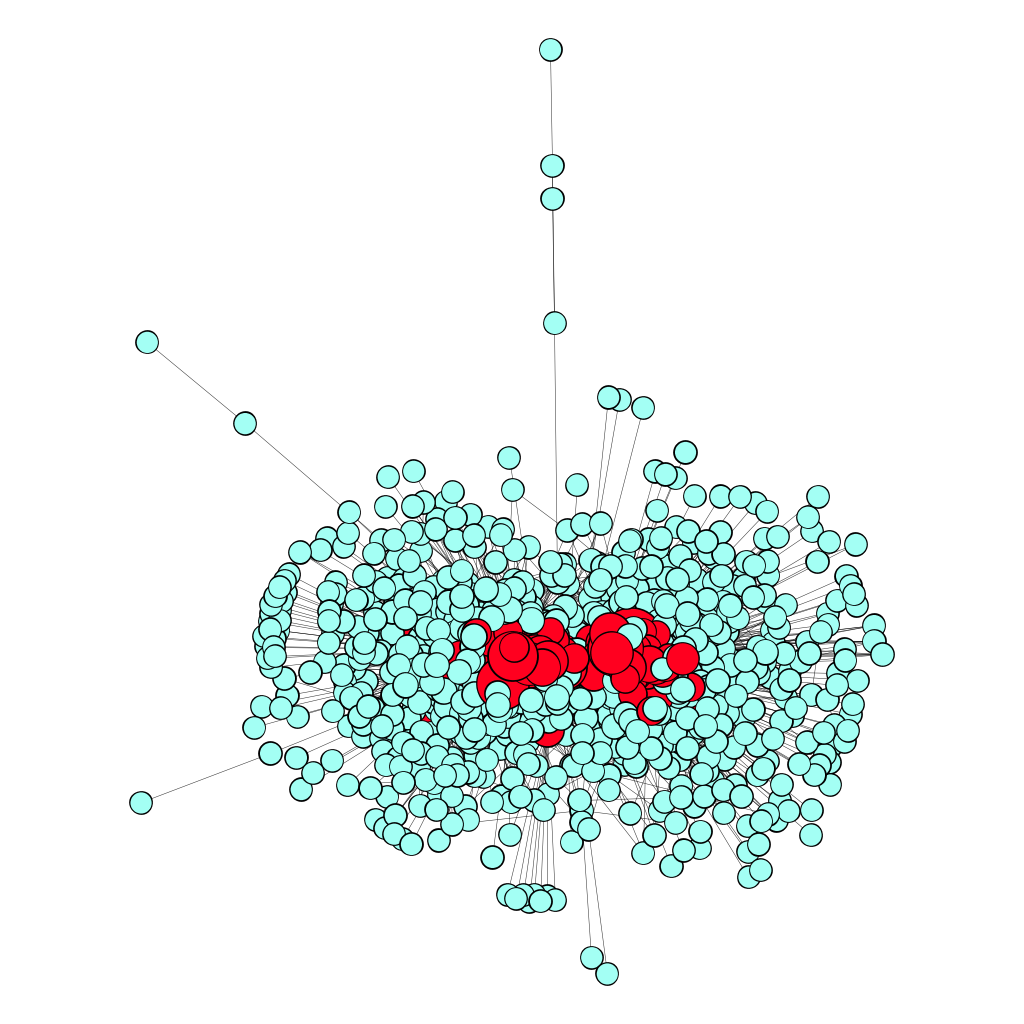}}
\hspace{0.01\linewidth}
\subfigure[SCP]{{}
\includegraphics[width=0.305 \linewidth]{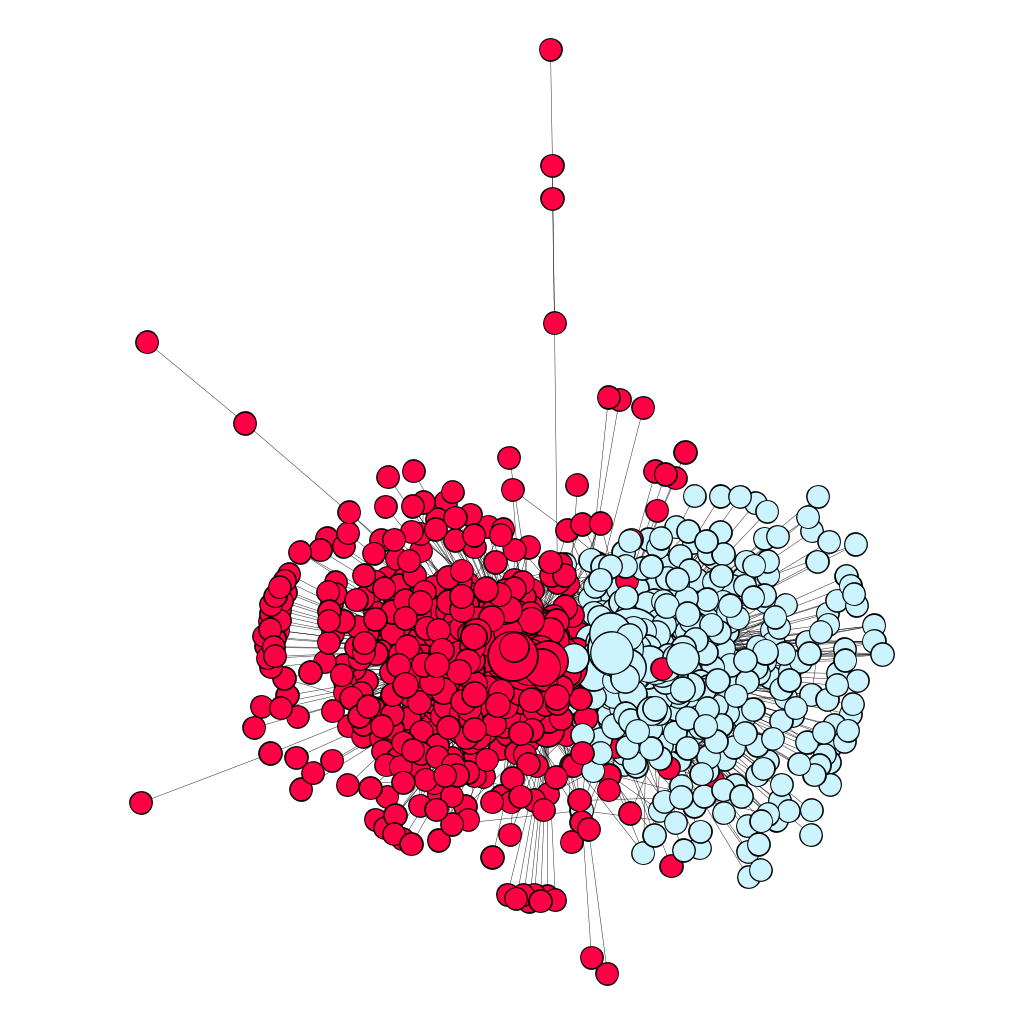}}
\hspace{0.01\linewidth}
\subfigure[CPL]{{}
\includegraphics[width=0.305 \linewidth]{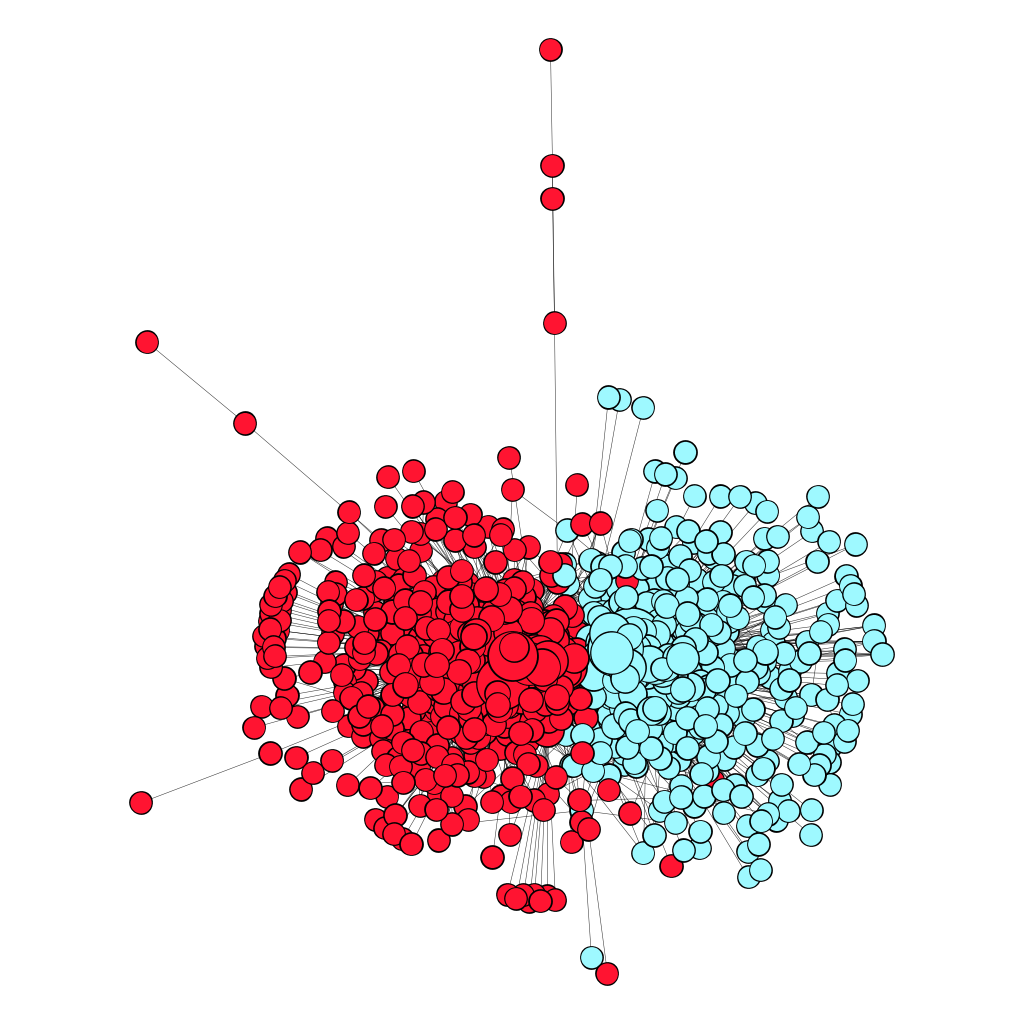}}
\hspace{0.01\linewidth}
\subfigure[DC-PPL]{{}
\includegraphics[width=0.305 \linewidth]{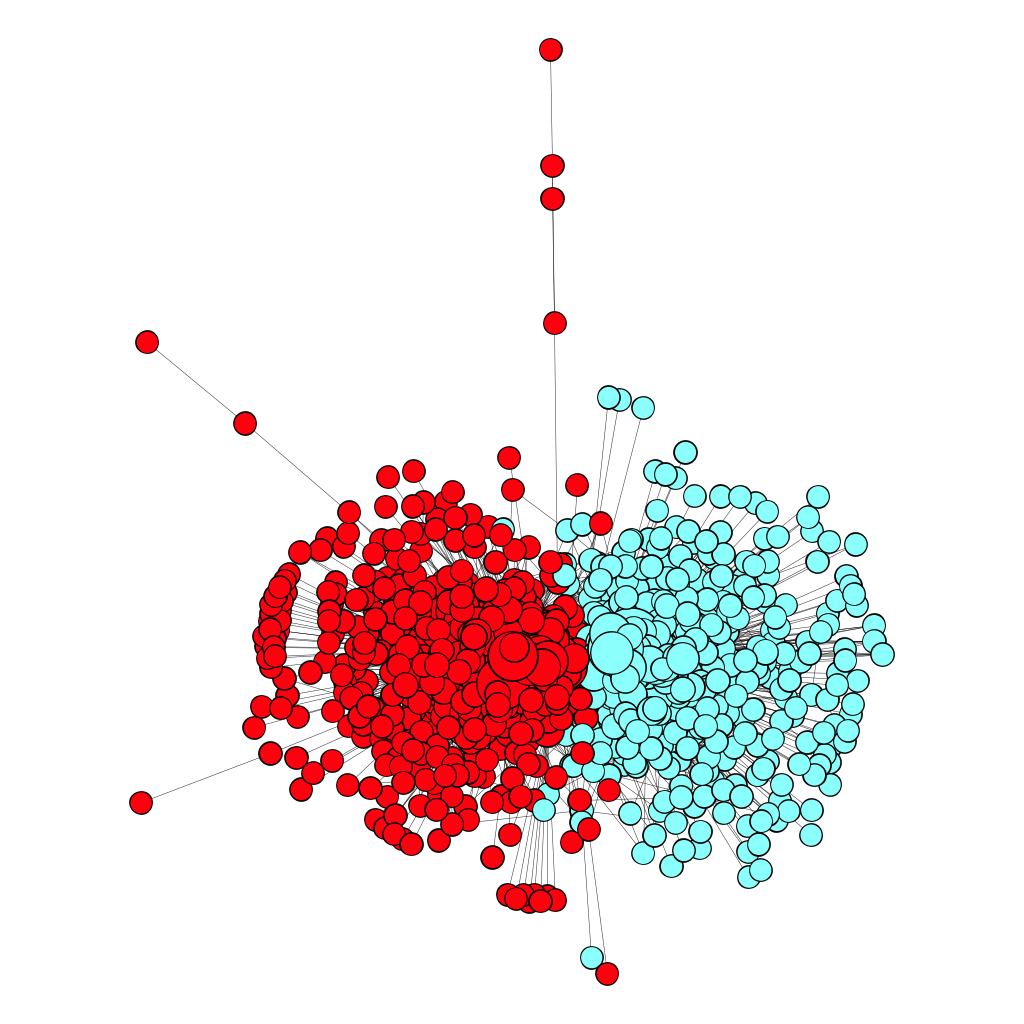}}
\hspace{0.01\linewidth}
\caption{Community detection on the political blogs data using \texttt{PL}, \texttt{PPL}, \texttt{SCP}, \texttt{CPL}, and \texttt{DC-PPL} respectively. The sizes of nodes are proportional to the their degree, and the color corresponds to different community labels.}
\label{fig:political-bolgs}
\end{figure}

To perform community detection, we consider five different methods, namely, \texttt{PL}, \texttt{PPL}, \texttt{SCP}, \texttt{CPL}, and \texttt{DC-PPL}.
We compute the NMI between the estimated community labels with the so-called ground truth labels.
Figure \ref{fig:political-bolgs} shows the community detection results from the five different methods.
It is seen that \texttt{PPL} and \texttt{PL} divide the nodes into two communities, with low degree and high degree nodes, respectively.
Both the \texttt{PPL} and \texttt{PL} estimates have NMI close to zero as neither of these two methods take into consideration the degree heterogeneity.
The partition obtained using \texttt{SCP} has NMI=0.653, while that from the \texttt{CPL} has NMI=0.722 and that from the \texttt{DC-PPL} has NMI=0.727.
Both \texttt{CPL} and \texttt{DC-PPL} achieve good performance in this application.

\subsection{International trade data}
\noindent
In this subsection, we apply our proposed method to the network of international trades.
The data contain yearly international trades among $n=58$ countries from 1981--2000 \citep{westveld2011mixed}.
Each node in the network corresponds to a country and an edge $(i,j)$ measures the amount of exports from country $i$ to country $j$ for a given year; see \cite{westveld2011mixed} for details.
Following \cite{saldana2017many}, we focus on the international trade network in 1995 and transform the directed and weighted adjacency network to an undirected binary network. Specifically, let $W_{ij}=\textrm{Trade}_{ij} + \textrm{Trade}_{ji}$, and set $A_{ij}=1$ if $W_{ij}\geq W_{0.5}$, and $A_{ij}=0$ otherwise. Here $\textrm{Trade}_{ij}$ records the amount of exports from country $i$ to country $j$ and $W_{0.5}$ denotes the 50th percentile of $\{W_{ij}\}_{1\leq i<j\leq n}$.
\textcolor{black}{Using different model selection procedures, both \cite{saldana2017many} and \cite{hu2020corrected} selected the number of SBM communities to be $K=3$ for this data set.
\cite{saldana2017many} suggested that larger community numbers such as $K=7$ are also reasonable and they tended to provide finer solutions.
We apply \texttt{PPL} to this network with $K=3$ and the community detection result is summarized in Table \ref{tab:trade}. It is seen that the three communities mostly correspond to developing countries in South America with low GDPs, countries with high GDPs and industrialized European and Asian countries with medium-level GDPs, respectively.}

\begin{table}[!t]
\textcolor{black}{\begin{tabular}{c|c}
\hline  Group & Countries \\\hline
1  &  \begin{tabular}[c]{@{}l@{}}Algeria, Barbados, Bolivia, Costa Rica, Cyprus, Ecuador, El Salvador, \\
Guatemala, Honduras, Iceland, Jamaica, Mauritius, Nepal, Oman, Panama, \\
Paraguay, Peru, Trinidad and Tobago, Tunisia, Uruguay, Venezuela\end{tabular} \\\hline
2  &  \begin{tabular}[c]{@{}l@{}}Belgium, Brazil, Canada, France, Germany, Italy, Japan, South Korea, \\
Mexico, Netherlands, Spain, Switzerland, United Kingdom, United States \end{tabular}\\\hline
3  &  \begin{tabular}[c]{@{}l@{}}Argentina, Australia, Austria, Chile, Colombia, Denmark, Egypt, Finland, \\
Greece, India, Indonesia, Ireland, Israel, Malaysia, Morocco, New Zealand,\\
Norway, Philippines, Portugal, Singapore, Sweden, Thailand, Turkey \end{tabular} \\\hline
\end{tabular}
\caption{\label{tab:trade} Community detection result on the international trade data using \texttt{PPL} with $K=3$.}}
\end{table}

\textcolor{black}{To evaluate goodness of fit, we consider the maximum entry-wise deviation based testing procedure \citep{hu2019using} that we investigated in Section \ref{section:inferencesim}.
%, which can be used to test the goodness of fit of a community label vector. The authors showed that the distribution of the test statistic, calculated with a strongly consistent community label, converges to a Gumbel distribution.
The community labels identified using \texttt{SCP} under $K=3$ gives a test statistic value of 52.13 with a $p$-value less than $10^{-10}$, suggesting a lack of fit. On the other hand, the community labels identified by \texttt{PPL}, initialized using \texttt{SCP} under $K=3$, gives a test statistic of 4.59 with a $p$-value of 0.03. Therefore, the goodness of fit test for \texttt{PPL} under $K=3$ is not rejected at the significance level of 0.01. It is also worth noting that when $K=4$, \texttt{PPL} gives a test statistic of 2.38 with a $p$-value of 0.08 while \texttt{SCP} gives a $p$-value less than $10^{-3}$.
It is seen through this data example that refinement of the initial clustering solution can be useful in inferential tasks such as the goodness of fit test.
}

\section{Discussion}\label{section:discuss}
\noindent
In this paper, we propose a new profile-pseudo likelihood method for fitting SBMs to large networks. Specifically, we consider a novel approach that decouples the membership labels of the rows and columns in the likelihood function, and treat the row labels as a vector of latent variables. Correspondingly, the likelihood can be maximized in an alternating fashion over the block model parameters and over the column community labels. Our proposed method retains and improves on the computational efficiency of the pseudo likelihood method, performs well for both small and large scale networks, and has provable convergence guarantee. We show that the community labels (i.e., column labels) estimated from our proposed method enjoy strong consistency, as long as the initial labels have an overlap with the truth beyond that of random guessing.

\textcolor{black}{In our approach, we consider spectral clustering as the initialization method, which requires computing $K$ leading eigenvectors. In real world applications, many implementations of eigen-decomposition are scalable, such as the PageRank algorithm adopted in Google search \citep{page1999pagerank}. We also note that our method needs not to limit the initialization algorithm to spectral clustering. For large-scale networks, one may consider the FastGreedy method by \citet{clauset2004finding}, which has a complexity of $O(n\log^2 n)$
or the Louvain algorithm by \citet{blondel2008fast}, which has a complexity of $O(n\log n)$ \citep{yang2016comparative}.
These fast algorithms, to our best knowledge, may not have theoretical guarantees on their performances. However, they have been validated empirically by many across various fields \citep{yang2016comparative} and can be considered as an initialization method when spectral clustering is not feasible.}
%Finding a computationally efficient initialization method with theoretical guarantees can be a direction for future research.}

Although we focus on SBMs and DCSBMs in this work, we envision the idea of simplifying the block model likelihoods by decoupling the membership labels of rows and columns can be applied to other network block model problems, such as mixed membership SBMs \citep{airoldi2008mixed}, block models with additional node features \citep{zhang2016community} and SBMs with dependent edges \citep{yuan2018community}. We plan to investigate these directions in our future work.

The code is publicly available on Github (https://github.com/WangJiangzhou/Fast-Network-Community-Detection-with-Profile-Pseudo-Likelihood-Methods).

\section*{Acknowledgment}
\noindent
Wang, Liu and Guo's research are supported by NSFC grants 11690012 and 11571068, the Fundamental Research Funds for the Central Universities grant 2412017BJ002, the Key Laboratory of Applied Statistics of MOE (KLAS) grants 130026507 and 130028612, {\color{black} the Special Fund for Key Laboratories of Jilin Province, China grant 20190201285JC}. Zhang's research is supported by NSF DMS-2015190 and Zhu's research is supported by NSF DMS-1821243.

\bibliographystyle{asa}
\begingroup
\baselineskip=16.5pt
\bibliography{ref}
\endgroup

\newpage
\renewcommand{\thesubsection}{A\arabic{subsection}}
\renewcommand{\theequation}{S\arabic{equation}}
\setcounter{equation}{0}
\setcounter{table}{0}
\setcounter{subsection}{0}
\setcounter{page}{1}
\def\eop
{\hfill $\Box$
}

\begin{center}
{\Large\bf Supplementary Materials} \\
{\large\bf Fast Network Community Detection with Profile-Pseudo Likelihood Methods}
\end{center}

\begin{center}
{Jiangzhou Wang, Jingfei Zhang, Binghui Liu, Ji Zhu, and Jianhua Guo}
\end{center}

\subsection{Proof of Theorem 1}
\noindent
To prove Theorem 1, it suffices to show
\begin{eqnarray}
L_{\textrm{PL}}({{\bOmega}}^{(s)}, {\e}^{(s)};  \{\a_i\})
 &\leq& L_{\textrm{PL}}({{\bOmega}}^{(s+1)}, {\e}^{(s)};  \{\a_i\}), \label{Th1_1}\\
L_{\textrm{PL}}({{\bOmega}}^{(s+1)}, {\e}^{(s)};  \{\a_i\})
 &\leq& L_{\textrm{PL}}({{\bOmega}}^{(s+1)}, {\e}^{(s+1)};  \{\a_i\}). \label{Th1_2}
\end{eqnarray}

Consider \eqref{Th1_1}. The updating procedure from $\{{{\bOmega}}^{(s)}, {\e}^{(s)}\}$ to $\{{{\bOmega}}^{(s+1)}, {\e}^{(s)}\}$ can be seen as a procedure of fitting some mixture model, thus the inequality \eqref{Th1_1} holds by the ascent property of the EM algorithm \citep{wu1983convergence}.

Consider \eqref{Th1_2}. It is equivalent to
\begin{eqnarray}
\ell_{\textrm{PL}}({{\bOmega}}^{(s+1)}, {\e}^{(s)};  \{\a_i\})\leq \ell_{\textrm{PL}}({{\bOmega}}^{(s+1)}, {\e}^{(s+1)};  \{\a_i\}).
\label{Th1_3}
\end{eqnarray}
We have
{\scriptsize
\begin{eqnarray*}
& &\ell_{\textrm{PL}}({{\bOmega}}^{(s+1)}, {{\e}}^{(s+1)};  \{\a_i\})-\ell_{\textrm{PL}}(\bOmega^{(s+1)}, {\e}^{(s)};  \{\a_i\})\\
&=&\sum\limits_{i=1}^{n}\mathrm{log}\left[\sum\limits_{l=1}^{K}{\pi}^{(s+1)}_{l}\prod\limits_{j=1}^{n}
{\left\{{{P}}^{(s+1)}_{l{e}_{j}^{(s+1)}}\right\}}^{A_{ij}}\left\{1-{{{P}}^{(s+1)}_{l{e}_{j}^{(s+1)}}}\right\}^{1-A_{ij}}\right]-\sum\limits_{i=1}^{n}\mathrm{log}\left[\sum\limits_{l=1}^{K}{\pi}^{(s+1)}_{l}\prod\limits_{j=1}^{n}
{\left\{{{P}}^{(s+1)}_{l{e}_{j}^{(s)}}\right\}}^{A_{ij}}\left\{1-{{{P}}^{(s+1)}_{l{e}_{j}^{(s)}}}\right\}^{1-A_{ij}}\right]\\
&=&\sum\limits_{i=1}^{n}\mathrm{log}\left[\sum\limits_{l=1}^{K}\frac {{\pi}^{(s+1)}_{l}\prod\limits_{j=1}^{n}{\left\{{{P}}^{(s+1)}_{le_{j}^{(s+1)}}\right\}}^{A_{ij}}\left\{1-{{{P}}^{(s+1)}_{l{e}_{j}^{(s+1)}}}\right\}^{1-A_{ij}}
}{{\pi}^{(s+1)}_{l}\prod\limits_{j=1}^{n}{\left\{{{P}}^{(s+1)}_{le_{j}^{(s)}}\right\}}^{A_{ij}}\left\{1-{{{P}}^{(s+1)}_{l{e}_{j}^{(s)}}}\right\}^{1-A_{ij}}
}\frac {{\pi}^{(s+1)}_{l}\prod\limits_{j=1}^{n}{\left\{{{P}}^{(s+1)}_{l{e}_{j}^{(s)}}\right\}}^{A_{ij}}\left\{1-{{{P}}^{(s+1)}_{l{e}_{j}^{(s)}}}\right\}^{1-A_{ij}}
}{\sum\limits_{l=1}^{K}{\pi}^{(s+1)}_{l}\prod\limits_{j=1}^{n}{\left\{{{P}}^{(s+1)}_{l{e}_{j}^{(s)}}\right\}}^{A_{ij}}\left\{1-{{{P}}^{(s+1)}_{l{e}_{j}^{(s)}}}\right\}^{1-A_{ij}}
}\right] \\
&\geq& \sum\limits_{i=1}^{n}\sum\limits_{l=1}^{K}\mathrm{log}\left[\frac {\prod\limits_{j=1}^{n}{\left\{{{P}}^{(s+1)}_{le_{j}^{(s+1)}}\right\}}^{A_{ij}}\left\{1-{{{P}}^{(s+1)}_{l{e}_{j}^{(s+1)}}}\right\}^{1-A_{ij}}
}{\prod\limits_{j=1}^{n}{\left\{{{P}}^{(s+1)}_{l{e}_{j}^{(s)}}\right\}}^{A_{ij}}\left\{1-{{{P}}^{(s+1)}_{l{e}_{j}^{(s)}}}\right\}^{1-A_{ij}}}\right]
\frac {{\pi}^{(s+1)}_{l}\prod\limits_{j=1}^{n}{\left\{{{P}}^{(s+1)}_{le_{j}^{(s)}}\right\}}^{A_{ij}}\left\{1-{{{P}}^{(s+1)}_{l{e}_{j}^{(s)}}}\right\}^{1-A_{ij}}
}{\sum\limits_{l=1}^{K}{\pi}^{(s+1)}_{l}\prod\limits_{j=1}^{n}{\left\{{{P}}^{(s+1)}_{le_{j}^{(s)}}\right\}}^{A_{ij}}\left\{1-{{{P}}^{(s+1)}_{l{e}_{j}^{(s)}}}\right\}^{1-A_{ij}}
}\\
&=&\sum\limits_{j=1}^{n}\sum\limits_{i=1}^{n}\sum\limits_{l=1}^{K}{\tau}^{(s+1)}_{il}\log\left[{\left\{{{P}}^{(s+1)}_{le_{j}^{(s+1)}}\right\}}^{A_{ij}}\left\{1-{{P}^{(s+1)}_{le_{j}^{(s+1)}}}\right\}^{1-A_{ij}}\right]
-\sum\limits_{j=1}^{n}\sum\limits_{i=1}^{n}\sum\limits_{l=1}^{K}{\tau}^{(s+1)}_{il}\log\left[{\left\{{{P}}^{(s+1)}_{le_{j}^{(s)}}\right\}}^{A_{ij}}\left\{1-{{P}^{(s+1)}_{le_{j}^{(s)}}}\right\}^{1-A_{ij}}\right]\nonumber\\
&=&\sum\limits_{j=1}^{n}\left(\sum\limits_{i=1}^{n}\sum\limits_{l=1}^{K}{\tau}^{(s+1)}_{il}\log\left[{\left\{{{P}}^{(s+1)}_{le_{j}^{(s+1)}}\right\}}^{A_{ij}}\left\{1-{{P}^{(s+1)}_{le_{j}^{(s+1)}}}\right\}^{1-A_{ij}}\right]-\sum\limits_{i=1}^{n}\sum\limits_{l=1}^{K}{\tau}^{(s+1)}_{il}\log\left[{\left\{{{P}}^{(s+1)}_{le_{j}^{(s)}}\right\}}^{A_{ij}}\left\{1-{{P}^{(s+1)}_{le_{j}^{(s)}}}\right\}^{1-A_{ij}}\right]\right)
\nonumber\\
&\geq& 0,
\end{eqnarray*}}
where the first inequality is due to Jensen's inequality, and the second inequality is due to the update strategy for ${{\e}}^{(s)}$ in Algorithm 1. The proof is completed.

\subsection{Proof of Theorem 2}
\noindent
We focus on the case of $\gamma\in (\frac{1}{2},1)$ and $a>b$.
For the remaining three cases of (i) $\gamma\in (\frac{1}{2},1)$, $a<b$, (ii) $\gamma\in (0, \frac{1}{2})$, $a>b$, and (iii) $\gamma\in (0,\frac{1}{2})$, $a<b$, the proofs are similar.

For any $(\hat{a}, \hat{b}) \in \mathcal{P}_{a, b}^{\delta}$, we have $\hat{a} > \hat{b}$. The \texttt{PPL} estimate can be written as follows:
\begin{eqnarray}
\hat{c}_{j}\{\e^{(0)}\}=\arg\max_{k\in \{1,2\}}\sum\limits_{i=1}^{n}\sum\limits_{l=1}^{2}\left[  \log\left\{(\widehat{P}_{lk})^{\widetilde{A}_{ij}}(1-\widehat{P}_{lk})^{1-\widetilde{A}_{ij}} \right\}\right]\hat{\tau}_{il}\{\e^{(0)}\} \nonumber.
\end{eqnarray}
Consider $j\in \{1, 2, \ldots, m\}$. Then $\hat{c}_{j}\{\e^{(0)}\}=1$ if
\begin{eqnarray}
\sum\limits_{i=1}^{n}\sum\limits_{l=1}^{2}\left[  \log\left\{(\widehat{P}_{l1})^{\widetilde{A}_{ij}}(1-\widehat{P}_{l1})^{(1-\widetilde{A}_{ij})}\right\} \right]\hat{\tau}_{il}\{\e^{(0)}\}>\sum\limits_{i=1}^{n}\sum\limits_{l=1}^{2}\left[  \log\left\{(\widehat{P}_{l2})^{\widetilde{A}_{ij}}(1-\widehat{P}_{l2})^{(1-\widetilde{A}_{ij})}\right\} \right]\hat{\tau}_{il}\{\e^{(0)}\}\nonumber,
\end{eqnarray}
which is equivalent to
\begin{eqnarray}
\begin{split}
\sum\limits_{l=1}^{2}\left\{\sum\limits_{i=1}^{n}\widetilde{A}_{ij}\hat{\tau}_{il}\{\e^{(0)}\}\mathrm{log}\widehat{P}_{l1}+
\sum\limits_{i=1}^{n}(1-\widetilde{A}_{ij})\hat{\tau}_{il}\{\e^{(0)}\}\mathrm{log}(1-\widehat{P}_{l1})\right\}>\\
\sum\limits_{l=1}^{2}\left\{\sum\limits_{i=1}^{n}\widetilde{A}_{ij}\hat{\tau}_{il}\{\e^{(0)}\}\mathrm{log}\widehat{P}_{l2}+
\sum\limits_{i=1}^{n}(1-\widetilde{A}_{ij})\hat{\tau}_{il}\{\e^{(0)}\}\mathrm{log}(1-\widehat{P}_{l2})\right\}.
\label{pmest}
\end{split}
\end{eqnarray}
We let $\widetilde{B}_{lj}^{'}\triangleq\sum\limits_{i=1}^{n}\widetilde{A}_{ij}\hat{\tau}_{il}\{\e^{(0)}\}$ and $n'_{l}\triangleq\sum\limits_{i=1}^{n}\hat{\tau}_{il}\{\e^{(0)}\}$ for all $j=1, 2, \ldots, n$ and $l=1, 2$, and recall
\begin{eqnarray}
\widehat{P}=\left(
    \begin{array}{cc}
     \widehat{P}_{11} & \widehat{P}_{12}\\
     \widehat{P}_{21} & \widehat{P}_{22}\\
    \end{array}
    \right)=\frac{1}{m}\left(
    \begin{array}{cc}
     \hat{a} & \hat{b}\\
     \hat{b} & \hat{a}\\
    \end{array}
    \right).
\label{Phat}
\end{eqnarray}
By simplifying \eqref{pmest}, we can restate that $\hat{c}_{j}\{\e^{(0)}\}=1$ if
\begin{eqnarray}
\left(\widetilde{B}_{1j}^{'}-\widetilde{B}_{2j}^{'}\right)\mathrm{log}\frac{\widehat{P}_{11}}{\widehat{P}_{12}}+
\left\{\widetilde{B}_{1j}^{'}-\widetilde{B}_{2j}^{'}-(n'_{1}-n'_{2})\right\}\mathrm{log}
\left(\frac{1-\widehat{P}_{12}}{1-\widehat{P}_{11}}\right)>0.
\label{pmj}
\end{eqnarray}
Since $\hat{a}>\hat{b}$, we have $\widehat{P}_{11}>\widehat{P}_{12}$. Thus by \eqref{pmj}, we have
\begin{eqnarray}
\mathbb{P}\left[\hat{c}_{j}\{\e^{(0)}\}\neq1\right] &\leq& \mathbb{P}\left[\left\{\widetilde{B}_{1j}^{'}-\widetilde{B}_{2j}^{'} \leq 0\right\} \bigcup \left\{\widetilde{B}_{1j}^{'}-\widetilde{B}_{2j}^{'}-\left(n'_{1}-n'_{2}\right) \leq 0\right\}  \right]\nonumber\\
&\leq& \mathbb{P}\left[\left\{\widetilde{B}_{1j}^{'}-\widetilde{B}_{2j}^{'} \leq \epsilon\right\} \bigcup \left\{\left|n'_{1}-n'_{2}\right|\geq \epsilon\right\}\right]\nonumber\\
&\leq& \mathbb{P}\left[\widetilde{B}_{1j}^{'}-\widetilde{B}_{2j}^{'} \leq \epsilon\right]+\mathbb{P}\left[\left|n'_{1}-n'_{2}\right| \geq \epsilon\right].
\label{cjew}
\end{eqnarray}
Next we upper bound the two terms $\mathbb{P}\left[\widetilde{B}_{1j}^{'}-\widetilde{B}_{2j}^{'} \leq \epsilon\right]$ and $\mathbb{P}\left[\left|n'_{1}-n'_{2}\right| \geq \epsilon\right]$ separately.

Firstly, we have
\begin{eqnarray}
&&\mathbb{P}\left[\widetilde{B}_{1j}^{'}-\widetilde{B}_{2j}^{'} \leq \epsilon\right]\nonumber\\
&=&\mathbb{P}\left[\sum\limits_{i=1}^{n}\widetilde{A}_{ij}\hat{\tau}_{i1}\{\e^{(0)}\}-\sum\limits_{i=1}^{n}\widetilde{A}_{ij}\hat{\tau}_{i2}\{\e^{(0)}\}\leq \epsilon \right]\nonumber\\
&=&\mathbb{P}\Big[\sum\limits_{i=1}^{n}\widetilde{A}_{ij}I(c_{i}=1)-\sum\limits_{i=1}^{n}\widetilde{A}_{ij}I(c_{i}=2)\nonumber\\
& &+\sum\limits_{i=1}^{n}\widetilde{A}_{ij}\left\{\hat{\tau}_{i1}\{\e^{(0)}\}-I(c_{i}=1)\right\}-\sum\limits_{i=1}^{n}\widetilde{A}_{ij}\left\{\hat{\tau}_{i2}\{\e^{(0)}\}-I(c_{i}=2)\right\}\leq \epsilon\Big]\nonumber\\
&\leq& \mathbb{P}\left[\sum\limits_{i=1}^{n}\widetilde{A}_{ij}I(c_{i}=1)-\sum\limits_{i=1}^{n}\widetilde{A}_{ij}I(c_{i}=2) + 2\sum\limits_{i=1}^{m}\widetilde{A}_{ij}\left\{\hat{\tau}_{i1}\{\e^{(0)}\}-I(c_{i}=1)\right\} \leq \epsilon \right]\nonumber\\
&\leq& \mathbb{P}\left[\left\{\sum\limits_{i=1}^{n}\widetilde{A}_{ij}I(c_{i}=1)-\sum\limits_{i=1}^{n}\widetilde{A}_{ij}I(c_{i}=2)\leq 2\epsilon\right\} \bigcup \left\{2\sum\limits_{i=1}^{m}\widetilde{A}_{ij}\left\{\hat{\tau}_{i1}\{\e^{(0)}\}-I(c_{i}=1)\right\}\leq -\epsilon\right\}\right]\nonumber\\
&\leq& \mathbb{P}\left[\sum\limits_{i=1}^{n}\widetilde{A}_{ij}I(c_{i}=1)-\sum\limits_{i=1}^{n}\widetilde{A}_{ij}I(c_{i}=2)\leq 2\epsilon\right]+\mathbb{P}\left[2\sum\limits_{i=1}^{m}\widetilde{A}_{ij}\left\{\hat{\tau}_{i1}\{\e^{(0)}\}-I(c_{i}=1)\right\}\leq -\epsilon\right]\nonumber\\
&\leq& \mathbb{P}\left[\sum\limits_{i=1}^{n}\widetilde{A}_{ij}I(c_{i}=1)-\sum\limits_{i=1}^{n}\widetilde{A}_{ij}I(c_{i}=2)\leq 2\epsilon\right]+\sum\limits_{i=1}^{m}\mathbb{P}\left[\left|\hat{\tau}_{i1}\{\e^{(0)}\}-I(c_{i}=1)\right| \geq \frac{\epsilon}{n}\right].
\label{BBw}
\end{eqnarray}
Next, we have
\begin{eqnarray}
&&\mathbb{P}\left[\left|n'_{1}-n'_{2}\right| \geq \epsilon\right]\nonumber\\
&\leq& \mathbb{P}\left[\{n'_{1} \leq \frac{n}{2}-\frac{\epsilon}{2}\} \bigcup \{n'_{2} \leq \frac{n}{2}-\frac{\epsilon}{2}\}\right]\nonumber\\
&\leq& \mathbb{P}\left[n'_{1} \leq \frac{n}{2}-\frac{\epsilon}{2}\right]+\mathbb{P}\left[n'_{2} \leq \frac{n}{2}-\frac{\epsilon}{2}\right]\nonumber\\
&\leq& \mathbb{P}\left[\bigcup_{i \in \{1,2,\ldots,m\}}\left\{|\hat{\tau}_{i1}\{\e^{(0)}\}-I(c_{i}=1)| \geq \frac{\epsilon}{n}\right\}\right]+
\mathbb{P}\left[\bigcup_{i \in \{m+1,m+2,\ldots,n\}}\left\{|\hat{\tau}_{i2}\{\e^{(0)}\}-I(c_{i}=2)| \geq \frac{\epsilon}{n}\right\}\right]\nonumber\\
&\leq& \sum\limits_{i=1}^{m}\mathbb{P}\left[|\hat{\tau}_{i1}\{\e^{(0)}\}-I(c_{i}=1)| \geq \frac{\epsilon}{n}\right]+\mathbb{P}\sum\limits_{i=m+1}^{n}\left[|\hat{\tau}_{i2}\{\e^{(0)}\}-I(c_{i}=2)| \geq \frac{\epsilon}{n}\right].
\label{nn}
\end{eqnarray}
Similar to Lemma 1 in \citet{amini2013pseudo}, we can upper bound the term $\mathbb{P}\Big[\sum\limits_{i=1}^{n}\widetilde{A}_{ij}I(c_{i}=1)-
\sum\limits_{i=1}^{n}\widetilde{A}_{ij}I(c_{i}=2)\leq 2\epsilon\Big]$ as follows.
Let\[
\widetilde{\eta}_{j}\big(\sigma(\c)\big)=\sum\limits_{i=1}^{n}\widetilde{A}_{ij}I(c_{i}=1)-
\sum\limits_{i=1}^{n}\widetilde{A}_{ij}I(c_{i}=2)
\triangleq \sum\limits_{i=1}^{n}\widetilde{A}_{ij}\sigma_{i}(\c),
\]
where
$\sigma_{i}(\c)=
\left\{
\begin{aligned}
1, \; c_{i}=1\\
-1, \; c_{i}=2
\end{aligned}\right.$, and $\sigma(\c)=\left(\sigma_{1}(\c), \sigma_{2}(\c), \ldots, \sigma_{n}(\c)\right)$.
Let $\widetilde{\alpha}_{ij}=\mathbb{E}[\widetilde{A}_{ij}]$. Since $\left|\widetilde{A}_{ij}\sigma_{j}(\c)-\mathbb{E}\left[\widetilde{A}_{ij}\sigma_{j}(\c)\right]\right|\leq \mathrm{max}\{\widetilde{\alpha}_{ij}, 1-\widetilde{\alpha}_{ij}\}\leq 1$, we have, for $j=1, 2, \ldots, m$,
\begin{eqnarray}
&&\mathbb{E}\left[\widetilde{\eta}_{j}\left(\sigma\left(\c\right)\right)\right]=m\times\frac{a}{b}-m\times\frac{b}{m}=(a-b)\nonumber,\\
&&\upsilon=\text{Var}\left(-\widetilde{\eta}_{j}\left(\sigma\left(\c\right)\right)\right)=\sum\limits_{i=1}^{n}Var(\widetilde{A}_{ij})\leq
\sum\limits_{i=1}^{n}\mathbb{E}\left[\widetilde{A}_{ij}^{2}\right]=\sum\limits_{i=1}^{n}\mathbb{E}\left[\widetilde{A}_{ij}\right]=(a+b)\nonumber.
\end{eqnarray}
Then by applying the Bernstein inequality to $-\widetilde{\eta}_{j}\big(\sigma(\c)\big)$, we have
\small
\begin{eqnarray}
\mathbb{P}\left[\widetilde{\eta}_{j}\big(\sigma(\c)\big) \leq \mathbb{E}\left[\widetilde{\eta}_{j}\big(\sigma(\c)\big)\right]-t\right]=
\mathbb{P}\left[-\widetilde{\eta}_{j}\big(\sigma(\c)\big) \geq -\mathbb{E}[\widetilde{\eta}_{j}\big(\sigma(\c)\big)]+t\right]\leq
e^{-\frac{t^{2}}{2(v+t/3)}}, \hspace{0.2cm} \forall t\geq 0.
\label{etaj}
\end{eqnarray}
\normalsize
Note that for $t \in [0, 3(a+b)]$, we have $2(\upsilon+t/3)\leq 4(a+b)$. It follows from \eqref{etaj} that
\begin{eqnarray}
\mathbb{P}\left[\widetilde{\eta}_{j}\big(\sigma(\c)\big) \leq (a-b)-t \right] \leq  e^{-\frac{t^{2}}{4(a+b)}}, \hspace{0.5cm} \forall t \in [0, 3(a+b)].
\label{etaj1}
\end{eqnarray}
In order to bound $\mathbb{P}\Big[\widetilde{\eta}_{j}\big(\sigma(\c)\big)\leq 2\epsilon\Big]$, we take $t=(a-b)-2\epsilon$. Then $t\in [0,3(a+b)]$ when $n$ is large enough as $\frac{\left(a-b\right)^{2}}{\left(a+b\right)}\ge C\log n$ for a sufficiently large $C$. Thus we have
\begin{eqnarray}
\mathbb{P}\left[\widetilde{\eta}_{j}\left(\sigma(\c)\right)\leq 2\epsilon\right]\leq
e^{-\frac{\left\{(a-b)-2\epsilon\right\}^{2}}{4(a+b)}}=
e^{-\frac{(a-b)^2-4\epsilon(a-b)+4\epsilon^{2}}{4(a+b)}}.
\label{etajj}
\end{eqnarray}

To obtain upper bounds of \eqref{BBw} and \eqref{nn}, we need to upper bound $\mathbb{P}\Big[|\hat{\tau}_{i1}\{\e^{(0)}\}-I(c_{i}=1)|\geq\frac{\epsilon}{n}\Big], \hspace{0.3cm} \forall i \in \{1, 2, \ldots, m\}$ and $\mathbb{P}\Big[|\hat{\tau}_{i2}\{\e^{(0)}\}-I(c_{i}=2)|\geq\frac{\epsilon}{n}\Big], \hspace{0.3cm} \forall i \in \{m+1, m+2, \ldots, n\}$. Firstly, we consider the case of $i \in \{1, 2, \ldots, m\}$.
With $( \hat{a}, \hat{b})\in \mathcal{P}_{a,b}^{\delta}$ and \eqref{Phat}, we have
\begin{eqnarray}
&&\frac{\frac{\widehat{P}_{11}}{1-\widehat{P}_{11}}}
{\frac{\widehat{P}_{12}}{1-\widehat{P}_{12}}}=\frac{\frac{\hat{a}}{m-\hat{a}}}{\frac{\hat{b}}{m-\hat{b}}}\geq
\frac{\frac{\hat{a}}{m-\hat{b}}}{\frac{\hat{b}}{m-\hat{b}}}=\frac{\hat{a}}{\hat{b}}\geq \delta.\nonumber
\end{eqnarray}
Let $\widetilde{B}_{ik}=\sum\limits_{j=1}^{n}\widetilde{A}_{ij}I(e_{j}=k)$ and $n_{k}=\sum\limits_{i=1}^{n}I(e_{j}=k)$, we then have
\begin{eqnarray}
&&\mathbb{P}\left[|\hat{\tau}_{i1}\{\e^{(0)}\}-I(c_{i}=1)|\geq\frac{\epsilon}{n}\right]\nonumber\\
&=&\mathbb{P}\left[\frac{\hat{\tau}_{i1}\{\e^{(0)}\}}{\hat{\tau}_{i2}\{\e^{(0)}\}} \leq \frac{1-\epsilon/n}{\epsilon/n}\right]\nonumber\\
&=&\mathbb{P}\left[\frac{(\widehat{P}_{11})^{\widetilde{B}_{i1}}(\widehat{P}_{12})^{\widetilde{B}_{i2}}(1-\widehat{P}_{11})^{n_{1}-\widetilde{B}_{i1}}(1-\widehat{P}_{12})^{n_{2}-\widetilde{B}_{i2}}}
{(\widehat{P}_{21})^{\widetilde{B}_{i1}}(\widehat{P}_{22})^{\widetilde{B}_{i2}}(1-\widehat{P}_{21})^{n_{1}-\widetilde{B}_{i1}}(1-\widehat{P}_{22})^{n_{2}-\widetilde{B}_{i2}}}
\leq \frac{1-\epsilon/n}{\epsilon/n}\right]\nonumber\nonumber\\
&=&\mathbb{P}\left[\left(\frac{\frac{\widehat{P}_{11}}{1-\widehat{P}_{11}}}
{\frac{\widehat{P}_{12}}{1-\widehat{P}_{12}}}\right)^{\widetilde{B}_{i1}-\widetilde{B}_{i2}}
\leq \frac{1-\epsilon/n}{\epsilon/n} \right]\nonumber\\
&=&\mathbb{P}\left[\left\{\left(\frac{\frac{\widehat{P}_{11}}{1-\widehat{P}_{11}}}
{\frac{\widehat{P}_{12}}{1-\widehat{P}_{12}}}\right)^{\widetilde{B}_{i1}-\widetilde{B}_{i2}}
\leq \frac{1-\epsilon/n}{\epsilon/n}\right\} \bigcap \left\{ \left\{\widetilde{B}_{i1}-\widetilde{B}_{i2}\geq 0\right\} \bigcup  \left\{\widetilde{B}_{i1}-\widetilde{B}_{i2}<0\right\} \right\} \right]\nonumber\\
&\leq& \mathbb{P}\left[\delta^{\widetilde{B}_{i1}-\widetilde{B}_{i2}}\leq \frac{1-\epsilon/n}{\epsilon/n}\right]+\mathbb{P}\left[\widetilde{B}_{i1}-\widetilde{B}_{i2}< 0\right]
\hspace{2cm} \nonumber\\
&=& \mathbb{P}\bigg[\widetilde{B}_{i1}-\widetilde{B}_{i2} \leq \frac{1}{\mathrm{log}\delta}\mathrm{log}\Big(\frac{1-\epsilon/n}{\epsilon/n}\Big)\bigg]+
\mathbb{P}\Big[\widetilde{B}_{i1}-\widetilde{B}_{i2}< 0\Big]\nonumber\\
&\leq& 2\mathbb{P}\bigg[\widetilde{B}_{i1}-\widetilde{B}_{i2} \leq \frac{1}{\mathrm{log}\delta}\mathrm{log}\Big(\frac{1-\epsilon/n}{\epsilon/n}\Big)\bigg].
\label{pi1}
\end{eqnarray}
Let
\[\widetilde{\xi}_{i}\big(\sigma\{\e^{(0)}\}\big)=\widetilde{B}_{i1}-\widetilde{B}_{i2}
\triangleq \sum\limits_{j=1}^{n}\widetilde{A}_{ij}\sigma_{j}\{\e^{(0)}\},
\]
where
$\sigma_{j}\{\e^{(0)}\}=
\left\{
\begin{aligned}
1, \; e_{j}=1\nonumber\\
-1, \; e_{j}=2\nonumber
\end{aligned}
\right.$, and $\sigma\{\e^{(0)}\}=\big(\sigma_{1}\{\e^{(0)}\}, \sigma_{2}\{\e^{(0)}\}, \ldots, \sigma_{n}\{\e^{(0)}\}\big)$.
Note that $\left|\widetilde{A}_{ij}\sigma_{j}\{\e^{(0)}\}-\mathbb{E}\left[\widetilde{A}_{ij}\sigma_{j}\{\e^{(0)}\}\right]\right|\leq \mathrm{max}\left\{\widetilde{\alpha}_{ij}, 1-\widetilde{\alpha}_{ij}\right\}\leq 1$. For $i \in \{1, 2, \ldots, m\}$, we have
\small
\begin{eqnarray}
&&\mathbb{E}\left[\widetilde{\xi}_{i}\big(\sigma\{\e^{(0)}\}\big)\right]=\gamma m\cdot \frac{a}{m}+(1-\gamma)m\cdot\frac{b}{m}-\left\{(1-\gamma)m\cdot\frac{a}{m}+\gamma m\cdot\frac{b}{m} \right\}=(2\gamma-1)(a-b)\nonumber,\\
&&\upsilon=\text{Var}\left(-\widetilde{\xi}_{i}\big(\sigma\{\e^{(0)}\}\big)\right)=\sum\limits_{j=1}^{n}Var\left(\widetilde{A}_{ij}\right)\leq
\sum\limits_{j=1}^{n}\mathbb{E}\left[\widetilde{A}_{ij}^{2}\right]=\sum\limits_{i=1}^{n}\mathbb{E}\left[\widetilde{A}_{ij}\right]=(a+b)\nonumber.
\end{eqnarray}
\normalsize
Then by applying the Bernstein inequality to $-\widetilde{\xi}_{i}\big(\sigma\{\e^{(0)}\}\big)$, we have
\small
\begin{eqnarray}
\mathbb{P}\left[\widetilde{\xi}_{i}\left(\sigma\{\e^{(0)}\}\right) \leq \mathbb{E}\left[\widetilde{\xi}_{i}\left(\sigma\{\e^{(0)}\}\right)\right]-t\right]=
\mathbb{P}\left[-\widetilde{\xi}_{i}\left(\sigma\{\e^{(0)}\}\right) \geq -\mathbb{E}\left[\widetilde{\xi}_{i}\left(\sigma\{\e^{(0)}\}\right)\right]+t\right]\leq
e^{-\frac{t^{2}}{2(v+t/3)}}, \hspace{0.2cm} \forall t\geq 0.
\label{xii}
\end{eqnarray}
\normalsize
Note that for $t \in [0, 3(a+b)]$, we have $2(\upsilon+t/3)\leq 4(a+b)$. It follows from \eqref{xii} that
\begin{eqnarray}
\mathbb{P}\Big[\widetilde{\xi}_{i}\big(\sigma\{\e^{(0)}\}\big) \leq (2\gamma-1)(a-b)-t \Big] \leq  e^{-\frac{t^{2}}{4(a+b)}}, \hspace{0.5cm} \forall t \in [0, 3(a+b)].
\label{xit}
\end{eqnarray}
In order to bound $\mathbb{P}\Big[\widetilde{\xi}_{i}\big(\sigma\{\e^{(0)}\}\big)\leq \frac{1}{\mathrm{log}\delta}\mathrm{log}\Big(\frac{1-\epsilon/n}{\epsilon/n}\Big) \Big]$, we take $t=(2\gamma-1)(a-b)-\frac{1}{\mathrm{log}\delta}\mathrm{log}\Big(\frac{1-\epsilon/n}{\epsilon/n}\Big)$. Then $t\in [0,3(a+b)]$ when $n$ is large enough. Thus, we have
\begin{eqnarray}
&&\mathbb{P}\left[\widetilde{\xi}_{i}\left(\sigma\{\e^{(0)}\}\right)
\leq\frac{1}{\mathrm{log}\delta}\mathrm{log}\left(\frac{1-\epsilon/n}{\epsilon/n}\right)\right]\nonumber\\
&\leq& e^{-\frac{\left\{(2\gamma-1)(a-b)-\frac{1}{\mathrm{log}\delta}\mathrm{log}\big(\frac{1-\epsilon/n}{\epsilon/n}\big)\right\}^{2}}{4(a+b)}}\nonumber\\
&\leq&
e^{-\left[\frac{(2\gamma-1)^2(a-b)^2}{4(a+b)}-\frac{2(2\gamma-1)(a-b)\frac{1}{\mathrm{log}\delta}\mathrm{log}\big(\frac{1-\epsilon/n}{\epsilon/n}\big)}{4(a+b)}+\left\{\frac{1}{\mathrm{log}\delta}\mathrm{log}(\frac{1-\epsilon/n}{\epsilon/n})\right\}^{2}\right]}\nonumber\\
&\leq& e^{-\frac{(2\gamma-1)^{2}(a-b)^{2}}{8(a+b)}} \hspace{1cm} (\mathrm{when}\ n\ \mathrm{is}\ \mathrm{large}\ \mathrm{enough}). \nonumber
\end{eqnarray}
It follows from \eqref{pi1} that (when $n$ is large enough)
\begin{eqnarray}
\mathbb{P}\left[\left|\hat{\tau}_{i1}\{\e^{(0)}\}-I(c_{i}=1)\right|\geq\frac{\epsilon}{n}\right] \leq 2e^{-\frac{(2\gamma-1)^{2}(a-b)^{2}}{8(a+b)}} \hspace{0.5cm} \forall i \in \{1, 2, \ldots, m\}.
\label{pil1}
\end{eqnarray}
Similar results for $\mathbb{P}\left[|\hat{\tau}_{i2}\{\e^{(0)}\}-I(c_{i}=2)|\geq\frac{\epsilon}{n}\right]$ can be obtained by using similar arguments.
Specifically, we have
\begin{eqnarray}
\mathbb{P}\left[\left|\hat{\tau}_{i2}\{\e^{(0)}\}-I(c_{i}=2)\right|\geq\frac{\epsilon}{n}\right] \leq 2e^{-\frac{(2\gamma-1)^{2}(a-b)^{2}}{8(a+b)}} \hspace{0.5cm} \forall i \in \{m+1, m+2, \ldots, n\}.
\label{pil2}
\end{eqnarray}
Thus by \eqref{BBw}, \eqref{etajj}, \eqref{pil1} and \eqref{pil2}, for $j=1, 2, \ldots, m$, we have
\begin{eqnarray}
\mathbb{P}\left[\widetilde{B}_{1j}^{'}-\widetilde{B}_{2j}^{'}\leq\epsilon\right]\leq
e^{-\frac{(a-b)^2-4\epsilon(a-b)+4\epsilon^{2}}{4(a+b)}}+ne^{-\frac{(2\gamma-1)^{2}(a-b)^{2}}{8(a+b)}}.
\label{Bsmall}
\end{eqnarray}

For $j=m+1,m+2, \ldots, n$, the term $\mathbb{P}\left[\widetilde{B}_{2j}^{'}-\widetilde{B}_{1j}^{'}\leq\epsilon\right]$ can be bounded as follows,
\begin{eqnarray}
\mathbb{P}\left[\widetilde{B}_{2j}^{'}-\widetilde{B}_{1j}^{'}\leq\epsilon\right]\leq
e^{-\frac{(a-b)^2-4\epsilon(a-b)+4\epsilon^{2}}{4(a+b)}}+ne^{-\frac{(2\gamma-1)^{2}(a-b)^{2}}{8(a+b)}}.
\label{Bbig}
\end{eqnarray}
According to \eqref{nn}, \eqref{pil1}, and \eqref{pil2}, we have
\begin{eqnarray}
&&\mathbb{P}\left[\left|n'_{1}-n'_{2}\right| \geq \epsilon\right]\leq
2ne^{-\frac{(2\gamma-1)^{2}(a-b)^{2}}{8(a+b)}}.
\label{nn1}
\end{eqnarray}
Finally, by \eqref{Bsmall}, \eqref{Bbig}, and \eqref{nn1}, we have
\begin{eqnarray}
&&\mathbb{P}\left[\hat{\c}\{\e^{(0)}\}\neq \c\right]\nonumber\\
&=&\mathbb{P}\left[\bigcup_{j \in \{1,2,\ldots,n \}}\left\{\hat{c}_{j}\{\e^{(0)}\}\neq c_{j}\right\} \right]\nonumber\\
&\leq&\mathbb{P}\left[\left\{\bigcup_{j \in \{1,2,\ldots,m \}}\left\{\widetilde{B}_{1j}^{'}-\widetilde{B}_{2j}^{'}\leq\epsilon\right\}\right\} \bigcup \left\{\bigcup_{j \in \{m+1,m+2,\ldots,n \}} \left\{ \widetilde{B}_{2j}^{'}-\widetilde{B}_{1j}^{'}\leq\epsilon\right\}\right\}\bigcup\left\{\left|n'_{1}-n'_{2}\right| \geq \epsilon\right\}\right]\nonumber\\
&\leq& \sum\limits_{j=1}^{m}\mathbb{P}\left[\widetilde{B}_{1j}^{'}-\widetilde{B}_{2j}^{'}\leq\epsilon\right]+
\sum\limits_{j=m+1}^{n}\left[\widetilde{B}_{2j}^{'}-\widetilde{B}_{1j}^{'}\leq\epsilon\right]+\mathbb{P}\left[\left|n'_{1}-n'_{2}\right| \geq \epsilon\right]\nonumber\\
&=&ne^{-\frac{(a-b)^2-4\epsilon(a-b)+4\epsilon^{2}}{4(a+b)}}+
n(n+2)e^{-\frac{(2\gamma-1)^{2}(a-b)^{2}}{8(a+b)}}.\nonumber
\end{eqnarray}
Therefore, we have that
\begin{eqnarray}
&&\mathbb{P}\left[\hat{\c}\{\e^{(0)}\}= \c\right]= 1-\mathbb{P}\left[\hat{\c}\{\e^{(0)}\}\neq \c\right]\nonumber\\
&\geq& 1-\left[ne^{-\frac{(a-b)^2-4\epsilon(a-b)+4\epsilon^{2}}{4(a+b)}}+
n(n+2)e^{-\frac{(2\gamma-1)^{2}(a-b)^{2}}{8(a+b)}}\right].\hspace{2cm} \nonumber
\end{eqnarray}

\subsection{Proof of Theorem 3}
\noindent
Recall that $\bm A$ and $\widetilde{\bm A}$ are the adjacency matrices of undirected and directed networks, respectively. Similar to the technique in \citet{amini2013pseudo}, we introduce a deterministic coupling between $\bm A$ and $\widetilde{\bm A}$, which allows us to carry over the results from the directed SBM.
Let
\begin{eqnarray}
\bm A=T\left(\widetilde{\bm A}\right), \hspace{1cm} \left[T\left(\widetilde{\bm A}\right)\right]=
\left\{
\begin{aligned}
&0, \;\hspace{0.5cm} \widetilde{A}_{ij}=\widetilde{A}_{ji}=0\\
&1, \;\hspace{0.5cm} otherwise
\end{aligned}
\right..
\label{cpAA}
\end{eqnarray}
That is, the graph of $\bm A$ is obtained from that of $\widetilde{
\bm A}$ by removing directions. Note that
\begin{center}
$P_{kl}=\mathbb{P}\left(A_{ij}=1\right)=1-\mathbb{P}\left(\widetilde{A}_{ij}=0\right)\mathbb{P}\left(\widetilde{A}_{ji}=0\right)=
2\widetilde{P}_{kl}-(\widetilde{P}_{kl})^{2}$,
\end{center}
which matches the relationship between \eqref{dirP} and \eqref{undirP}. From
\eqref{cpAA}, it is not difficult to see that
\begin{eqnarray}
A_{ij}\geq \widetilde{A}_{ij} \hspace{1.2cm} \forall\ i, j \in \left\{ 1, 2, \ldots, n \right\}.\nonumber
\end{eqnarray}
We focus on the case of $\gamma\in (\frac{1}{2},1)$ and $a>b$.
For the remaining three cases of (i) $\gamma\in (\frac{1}{2},1)$, $a<b$, (ii) $\gamma\in (0, \frac{1}{2})$, $a>b$, and (iii) $\gamma\in (0,\frac{1}{2})$, $a<b$, the proofs are similar.
For any $(\hat{a}, \hat{b}) \in \mathcal{P}_{a, b}^{\delta}$, we have $\hat{a}>\hat{b}$. The \texttt{PPL} estimate can be written as
\begin{eqnarray}
\hat{c}_{j}\{\e^{(0)}\}=\mathrm{arg}\max_{k\in \{1,2\}}\sum\limits_{i=1}^{n}\sum\limits_{l=1}^{2}\left[  \log\left\{{\widehat{P}_{lk}}^{A_{ij}}\left(1-\widehat{P}_{lk}\right)^{(1-A_{ij})}\right\} \right]\hat{\tau}_{il}\{\e^{(0)}\}.
\end{eqnarray}
We first consider $j\in \{1, 2, \ldots, m\}$. Then $\hat{c}_{j}\{\e^{(0)}\}=1$ if
\begin{center}
$\sum\limits_{i=1}^{n}\sum\limits_{l=1}^{2}\left[  \log\left\{{\widehat{P}_{l1}}^{A_{ij}}\left(1-\widehat{P}_{l1}\right)^{\left(1-A_{ij}\right)}\right\} \right]\hat{\tau}_{il}\{\e^{(0)}\}>\sum\limits_{i=1}^{n}\sum\limits_{l=1}^{2}\left[  \log\left\{{\widehat{P}_{l2}}^{A_{ij}}\left(1-\widehat{P}_{l2}\right)^{\left(1-A_{ij}\right)}\right\} \right]\hat{\tau}_{il}\{\e^{(0)}\}$,
\end{center}
which is equivalent to
\begin{eqnarray}
\begin{split}
\sum\limits_{l=1}^{2}\left\{\sum\limits_{i=1}^{n}A_{ij}\hat{\tau}_{il}\{\e^{(0)}\}\mathrm{log}\widehat{P}_{l1}+
\sum\limits_{i=1}^{n}\left(1-A_{ij}\right)\hat{\tau}_{il}\{\e^{(0)}\}\mathrm{log}\left(1-\widehat{P}_{l1}\right)\right\}>\\
\sum\limits_{l=1}^{2}\left\{\sum\limits_{i=1}^{n}A_{ij}\hat{\tau}_{il}\{\e^{(0)}\}\mathrm{log}\widehat{P}_{l2}+
\sum\limits_{i=1}^{n}\left(1-A_{ij}\right)\hat{\tau}_{il}\{\e^{(0)}\}\mathrm{log}\left(1-\widehat{P}_{l2}\right)\right\}.
\label{ppmest}
\end{split}
\end{eqnarray}
Let $B_{lj}^{'}=\sum\limits_{i=1}^{n}A_{ij}\hat{\tau}_{il}\{\e^{(0)}\}$ and $n'_{l}=\sum\limits_{i=1}^{n}\hat{\tau}_{il}\{\e^{(0)}\}$ for all $j\in \{1, 2, \ldots, n\}$ and $l \in \{1, 2\}$. We have
\begin{eqnarray}
\widehat{P}=\left(
    \begin{array}{cc}
     \widehat{P}_{11} & \widehat{P}_{12}\\
     \widehat{P}_{21} & \widehat{P}_{22}\\
    \end{array}
    \right)=\frac{2}{m}\left(
    \begin{array}{cc}
     \hat{a} & \hat{b}\\
     \hat{b} & \hat{a}\\
    \end{array}
    \right)
    -\frac{1}{m^{2}}\left(
    \begin{array}{cc}
     {\hat{a}}^{2} & {\hat{b}}^{2}\\
     {\hat{b}}^{2} & {\hat{a}}^{2}\\
    \end{array}
    \right).
\label{PPhat}
\end{eqnarray}
By simplifying \eqref{ppmest}, we can restate that $\hat{c}_{j}\{\e^{(0)}\}=1$ if
\begin{eqnarray}
\left(B_{1j}^{'}-B_{2j}^{'}\right)\mathrm{log}\frac{\widehat{P}_{11}}{\widehat{P}_{12}}+
\left\{B_{1j}^{'}-B_{2j}^{'}-\left(n'_{1}-n'_{2}\right)\right\}\mathrm{log}
\left(\frac{1-\widehat{P}_{12}}{1-\widehat{P}_{11}}\right)>0.
\label{ppmj}
\end{eqnarray}
Since $\hat{a}>\hat{b}$, we have $\widehat{P}_{11}>\widehat{P}_{12}$. Thus by \eqref{ppmj}, we have
\begin{eqnarray}
\mathbb{P}\left[\hat{c}_{j}\{\e^{(0)}\}\neq1\right] &\leq& \mathbb{P}\left[\{B_{1j}^{'}-B_{2j}^{'} \leq 0\} \bigcup \{B_{1j}^{'}-B_{2j}^{'}-\left(n'_{1}-n'_{2}\right) \leq 0\}  \right]\nonumber\\
&\leq& \mathbb{P}\left[\{B_{1j}^{'}-B_{2j}^{'} \leq \epsilon\} \bigcup \left\{|n'_{1}-n'_{2}|\geq \epsilon\right\}\right]\nonumber\\
&\leq& \mathbb{P}\left[B_{1j}^{'}-B_{2j}^{'} \leq \epsilon\right]+\mathbb{P}\left[\left|n'_{1}-n'_{2}\right| \geq \epsilon\right].
\label{cje}
\end{eqnarray}
Now we bound $\mathbb{P}\left[B_{1j}^{'}-B_{2j}^{'} \leq \epsilon\right]$ and $\mathbb{P}\left[\left|n'_{1}-n'_{2}\right| \geq \epsilon\right]$ separately. Firstly,
\begin{eqnarray}
&&\mathbb{P}\left[B_{1j}^{'}-B_{2j}^{'} \leq \epsilon\right]\nonumber\\
&=&\mathbb{P}\left[\sum\limits_{i=1}^{n}A_{ij}\hat{\tau}_{i1}\{\e^{(0)}\}-\sum\limits_{i=1}^{n}A_{ij}\hat{\tau}_{i2}\{\e^{(0)}\}\leq \epsilon \right]\nonumber\\
&=&\mathbb{P}\Bigg[\sum\limits_{i=1}^{n}A_{ij}I\left(c_{i}=1\right)-\sum\limits_{i=1}^{n}A_{ij}I\left(c_{i}=2\right)+
\sum\limits_{i=1}^{n}A_{ij}\left\{\hat{\tau}_{i1}\{\e^{(0)}\}-I\left(c_{i}=1\right)\right\}-\nonumber\\& &\sum\limits_{i=1}^{n}A_{ij}\left\{\hat{\tau}_{i2}\{\e^{(0)}\}-I\left(c_{i}=2\right)\right\}\leq \epsilon\Bigg]\nonumber\\
&\leq& \mathbb{P}\left[\sum\limits_{i=1}^{n}A_{ij}I\left(c_{i}=1\right)-\sum\limits_{i=1}^{n}A_{ij}I\left(c_{i}=2\right) + 2\sum\limits_{i=1}^{m}A_{ij}\left\{\hat{\tau}_{i1}\{\e^{(0)}\}-I\left(c_{i}=1\right)\right\} \leq \epsilon \right]\nonumber\\
&\leq& \mathbb{P}\left[\left\{\sum\limits_{i=1}^{n}A_{ij}I\left(c_{i}=1\right)-\sum\limits_{i=1}^{n}A_{ij}I\left(c_{i}=2\right)\leq 2\epsilon\right\} \bigcup \left\{2\sum\limits_{i=1}^{m}A_{ij}\left\{\hat{\tau}_{i1}\{\e^{(0)}\}-I\left(c_{i}=1\right)\right\}\leq -\epsilon\right\}\right]\nonumber\\
&\leq& \mathbb{P}\left[\sum\limits_{i=1}^{n}A_{ij}I\left(c_{i}=1\right)-\sum\limits_{i=1}^{n}A_{ij}I\left(c_{i}=2\right)\leq 2\epsilon\right]+\mathbb{P}\left[2\sum\limits_{i=1}^{m}A_{ij}\left\{\hat{\tau}_{i1}\{\e^{(0)}\}-I\left(c_{i}=1\right)\right\}\leq -\epsilon\right]\nonumber\\
&\leq& \mathbb{P}\left[\sum\limits_{i=1}^{n}A_{ij}I\left(c_{i}=1\right)-\sum\limits_{i=1}^{n}A_{ij}I\left(c_{i}=2\right)\leq 2\epsilon\right]+\sum\limits_{i=1}^{m}\mathbb{P}\left[\left|\hat{\tau}_{i1}\{\e^{(0)}\}-I\left(c_{i}=1\right)\right| \geq \frac{\epsilon}{n}\right].
\label{BB}
\end{eqnarray}
Secondly,
\begin{eqnarray}
&&\mathbb{P}\left[\left|n'_{1}-n'_{2}\right| \geq \epsilon\right]\nonumber\\
&\leq& \mathbb{P}\left[\left\{n'_{1} \leq \frac{n}{2}-\frac{\epsilon}{2}\right\} \bigcup \left\{n'_{2} \leq \frac{n}{2}-\frac{\epsilon}{2}\right\}\right]\nonumber\\
&\leq& \mathbb{P}\left[n'_{1} \leq \frac{n}{2}-\frac{\epsilon}{2}\right]+\mathbb{P}\left[n'_{2} \leq \frac{n}{2}-\frac{\epsilon}{2}\right]\nonumber\\
&\leq& \mathbb{P}\left[\bigcup_{i \in \{1,2,\ldots,m\}}\left\{\left|\hat{\tau}_{i1}\{\e^{(0)}\}-I\left(c_{i}=1\right)\right| \geq \frac{\epsilon}{n}\right\}\right]+
\mathbb{P}\left[\bigcup_{i \in \{m+1,m+2,\ldots,n\}}\left\{\left|\hat{\tau}_{i2}\{\e^{(0)}\}-I\left(c_{i}=2\right)\right| \geq \frac{\epsilon}{n}\right\}\right]\nonumber\\
&\leq& \sum\limits_{i=1}^{m}\mathbb{P}\left[\left|\hat{\tau}_{i1}\{\e^{(0)}\}-I\left(c_{i}=1\right)\right| \geq \frac{\epsilon}{n}\right]+\sum\limits_{i=m+1}^{n}\mathbb{P}\left[\left|\hat{\tau}_{i2}\{\e^{(0)}\}-I\left(c_{i}=2\right)\right| \geq \frac{\epsilon}{n}\right].
\label{nn2}
\end{eqnarray}
Similar to Lemma 1 in \citet{amini2013pseudo}, we upper bound $\mathbb{P}\left[|\hat{\tau}_{i1}\{\e^{(0)}\}-I\left(c_{i}=1\right)| \geq \frac{\epsilon}{n}\right],\ \ \forall i\in\{1,2, \ldots, m\}$ and $\mathbb{P}\left[\left|\hat{\tau}_{i2}\{\e^{(0)}\}-I\left(c_{i}=2\right)\right| \geq \frac{\epsilon}{n}\right],\ \ \forall i\in\left\{m+1,m+2, \ldots, n\right\}$ as follows.
With \eqref{PPhat}, it can be deduced that
$\left(\frac{\widehat{P}_{11}}{1-\widehat{P}_{11}}\right)/
\left(\frac{\widehat{P}_{12}}{1-\widehat{P}_{12}}\right)\geq \delta^{2}/\left(1-\delta\right)>1$.
Let $B_{ik}=\sum\limits_{j=1}^{n}A_{ij}I(e_{j}=k)$, $n_{k}=\sum\limits_{i=1}^{n}I(e_{j}=k)$ and $\tilde{\delta}=\delta^{2}/\left(1-\delta\right)$, we have
\begin{eqnarray}
&&\mathbb{P}\left[\left|\hat{\tau}_{i1}\{\e^{(0)}\}-I\left(c_{i}=1\right)\right|\geq\frac{\epsilon}{n}\right]\nonumber\\
&=&\mathbb{P}\left[\frac{\hat{\tau}_{i1}\{\e^{(0)}\}}{\hat{\tau}_{i2}\{\e^{(0)}\}}\leq\frac{1-\epsilon/n}{\epsilon/n}\right]\nonumber\\
&=&\mathbb{P}\left[\frac{{\widehat{P}_{11}}^{B_{i1}}{\widehat{P}_{12}}^{B_{i2}}\left(1-\widehat{P}_{11}\right)^{n_{1}-B_{i1}}\left(1-\widehat{P}_{12}\right)^{n_{2}-B_{i2}}}
{{\widehat{P}_{21}}^{B_{i1}}{\widehat{P}_{22}}^{B_{i2}}\left(1-\widehat{P}_{21}\right)^{n_{1}-B_{i1}}\left(1-\widehat{P}_{22}\right)^{n_{2}-B_{i2}}}
\leq \frac{1-\epsilon/n}{\epsilon/n}\right]\nonumber\nonumber\\
&=&\mathbb{P}\left[\left\{\left(\frac{\frac{\widehat{P}_{11}}{1-\widehat{P}_{11}}}
{\frac{\widehat{P}_{12}}{1-\widehat{P}_{12}}}\right)^{B_{i1}-B_{i2}}
\leq \frac{1-\epsilon/n}{\epsilon/n}\right\} \bigcap \left\{ \left\{B_{i1}-B_{i2}\geq 0\right\} \bigcup  \left\{B_{i1}-B_{i2}<0\right\} \right\} \right]\nonumber\\
&\leq& \mathbb{P}\left[\tilde{\delta}^{B_{i1}-B_{i2}}\leq \frac{1-\epsilon/n}{\epsilon/n}\right]+\mathbb{P}\left[B_{i1}-B_{i2}< 0\right]\nonumber\\
&=& \mathbb{P}\left[B_{i1}-B_{i2} \leq \frac{1}{\mathrm{log}\tilde{\delta}}\mathrm{log}\left(\frac{1-\epsilon/n}{\epsilon/n}\right)\right]+
\mathbb{P}\left[B_{i1}-B_{i2}< 0\right]\nonumber\\
&\leq& 2\mathbb{P}\left[B_{i1}-B_{i2} \leq \frac{1}{\mathrm{log}\tilde{\delta}}\mathrm{log}\left(\frac{1-\epsilon/n}{\epsilon/n}\right)\right].
\label{unpi}
\end{eqnarray}
Let $\xi_{i}\big(\sigma\{\e^{(0)}\}\big)=B_{i1}-B_{i2}$, and recall that $\widetilde{\xi}_{i}\big(\sigma\{\e^{(0)}\}\big)=\widetilde{B}_{i1}-\widetilde{B}_{i2}$. Then we have
\begin{eqnarray}
\xi_{i}\big(\sigma\{\e^{(0)}\}\big)-\widetilde{\xi}_{i}\big(\sigma\{\e^{(0)}\}\big)&=&(B_{i1}-\widetilde{B}_{i1})-(B_{i2}-\widetilde{B}_{i2})\nonumber\\
&=&\sum\limits_{j=1}^{n}(A_{ij}-\widetilde{A}_{ij})I(e_{j}=1)-\sum\limits_{j=1}^{n}(A_{ij}-\widetilde{A}_{ij})I(e_{j}=2)\nonumber\\
&\geq&-\sum\limits_{j=1}^{n}(A_{ij}-\widetilde{A}_{ij})I(e_{j}=2)\nonumber\\
&\geq&-\sum\limits_{j=1}^{n}(\widetilde{A}_{ij}+\widetilde{A}_{ji})I(e_{j}=2) \hspace{0.6cm} ( by\ A_{ij}-\widetilde{A}_{ij}\leq \widetilde{A}_{ij}+\widetilde{A}_{ji} ). \nonumber
\end{eqnarray}
Thus, we have shown that
\begin{eqnarray}
\xi_{i}\big(\sigma\{\e^{(0)}\}\big)\geq\widetilde{\xi}_{i}\big(\sigma\{\e^{(0)}\}\big)-\sum\limits_{j=1}^{n}\widetilde{A}_{ij}I(e_{j}=2)-\sum\limits_{j=1}^{n}\widetilde{A}_{ji}I(e_{j}=2).\nonumber
\end{eqnarray}
Consequently, we have
\begin{eqnarray}
&&\mathbb{P}\left[\xi_{i}\left(\sigma\{\e^{(0)}\}\right) \leq\frac{1}{\mathrm{log}\tilde{\delta}}\mathrm{log}\left(\frac{1-\epsilon/n}{\epsilon/n}\right)\right]\nonumber\\
&\leq&\mathbb{P}\left[\widetilde{\xi}_{i}\big(\sigma\{\e^{(0)}\}\big)-\sum\limits_{j=1}^{n}\widetilde{A}_{ij}I(e_{j}=2)-\sum\limits_{j=1}^{n}\widetilde{A}_{ji}I(e_{j}=2)\leq\frac{1}{\mathrm{log}\tilde{\delta}}\mathrm{log}\Big(\frac{1-\epsilon/n}{\epsilon/n}\Big)\right]\nonumber\\
&\leq&\mathbb{P}\left[\widetilde{\xi}_{i}\big(\sigma\{\e^{(0)}\}\big)\leq2(1+\epsilon)a_{\gamma}+\frac{1}{\mathrm{log}\tilde{\delta}}\mathrm{log}\Big(\frac{1-\epsilon/n}{\epsilon/n}\Big)\right]+
\mathbb{P}\left[\sum\limits_{j=1}^{n}\widetilde{A}_{ij}I\left(e_{j}=2\right)\geq\left(1+\epsilon\right)a_{\gamma}\right]\nonumber\\
&&+\mathbb{P}\left[\sum\limits_{j=1}^{n}\widetilde{A}_{ji}I\left(e_{j}=2\right)\geq\left(1+\epsilon\right)a_{\gamma}\right].
\label{xxii}
\end{eqnarray}
Now we consider the term $\mathbb{P}\left[\widetilde{\xi}_{i}\big(\sigma\{\e^{(0)}\}\big)\leq2(1+\epsilon)a_{\gamma}+\frac{1}{\mathrm{log}\tilde{\delta}}\mathrm{log}\Big(\frac{1-\epsilon/n}{\epsilon/n}\Big)\right]$. Recall that
\begin{eqnarray}
\widetilde{\xi}_{i}\big(\sigma\{\e^{(0)}\}\big)=\widetilde{B}_{i1}-\widetilde{B}_{i2}
= \sum\limits_{j=1}^{n}\widetilde{A}_{ij}\sigma_{j}\{\e^{(0)}\},\hspace{0.4cm} \text{where} \;
\sigma_{j}\{\e^{(0)}\}=
\left\{
\begin{aligned}
1, \; e_{j}=1\nonumber\\
-1, \; e_{j}=2\nonumber
\end{aligned}
\right..
\end{eqnarray}
We have shown in \eqref{xit} that
\begin{eqnarray}
\mathbb{P}\left[\widetilde{\xi}_{i}\left(\sigma\{\e^{(0)}\}\right) \leq (2\gamma-1)(a-b)-t \right] \leq  e^{-\frac{t^{2}}{4(a+b)}}, \hspace{0.5cm} \forall t \in \left[0, 3(a+b)\right].
\label{xitt}
\end{eqnarray}
Take $t=(2\gamma-1)(a-b)-\left\{2(1+\epsilon)a_{\gamma}+\frac{1}{\mathrm{log}\tilde{\delta}}\mathrm{log}\left(\frac{1-\epsilon/n}{\epsilon/n}\right)\right\}$. Recall $a_{\gamma}=(1-\gamma)a+\gamma b$, and $(a-b)\rightarrow \infty, \;n\rightarrow \infty$. Then when $n$ is large enough, we have
\begin{eqnarray}
\frac{1-\epsilon}{2}(2\gamma-1)(a-b)>\frac{1}{\mathrm{log}\tilde{\delta}}\mathrm{log}\Big(\frac{1-\epsilon/n}{\epsilon/n}\Big).
\label{eps1}
\end{eqnarray}
With the assumption that $\epsilon(2\gamma-1)(a-b)\geq2(1+\epsilon)a_{\gamma}$ and \eqref{eps1}, we have
\begin{eqnarray}
2(1+\epsilon)a_{\gamma}+\frac{1}{\mathrm{log}\tilde{\delta}}\mathrm{log}\Big(\frac{1-\epsilon/n}{\epsilon/n}\Big)
\leq \frac{1+\epsilon}{2}(2\gamma-1)(a-b).\nonumber
\end{eqnarray}
Thus, we have
\begin{eqnarray}
0<\frac{1-\epsilon}{2}(2\gamma-1)(a-b)\leq t\leq (2\gamma-1)(a-b) \leq 3(a+b).
\label{eq2}
\end{eqnarray}
By plugging $t=(2\gamma-1)(a-b)-\left\{2(1+\epsilon)a_{\gamma}+\frac{1}{\mathrm{log}\tilde{\delta}}\mathrm{log}\Big(\frac{1-\epsilon/n}{\epsilon/n}\Big)\right\}$ in \eqref{xitt}, it follows that
\begin{eqnarray}
\mathbb{P}\left[\widetilde{\xi}_{i}\big(\sigma\{\e^{(0)}\}\big)\leq2(1+\epsilon)a_{\gamma}+\frac{1}{\mathrm{log}\tilde{\delta}}\mathrm{log}\Big(\frac{1-\epsilon/n}{\epsilon/n}\Big)\right]
\leq e^{-\frac{t^{2}}{4(a+b)}}\leq e^{-\frac{\left(\frac{1-\epsilon}{2}\right)^{2}(2\gamma-1)^{2}(a-b)^{2}}{4(a+b)}}.
\label{xi2}
\end{eqnarray}

Next, we consider the terms $\mathbb{P}\left[\sum\limits_{j=1}^{n}\widetilde{A}_{ij}I(e_{j}=2)\geq(1+\epsilon)a_{\gamma}\right]$ and $\mathbb{P}\Bigg[\sum\limits_{j=1}^{n}\widetilde{A}_{ji}I(e_{j}=2)\geq(1+\epsilon)a_{\gamma}\Bigg]$.
Let $\widetilde{A}_{i\ast}\{\e^{(0)}\}=\sum\limits_{j=1}^{n}\widetilde{A}_{ij}I(e_{j}=2)$ and $\widetilde{A}_{\ast i}\{\e^{(0)}\}=\sum\limits_{j=1}^{n}\widetilde{A}_{ji}I(e_{j}=2)$. By symmetry, we have that
\begin{center}
$\mathbb{P}\left[\widetilde{A}_{i\ast}\{\e^{(0)}\}\geq(1+\epsilon)a_{\gamma}\right]=\mathbb{P}\left[\widetilde{A}_{\ast i}\{\e^{(0)}\}\geq(1+\epsilon)a_{\gamma}\right]$.
\end{center}
Note that since both $\widetilde{A}_{i\ast}\{\e^{(0)}\}$ and $\widetilde{A}_{\ast i}\{\e^{(0)}\}$ are sums of independent bounded random variables, we can apply the Bernstein inequality to obtain upper bounds. For $\widetilde{A}_{i\ast}\{\e^{(0)}\}$, we have \\
$\left|\widetilde{A}_{ij}I(e_{j}=2)-\mathbb{E}\widetilde{A}_{ij}I(e_{j}=2)\right|\leq1$, and
\begin{eqnarray}
&&\mathbb{E}\left[\widetilde{A_{i\ast}}\{\e^{(0)}\}\right]=(1-\gamma)m\cdot \frac{a}{m}+\gamma m\cdot\frac{b}{m}=(1-\gamma)a+\gamma b=a_{\gamma}\nonumber,\\
&&\upsilon=\text{Var}\left(\widetilde{A}_{i\ast}\{\e^{(0)}\}\right)=\sum\limits_{j=1}^{n}\text{Var}(\widetilde{A}_{ij})I(e_{j}=2)\leq
\sum\limits_{j=1}^{n}\mathbb{E}\left[\widetilde{A}_{ij}^{2}\right]I(e_{j}=2)=a_{\gamma}\nonumber.
\end{eqnarray}
Then by applying the Bernstein inequality to $\widetilde{A}_{i\ast}\{\e^{(0)}\}$, we have
\begin{eqnarray}
\mathbb{P}\left[\widetilde{A}_{i\ast}\{\e^{(0)}\}\geq \mathbb{E}\left[\widetilde{A}_{i\ast}\{\e^{(0)}\}\right]+t\right]\leq
e^{-\frac{t^{2}}{2(v+t/3)}}, \hspace{0.5cm} \forall t\geq 0.
\label{Aistar}
\end{eqnarray}
Let $t=\epsilon a_{\gamma}\geq0$ in \eqref{Aistar} and by noting that $v\leq a_{\gamma}$, we have
\begin{eqnarray}
\mathbb{P}\left[\widetilde{A}_{\ast i}\{\e^{(0)}\}\geq (1+\epsilon)a_{\gamma}\right]=\mathbb{P}\left[\widetilde{A}_{i\ast}\{\e^{(0)}\}\geq (1+\epsilon)a_{\gamma}\right]\leq
e^{-\frac{\epsilon^{2}/2}{1+\epsilon/3}a_{\gamma}}.
\label{eq3}
\end{eqnarray}
Thus, by \eqref{unpi}, \eqref{xxii}, \eqref{xi2} and \eqref{eq3}, it follows that for $i=1,2,\ldots,m$,
\begin{eqnarray}
\mathbb{P}\left[\left|\hat{\tau}_{i1}\{\e^{(0)}\}-I(c_{i}=1)\right|\geq\frac{\epsilon}{n}\right]\leq
2\left\{e^{-\frac{(\frac{1-\epsilon}{2})^{2}(2\gamma-1)^{2}(a-b)^{2}}{4(a+b)}}+
2e^{-\frac{\epsilon^{2}/2}{1+\epsilon/3}a_{\gamma}}\right\}.
\label{ppi1}
\end{eqnarray}
Similarly, we can also obtain, for $i=m+1, m+2, \ldots, n$,
\begin{eqnarray}
\mathbb{P}\left[\left|\hat{\tau}_{i2}\{\e^{(0)}\}-I(c_{i}=2)\right|\geq\frac{\epsilon}{n}\right]\leq
2\left\{e^{-\frac{(\frac{1-\epsilon}{2})^{2}(2\gamma-1)^{2}(a-b)^{2}}{4(a+b)}}+
2e^{-\frac{\epsilon^{2}/2}{1+\epsilon/3}a_{\gamma}}\right\}.
\label{ppi2}
\end{eqnarray}
By \eqref{nn2}, \eqref{ppi1} and \eqref{ppi2}, we have
\begin{eqnarray}
\mathbb{P}\left[\left|n'_{1}-n'_{2}\right| \geq \epsilon\right]\leq
2n\left\{e^{-\frac{(\frac{1-\epsilon}{2})^{2}(2\gamma-1)^{2}(a-b)^{2}}{4(a+b)}}+
2e^{-\frac{\epsilon^{2}/2}{1+\epsilon/3}a_{\gamma}}\right\}.
\label{nn3}
\end{eqnarray}

According to \eqref{BB}, we still need to obtain the upper bound of $\mathbb{P}\Bigg[\sum\limits_{i=1}^{n}A_{ij}I(c_{i}=1)-\sum\limits_{i=1}^{n}A_{ij}I(c_{i}=2)\leq 2\epsilon\Bigg]$. Let $\eta_{j}\big(\sigma(\c)\big)=\sum\limits_{i=1}^{n}A_{ij}I(c_{i}=1)-\sum\limits_{i=1}^{n}A_{ij}I(c_{i}=2)$. We then have
\begin{eqnarray}
\eta_{j}\big(\sigma(\c)\big)-\widetilde{\eta}_{j}\big(\sigma(\c)\big)&=&\sum\limits_{i=1}^{n}(A_{ij}-\widetilde{A}_{ij})I(c_{i}=1)-
\sum\limits_{i=1}^{n}(A_{ij}-\widetilde{A}_{ij})I(c_{i}=2)\nonumber\\
&\geq& -\sum\limits_{i=1}^{n}(A_{ij}-\widetilde{A}_{ij})I(c_{i}=2)\nonumber\\
&\geq& -\sum\limits_{i=1}^{n}(\widetilde{A}_{ij}+\widetilde{A}_{ji})I(c_{i}=2)
\hspace{0.6cm} ( by\ A_{ij}-\widetilde{A}_{ij}\leq \widetilde{A}_{ij}+\widetilde{A}_{ji} ).\nonumber
\end{eqnarray}
Let $\widetilde{A}_{\ast j}(\c)=\sum\limits_{i=1}^{n}\widetilde{A}_{ij}I(c_{i}=2)$ and $\widetilde{A}_{j\ast}(\c)=\sum\limits_{i=1}^{n}\widetilde{A}_{ji}I(c_{i}=2)$. We have
\begin{eqnarray}
\eta_{j}\big(\sigma(\c)\big)\geq\widetilde{\eta}_{j}\big(\sigma(\c)\big)-\widetilde{A}_{\ast j}(\c)-\widetilde{A}_{j\ast}(\c).\nonumber
\end{eqnarray}
From the assumption that $\epsilon(2\gamma-1)(a-b)\geq2(1+\epsilon)a_{\gamma}$, $\gamma \in (\frac{1}{2}, 1)$ and $a>b$, we can get that
\begin{eqnarray}
a\geq\frac{2(1+\epsilon)\gamma+\epsilon(2\gamma-1)}{\epsilon(2\gamma-1)-2(1+\epsilon)(1-\gamma)}b>b.
\nonumber
\end{eqnarray}
It is not difficult to check that there exists $\rho \in(0,1)$ such that
\begin{eqnarray}
\rho(a-b)-2(1+\epsilon)b>0.
\label{dlts}
\end{eqnarray}
Then, we have
\begin{eqnarray}
&&\mathbb{P}\left[\eta_{j}\big(\sigma(\c)\big)\leq2\epsilon\right]\nonumber\\
&\leq&\mathbb{P}\left[\widetilde{\eta}_{j}\left(\sigma(\c)\right)-\widetilde{A}_{\ast j}(\c)-\widetilde{A}_{j\ast}(\c)\leq 2\epsilon\right]\nonumber\\
&\leq&\mathbb{P}\left[\widetilde{\eta}_{j}\big(\sigma(\c)\big)\leq\frac{1-\rho}{2}(a-b)+2(1+\epsilon)b+2\epsilon\right]+
\mathbb{P}\left[\widetilde{A}_{\ast j}(\c)\geq(1+\epsilon)b+\frac{1-\rho}{4}(a-b)\right]\nonumber\\
&&+\mathbb{P}\left[\widetilde{A}_{j\ast}(\c)\geq(1+\epsilon)b+\frac{1-\rho}{4}(a-b)\right].
\label{etaa}
\end{eqnarray}
Consider the term $\mathbb{P}\left[\widetilde{\eta}_{j}\big(\sigma(\c)\big)\leq\frac{1-\rho}{2}(a-b)+2(1+\epsilon)b+2\epsilon\right]$. Recall that in \eqref{etaj1}, we have shown
\begin{eqnarray}
\mathbb{P}\left[\widetilde{\eta}_{j}\big(\sigma(\c)\big) \leq (a-b)-t \right] \leq  e^{-\frac{t^{2}}{4(a+b)}}, \hspace{0.5cm} \forall t \in [0, 3(a+b)].
\label{etaj2}
\end{eqnarray}
Then we can take
\begin{eqnarray}
t&=&(a-b)-\left\{\frac{1-\rho}{2}(a-b)+2(1+\epsilon)b+2\epsilon\right\}\nonumber\\
&=&\left\{\frac{1-\rho}{2}(a-b)-2\epsilon\right\}+\left\{\rho(a-b)-2(1+\epsilon)b\right\}.
\label{t1}
\end{eqnarray}
With \eqref{dlts}, \eqref{t1} and $(a-b)\rightarrow \infty$ as $n \rightarrow \infty$, it follows that when $n$ is large enough we have
\begin{eqnarray}
0<\frac{1-\rho}{4}(a-b)\leq t\leq 3(a+b).
\label{dlt1}
\end{eqnarray}
With \eqref{etaj2}, \eqref{t1}, \eqref{dlt1}, we get (when $n$ is large enough)
\begin{eqnarray}
\mathbb{P}\left[\widetilde{\eta}_{j}\big(\sigma(\c)\big)\leq\frac{1-\rho}{2}(a-b)+2(1+\epsilon)b+2\epsilon\right]
\leq e^{-\frac{\left(\frac{1-\rho}{4}\right)^{2}(a-b)^{2}}{4(a+b)}}.
\label{bd1}
\end{eqnarray}
To bound the term $\mathbb{P}\left[\widetilde{A}_{\ast j}(\c)\geq(1+\epsilon)b+\frac{1-\rho}{4}(a-b)\right]$, first recall that $\widetilde{A}_{\ast j}(\c)=\sum\limits_{i=1}^{n}\widetilde{A}_{ij}I(c_{i}=2)$ is the sum of independent random variables and $\Big|\widetilde{A}_{ij}I(c_{i}=2)-\mathbb{E}\widetilde{A}_{ij}I(c_{i}=2)\Big|\leq1$. Therefore we can also apply the Bernstein inequality. We have
\begin{eqnarray}
&&\mathbb{E}\left[\widetilde{A}_{\ast j}(\c)\right]=m\cdot\frac{b}{m}=b,\label{ebaa}\\
&&v=\text{Var}\left(\widetilde{A}_{\ast j}(\c)\right)=\sum\limits_{j=1}^{n}\text{Var}(\widetilde{A}_{ij})I(c_{j}=2)\leq
\sum\limits_{j=1}^{n}\mathbb{E}\left[\widetilde{A}_{ij}^{2}\right]I(c_{j}=2)=b.\nonumber
\end{eqnarray}
Thus, by applying the Bernstein inequality to $\widetilde{A}_{\ast j}(\c)$, we have
\begin{eqnarray}
\mathbb{P}\left[\widetilde{A}_{\ast j}(\c)\geq \mathbb{E}\left[\widetilde{A}_{\ast j}(\c)\right]+t\right]\leq
e^{-\frac{t^{2}/2}{v+t/3}}\leq e^{-\frac{t^{2}/2}{b+t/3}},\hspace{0.5cm} \forall t\geq0.
\label{bara}
\end{eqnarray}
Take $t=\epsilon b+\frac{1-\rho}{4}(a-b)$. With \eqref{bara} and \eqref{ebaa}, we have
\begin{eqnarray}
\mathbb{P}\left[\widetilde{A}_{\ast j}(\c)\geq(1+\epsilon)b+\frac{1-\rho}{4}(a-b)\right]\leq
e^{-\frac{\frac{1}{2}\left(\frac{1-\rho}{4}\right)^{2}(a-b)^{2}}{b+2a/3}}\leq
e^{-\frac{\left(\frac{1-\rho}{4}\right)^{2}(a-b)^2}{2(a+b)}}.
\label{bara1}
\end{eqnarray}
By symmetry, we also have
\begin{eqnarray}
\mathbb{P}\left[\widetilde{A}_{\ast j}(\c)\geq(1+\epsilon)b+\frac{1-\rho}{4}(a-b)\right]\leq
e^{-\frac{\frac{1}{2}\left(\frac{1-\rho}{4}\right)^{2}(a-b)^{2}}{b+2a/3}}\leq
e^{-\frac{\left(\frac{1-\rho}{4}\right)^{2}(a-b)^2}{2(a+b)}}.
\label{bara2}
\end{eqnarray}
Therefore, with \eqref{etaa}, \eqref{bd1}, \eqref{bara1} and \eqref{bara2}, it follows that
\begin{eqnarray}
\mathbb{P}\left[\eta_{j}\big(\sigma(\c)\big)\leq2\epsilon\right]\leq e^{-\frac{\left(\frac{1-\rho}{4}\right)^{2}(a-b)^{2}}{4(a+b)}}+2
e^{-\frac{\left(\frac{1-\rho}{4}\right)^{2}(a-b)^2}{2(a+b)}}\leq
3e^{-\frac{\left(\frac{1-\rho}{4}\right)^{2}(a-b)^{2}}{4(a+b)}}.
\label{etaj3}
\end{eqnarray}
With \eqref{BB}, \eqref{ppi1} and \eqref{etaj3}, we can get that, for $j=1, 2, \ldots, m$,
\begin{eqnarray}
\mathbb{P}\left[B_{1j}^{'}-B_{2j}^{'} \leq \epsilon\right]\leq
3e^{-\frac{\left(\frac{1-\rho}{4}\right)^{2}(a-b)^{2}}{4(a+b)}}+
n\left\{e^{-\frac{\left(\frac{1-\epsilon}{2}\right)^{2}(2\gamma-1)^{2}(a-b)^{2}}{4(a+b)}}+
2e^{-\frac{\epsilon^{2}/2}{1+\epsilon/3}a_{\gamma}}\right\}.
\label{Bt1}
\end{eqnarray}
Similarly, with the same arguments, we can get that for $j=m+1, m+2, \ldots, n$,
\begin{eqnarray}
\mathbb{P}\left[B_{2j}^{'}-B_{1j}^{'} \leq \epsilon\right]\leq
3e^{-\frac{\left(\frac{1-\rho}{4}\right)^{2}(a-b)^{2}}{4(a+b)}}+
n\left\{e^{-\frac{\left(\frac{1-\epsilon}{2}\right)^{2}(2\gamma-1)^{2}(a-b)^{2}}{4(a+b)}}+
2e^{-\frac{\epsilon^{2}/2}{1+\epsilon/3}a_{\gamma}}\right\}.
\label{Bt2}
\end{eqnarray}
Finally, with \eqref{nn3}, \eqref{Bt1} and \eqref{Bt2}, it follows that when $n$ is large enough, we have
\begin{eqnarray}
&&\mathbb{P}\left[\hat{\c}\{\e^{(0)}\}\neq \c\right]\nonumber\\
&=&\mathbb{P}\left[\bigcup_{j \in \{1,2,\ldots,n \}}\big\{\hat{c}_{j}\{\e^{(0)}\}\neq c_{j}\big\} \right]\nonumber\\
&\leq&\mathbb{P}\left[\left\{\bigcup_{j \in \{1,2,\ldots,m \}}\left\{B_{1j}^{'}-B_{2j}^{'}\leq\epsilon\right\}\right\} \bigcup \left\{\bigcup_{j \in \{m+1,m+2,\ldots,n \}} \left\{ B_{2j}^{'}-B_{1j}^{'}\leq\epsilon\right\}\right\}\bigcup\left\{\left|n'_{1}-n'_{2}\right| \geq \epsilon\right\}\right]\nonumber\\
&\leq& \sum\limits_{j=1}^{m}\mathbb{P}\left[B_{1j}^{'}-B_{2j}^{'}\leq\epsilon\right]+
\sum\limits_{j=m+1}^{n}\mathbb{P}\left[B_{2j}^{'}-B_{1j}^{'}\leq\epsilon\right]+\mathbb{P}\left[\left|n'_{1}-n'_{2}\right| \geq \epsilon\right]\nonumber\\
&=&3ne^{-\frac{\left(\frac{1-\rho}{4}\right)^{2}(a-b)^{2}}{4(a+b)}}+
n(n+2)\left\{e^{-\frac{\left(\frac{1-\epsilon}{2}\right)^{2}(2\gamma-1)^{2}(a-b)^{2}}{4(a+b)}}+
2e^{-\frac{\epsilon^{2}/2}{1+\epsilon/3}a_{\gamma}}\right\}.\nonumber
\end{eqnarray}
Therefore, we have
\begin{eqnarray}
&&\mathbb{P}\left[\hat{\c}\{\e^{(0)}\}= \c\right]= 1-\mathbb{P}\left[\hat{\c}\{\e^{(0)}\}\neq \c\right]\nonumber\\
&\geq& 1-\left[3ne^{-\frac{\left(\frac{1-\rho}{4}\right)^{2}(a-b)^{2}}{4(a+b)}}+
n(n+2)\left\{e^{-\frac{\left(\frac{1-\epsilon}{2}\right)^{2}(2\gamma-1)^{2}(a-b)^{2}}{4(a+b)}}+
2e^{-\frac{\epsilon^{2}/2}{1+\epsilon/3}a_{\gamma}}\right\}\right].\hspace{2cm} \nonumber
\end{eqnarray}
Thus we complete the proof of Theorem 3.

\subsection{Distributions of $\hat{\c}^{(\text{s})}$ and $\hat{\c}^{(\text{w})}$}
\noindent
\textcolor{black}{
We first show that $\hat{\c}^{(\text{w})}$ is weakly consistent to $\bm c$. Let
$X_{i}\triangleq 1(\hat{c}_{i}^{(\text{w})}\neq c_{i})-\mathbb{P}(\hat{c}_{i}^{(\text{w})}\neq c_{i})$, where $\mathbb{P}(\hat{c}_{i}^{(\text{w})}\neq c_{i})=(1+\pi_{1})p_{n}$ with $p_{n}=\frac{1}{\log n}$. Then, it can be seen that
\begin{eqnarray}
&& |X_{i}|\leq 1,\; \forall i=1, 2, \ldots, n, \nonumber\\
&& \mathbb{E}X_{i}=0,\; \forall i=1, 2, \ldots, n,\nonumber\\
&& \sum\limits_{i=1}^{n}\mathbb{E}X_{i}^{2}=n\left[(1+\pi_{1})p_{n}\left\{1-(1+\pi_{1})p_{n}\right\}\right].\nonumber
\end{eqnarray}
Thus, by applying Bernstein inequality for $\Sigma_{i=1}^{n}X_{i}$, we can get that
\begin{eqnarray}
\mathbb{P}\left\{\sum\limits_{i=1}^{n}X_{i}\geq t\right\} \leq
\exp\left(-\frac{t^{2}/2}{n\left[(1+\pi_{1})p_{n}\left\{1-(1+\pi_{1})p_{n}\right\}\right]
+\frac{1}{3}t}\right), \quad \forall t\geq 0.\label{eq_bn}
\end{eqnarray}
Recall $X_{i}\triangleq 1(\hat{c}_{i}^{(\text{w})}\neq c_{i})-\mathbb{P}(\hat{c}_{i}^{(\text{w})}\neq c_{i})$. We plug $t=n\epsilon-\sum\limits_{i=1}^{n}\mathbb{P}(\hat{c}_{i}^{(\text{w})}\neq c_{i})$ (which is nonnegative when $n$ is sufficient large) into  (\ref{eq_bn}) and get that
\begin{eqnarray}
&&\mathbb{P}\left\{\frac{1}{n}\sum\limits_{i=1}^{n}1(\hat{c}_{i}^{(\text{w})}\neq c_{i})\geq \epsilon\right\}\\
&=&\mathbb{P}\left\{\sum\limits_{i=1}^{n}1(\hat{c}_{i}^{(\text{w})}\neq c_{i})-\sum\limits_{i=1}^{n}\mathbb{P}(\hat{c}_{i}^{(\text{w})}\neq c_{i})\geq n\epsilon-\sum\limits_{i=1}^{n}\mathbb{P}(\hat{c}_{i}^{(\text{w})}\neq c_{i})\right\}\nonumber\\
&\leq& \exp\left(-\frac{\left\{n\epsilon-n(1+\pi_{1})p_{n}\right\}^{2}/2}{n\left[(1+\pi_{1})p_{n}\left\{1-(1+\pi_{1})p_{n}\right\}\right]
+\left\{n\epsilon-n(1+\pi_{1})p_{n}\right\}/3}\right).
\label{eq_up}
\end{eqnarray}
Thus, $\hat{\c}^{(\text{w})}$ is weakly consistent to $\bm c$.
Next, we show that $\hat{\c}^{(\text{w})}$ is not strongly consistent to $\bm c$. Specifically, we have
\begin{eqnarray}
\mathbb{P}(\hat{\c}^{(\text{w})}=\bm c)=\prod\limits_{i=1}^{n}\mathbb{P}(\hat{c}_{i}^{(\text{w})}= c_{i})
\leq\prod\limits_{i=1}^{n}\left(1-\frac{1}{\log n}\right)
=\left\{\left(1-\frac{1}{\log n}\right)^{-\log n}\right\}^{-\frac{n}{\log n}}.\label{eq_nAC}
\end{eqnarray}
Thus by \eqref{eq_nAC}, we know that $\hat{\c}^{(\text{w})}$ is not strongly consistent to $\bm c$.
}

\textcolor{black}{
By the classical central limit theorem for independent and identically distributed random variables, we have
\begin{eqnarray}
\sqrt{n}\left\{\frac{1}{n}\sum\limits_{i=1}^{n}1(c_{i}=1)-\pi_{1}\right\}\stackrel{d}{\longrightarrow} N\left\{0, \pi_{1}(1-\pi_{1})\right\}.
\label{eq_Tnorm}
\end{eqnarray}
We also have that
\begin{eqnarray}
&&\sqrt{n}\left\{\frac{1}{n}\sum\limits_{i=1}^{n}1(\hat{c}_{i}^{(s)}=1)-\pi_{1}\right\}-
\sqrt{n}\left\{\frac{1}{n}\sum\limits_{i=1}^{n}1(c_{i}=1)-\pi_{1}\right\}\nonumber\\
&=&\frac{1}{\sqrt{n}}\sum\limits_{i=1}^{n}\left\{1(\hat{c}_{i}^{(s)}=1)-1(c_{i}=1)\right\}=o_{p}(1),\nonumber
\end{eqnarray}
which is based on the fact that $\forall \epsilon>0$,
\begin{eqnarray}
P\left[\left|\frac{1}{\sqrt{n}}\sum\limits_{i=1}^{n}\left\{1(\hat{c}_{i}^{(s)}=1)-1(c_{i}=1)\right\}\right|\geq\epsilon\right]
\leq P(\bm c^{(s)}\neq \bm c)=o(1).\nonumber
\end{eqnarray}
Thus, $\sqrt{n}\left\{\frac{1}{n}\sum\limits_{i=1}^{n}1(\hat{c}_{i}^{(s)}=1)-\pi_{1}\right\}$ has the same limit distribution as $\sqrt{n}\left\{\frac{1}{n}\sum\limits_{i=1}^{n}1(c_{i}=1)-\pi_{1}\right\}$.}

\textcolor{black}{
Finally, we show that
\begin{eqnarray}
\sqrt{n}\left\{\frac{1}{n}\sum\limits_{i=1}^{n}1(\hat{c}_{i}^{(\text{w})}=1)-\left(\pi_{1}+\frac{1-3\pi_{1}}{\log n}\right)\right\}\stackrel{d}{\longrightarrow} N\left\{0, \pi_{1}(1-\pi_{1})\right\}.\nonumber
\end{eqnarray}
Let $X_{ni}\triangleq 1(\hat{c}_{i}^{(\text{w})}=1)-\mathbb{P}(\hat{c}_{i}^{(\text{w})}=1)$.  We have
\begin{eqnarray}
&& \mathbb{E}X_{ni}=0,\nonumber\\
&& s_{n}^{2}=\frac{1}{n}\sum\limits_{i=1}^{n}\mathbb{E}X_{ni}^{2}=(\pi_{1}-\pi_{1}^{2})-O(p_{n})\rightarrow s^{2}=\pi_{1}(1-\pi_{1})\neq 0, \; \mathrm{as}\; n\rightarrow \infty.\nonumber
\end{eqnarray}
We show the following Lindeberg condition. Specifically, note that $\mathbb{P}(\hat{c}_{i}^{(\text{w})}=1)=\pi_{1}+\frac{1-3\pi_{1}}{\log n}$, then for every $\epsilon>0$, we have
\begin{eqnarray}
&&\frac{1}{n}\sum\limits_{i=1}^{n}\mathbb{E}\left\{X_{ni}^{2}1\left(|X_{ni}|\geq \epsilon \sqrt{n}\right)\right\}\nonumber\\
&=&\mathbb{E}\left[|1\left(\hat{c}_{i}^{(\text{w})}=1\right)-\mathbb{P}(\hat{c}_{i}^{(\text{w})}=1)|^{2}1\left\{|1\left(\hat{c}_{i}^{(\text{w})}=1\right)-\mathbb{P}(\hat{c}_{i}^{(\text{w})}=1)|\geq \epsilon \sqrt{n}\right\}\right]\nonumber\\
&\leq& \mathbb{P}\left\{|1\left(\hat{c}_{i}^{(\text{w})}=1\right)-\mathbb{P}(\hat{c}_{i}^{(\text{w})}=1)|\geq \epsilon \sqrt{n}\right\}\nonumber\\
&=& \mathbb{P}\left\{|1\left(\hat{c}_{i}^{(\text{w})}=1\right)-(\pi_{1}+\frac{1-3\pi_{1}}{\log n})|\geq \epsilon \sqrt{n}\right\}.\label{eq_tp}
\end{eqnarray}
Also note that
\begin{eqnarray}
\mathbb{P}\left\{|1\left(\hat{c}_{i}^{(\text{w})}=1\right)-(\pi_{1}+\frac{1-3\pi_{1}}{\log n})|\geq \epsilon \sqrt{n}\right\}\leq \mathbb{P}\left(1\geq \epsilon\sqrt{n}\right)\rightarrow 0, \; \mathrm{as}\; n\rightarrow \infty.\label{eq_tz}
\end{eqnarray}
Thus, putting \eqref{eq_tp} and \eqref{eq_tz} together yields
\begin{eqnarray}
\frac{1}{n}\sum\limits_{i=1}^{n}\mathbb{E}\left\{X_{ni}^{2}1\left(|X_{ni}|\geq \epsilon \sqrt{n}\right)\right\}
\rightarrow 0,\quad \mathrm{as}\; n\rightarrow \infty.
\end{eqnarray}
By the Lindeberg-Feller central limit theorem, we can get that
\begin{eqnarray}
\sqrt{n}\left(\frac{1}{n}\sum\limits_{i=1}^{n}X_{ni}\right)
\stackrel{d}{\longrightarrow}N(0,s^{2}),\nonumber
\end{eqnarray}
which is also
\begin{eqnarray}
\sqrt{n}\left\{\frac{1}{n}\sum\limits_{i=1}^{n}1(\hat{c}_{i}^{(\text{w})}=1)-\left(\pi_{1}+\frac{1-3\pi_{1}}{\log n}\right)\right\}\stackrel{d}{\longrightarrow}N\left\{0,\pi_{1}(1-\pi_{1})\right\}.\label{eq_clt}
\end{eqnarray}
}

\baselineskip=24.5pt

\subsection{Extension to the Bipartite SBM}
\label{section:bisbm}
\noindent
The bipartite network is a ubiquitous class of networks, in which nodes are of two disjoint types and edges are only formed between nodes from different types. Bipartite networks can be used to characterize many real-world systems, such as authorship of papers and people attending events \citep{zhang2018modularity}.
Community detection in bipartite networks have been studied in many scientific fields, such as text mining \citep{bisson2008chi}, physics \citep{larremore2014efficiently}, and genetic studies \citep{madeira2010identification}.
In this section, we extend the proposed profile-pseudo likelihood method to the case of bipartite stochastic blockmodels (BiSBM).

Let $G(V_1,V_2,E)$ denote a bipartite network, where $V_{1}=\{1,\ldots, m\}$ and $V_{2}=\{1,\ldots, n\}$ are node sets of the two different types of nodes, respectively, and $E$ is the set of edges between nodes in $V_{1}$ and $V_{2}
$. The network $G(V_1,V_2,E)$ can be uniquely represented by the corresponding $m\times n$ bi-adjacency matrix $\A=[A_{ij}]$, where $A_{ij}=1$ if there is an edge from node $i$ of type 1 to node $j$ of type 2 and $A_{ij}=0$ otherwise.
Under the BiSBM, nodes in $V_1$ form $K_1$ blocks and nodes in $V_2$ form $K_2$ blocks.
Specifically, for nodes in $V_1$, the labels $\c_1=(c_{11}, c_{12}, \ldots, c_{1m})$ are drawn independently from a multinomial distribution with parameters $\bm \pi_1=(\pi_{11}, \pi_{12},\ldots, \pi_{1K_{1}})$, and for nodes in $V_2$, the labels $\c_2=(c_{21}, c_{22}, \ldots, c_{2n})$ are drawn independently from a multinomial distribution with parameters $\bm \pi_2=(\pi_{21}, \pi_{22},\ldots, \pi_{2K_{2}})$.
Conditional on $\c_1$ and $\c_2$, the edges $A_{ij}$'s are independent Bernoulli variables with
\begin{center}
$\mathbb{E}[A_{ij}|\c_1, \c_2]=P_{c_{1i}c_{2j}}$,
\end{center}
where $\bm P=[P_{kl}]$ is a $K_{1}\times K_{2}$ matrix. The goal of community detection is then to estimate the node labels $\c_1$ and $\c_2$ from the bi-adjacent matrix $\bm A$.
\begin{algorithm}[!t]
\caption{BiSBM Profile-Pseudo Likelihood Maximization Algorithm.}
\begin{algorithmic}
\STATE \textbf{Step 1}: Initialize $\e_{1}^{(0)}$ and $\e_{2}^{(0)}$ by applying {SCP} to $\A\A^\top$ and $\A^\top\A$, respectively.
\STATE \textbf{Step 2}: Calculate $\bOmega^{(0)}=(\bm \pi_{1}^{(0)}, \bm P^{(0)})$. That is, for $1\le k\le K_1$ and $1\le l\le K_2$,
\begin{eqnarray*}
&{\pi}^{(0)}_{1k}=\frac{1}{m}\sum\limits_{i=1}^{m}I({e}^{(0)}_{1i}=k),\quad
{{P}}^{(0)}_{kl}=\frac{\sum\limits_{i=1}^{m}\sum\limits_{j=1}^{n}A_{ij}I({e}^{(0)}_{1i}=k)I({e}^{(0)}_{2j}=l)}{\sum\limits_{i=1}^{m}\sum\limits_{j=1}^{n}I({e}^{(0)}_{1i}=k)I({e}^{(0)}_{2j}=l)}.
\end{eqnarray*}
\STATE \textbf{Step 3}: Initialize ${{\bOmega}}^{(0,0)}=({\bm \pi}^{(0,0)}_{1}, {{\bm P}}^{(0,0)})=({\bm \pi}^{(0)}_{1}, {{\bm P}}^{(0)})$.
\REPEAT
\REPEAT
\STATE \textbf{Step 4}: E-step: compute ${\tau}_{ik}^{(s,t+1)}$. That is, for $1\le k\le K_{1}$ and $1\le i\le m$,
\begin{eqnarray*}
&\tau_{ik}^{(s,t+1)}=\frac{{\pi}^{(s,t)}_{1k}\prod\limits_{j=1}^{n}\left\{{{P}}^{(s,t)}_{k{e}^{(s)}_{2j}}\right\}^{A_{ij}}\left\{1-{{P}}^{(s,t)}_{k{e}^{(s)}_{2j}}\right\}^{1-A_{ij}}}
{\sum\limits_{l=1}^{K_{1}}{\pi}^{(s,t)}_{1l}\prod\limits_{j=1}^{n}\left\{{{P}}^{(s,t)}_{l{e}^{(s)}_{2j}}\right\}^{A_{ij}}\left\{1-{{P}}^{(s,t)}_{l{e}^{(s)}_{2j}}\right\}^{1-A_{ij}}}.\nonumber
\end{eqnarray*}
\STATE \textbf{Step 5}: M-step: compute $\bm \pi^{(s,t+1)}_{1}$, $\bm P^{(s,t+1)}$. That is, for $1\le k\le K_{1}$ and $1\le l\le K_{2}$,
\begin{eqnarray*}
&\pi_{1k}^{(s,t+1)}=\frac{1}{n}\sum\limits_{i=1}^{m}\tau_{ik}^{(s,t+1)}, \quad
P_{kl}^{(s,t+1)}=\frac{\sum\limits_{i=1}^{m}\sum\limits_{j=1}^{n}A_{ij}\tau_{ik}^{(s,t+1)}I(e^{(s)}_{2j}=l)}{\sum\limits_{i=1}^{m}\sum\limits_{j=1}^{n}\pi_{ik}^{(s,t+1)}I(e^{(s)}_{2j}=l)}.
\end{eqnarray*}
\UNTIL{the EM algorithm converges.}
\STATE \textbf{Step 6}: Set $\bOmega^{(s+1)}$ to be the final EM update.
\STATE \textbf{Step 7}: Given ${{\bOmega}}^{(s+1)}$, update $e_{2j}^{(s+1)}$, $1\le j\le n$, using
\begin{eqnarray*}
&{e}^{(s+1)}_{2j}=\arg\max_{l\in \{1,2,\ldots,K_{2}\}}\sum\limits_{i=1}^{m}\sum\limits_{k=1}^{K_{1}}\tau_{ik}^{(s+1)}\left\{{A_{ij}}\log{P^{(s+1)}_{kl}}+(1-A_{ij})\log\left(1-P^{(s+1)}_{kl}\right) \right\}.
\end{eqnarray*}
\UNTIL{the profile-pseudo likelihood converges.}
\end{algorithmic}\label{algo3}
\end{algorithm}

Define $\bOmega=(\bm \pi_{1}, \bm P)$ and $\bm {e}_{2}=\left(e_{21}, e_{22}, \ldots, e_{2n}\right)$. To estimate the node labels $\c_2$ from the bi-adjacent matrix $\bm A$, we define the following log pseudo likelihood function
%\begin{eqnarray}
%L^{\textrm{B}}_{\textrm{PL}}(\bOmega, \bm{e}_{2};  \{\a_i\})=\prod\limits_{i=1}^{m}\left\{\sum\limits_{k=1}^{K_{1}}\pi_{1k}\prod\limits_{j=1}^{n}P_{ke_{2j}}^{A_{ij}}(1-P_{ke_{2j}})^{1-A_{ij}}\right\},\nonumber
%\end{eqnarray}
%with its logrithm as follows
\[
\ell^{\textrm{B}}_{\textrm{PL}}(\bOmega, \bm {e}_{2};  \{\a_i\})=\sum\limits_{i=1}^{m}\log\left\{\sum\limits_{k=1}^{K_{1}}\pi_{1k}\prod\limits_{j=1}^{n}P_{ke_{2j}}^{A_{ij}}(1-P_{ke_{2j}})^{1-A_{ij}}\right\}.
\]
A profile-pseudo likelihood algorithm that maximizes $\ell^{\textrm{B}}_{\textrm{PL}}(\bOmega, \bm {e}_{2};  \{\a_i\})$ is described in Algorithm \ref{algo3}. Note that $\c_{1}$ can be estimated similarly as that for $\c_{2}$, and we omit the details.

\begin{figure}[!t]
\centering
\includegraphics[trim=0 10mm 0 0, scale=0.45]{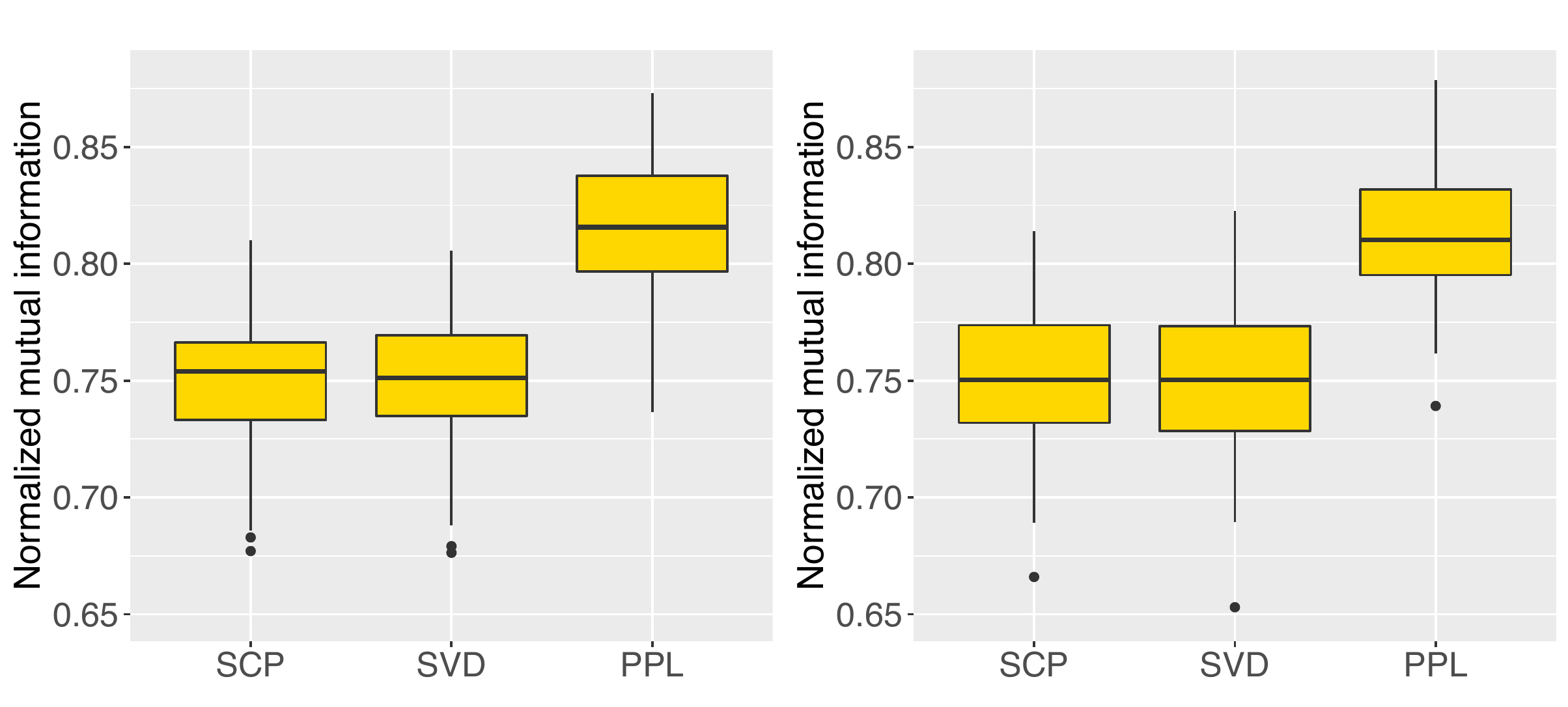}
\caption{Left: comparison of \texttt{PPL}, SVD and \texttt{SCP} for estimating $\c_1$ in BiSBM; right: comparison of \texttt{PPL}, SVD and \texttt{SCP} for estimating $\c_2$ in BiSBM.}
\label{sim4}
\end{figure}

We investigate the performance of the proposed profile-pseudo likelihood method for BiSBM. We fix $m=n=1200$, $K_{1}=K_{2}=2$, $\bm \pi_1=(1/2, 1/2)$, $\bm \pi_2=(1/2, 1/2)$ and edge probability between communities $k$ and $l$ $P_{kl}=0.1(1.2+0.4\times1(k=l))$ for all $k,l=1, 2$.
We compare \texttt{PPL} with two other clustering methods, namely the \texttt{SCP} and SVD \citep{Rohe2012Co, Sarkar2011Community}.
As for \texttt{SCP}, to deal with bipartite networks, we apply it to $\bm A\bm A^{T}$ to get an estimate of $\c_1$, and apply it to $\bm A^{T}\bm A$ to get the estimate of $\c_2$.
The result is summarized in Figure \ref{sim4}, based on 100 replications.
It is seen that \texttt{PPL} outperforms both \texttt{SCP} and SVD for community detection in bipartite networks.

\subsection{Additional Numerical Results}\label{sec:add}
\subsubsection{Running time for \texttt{SCP}}
\begin{figure}[H]
\centering
\includegraphics[scale=0.45]{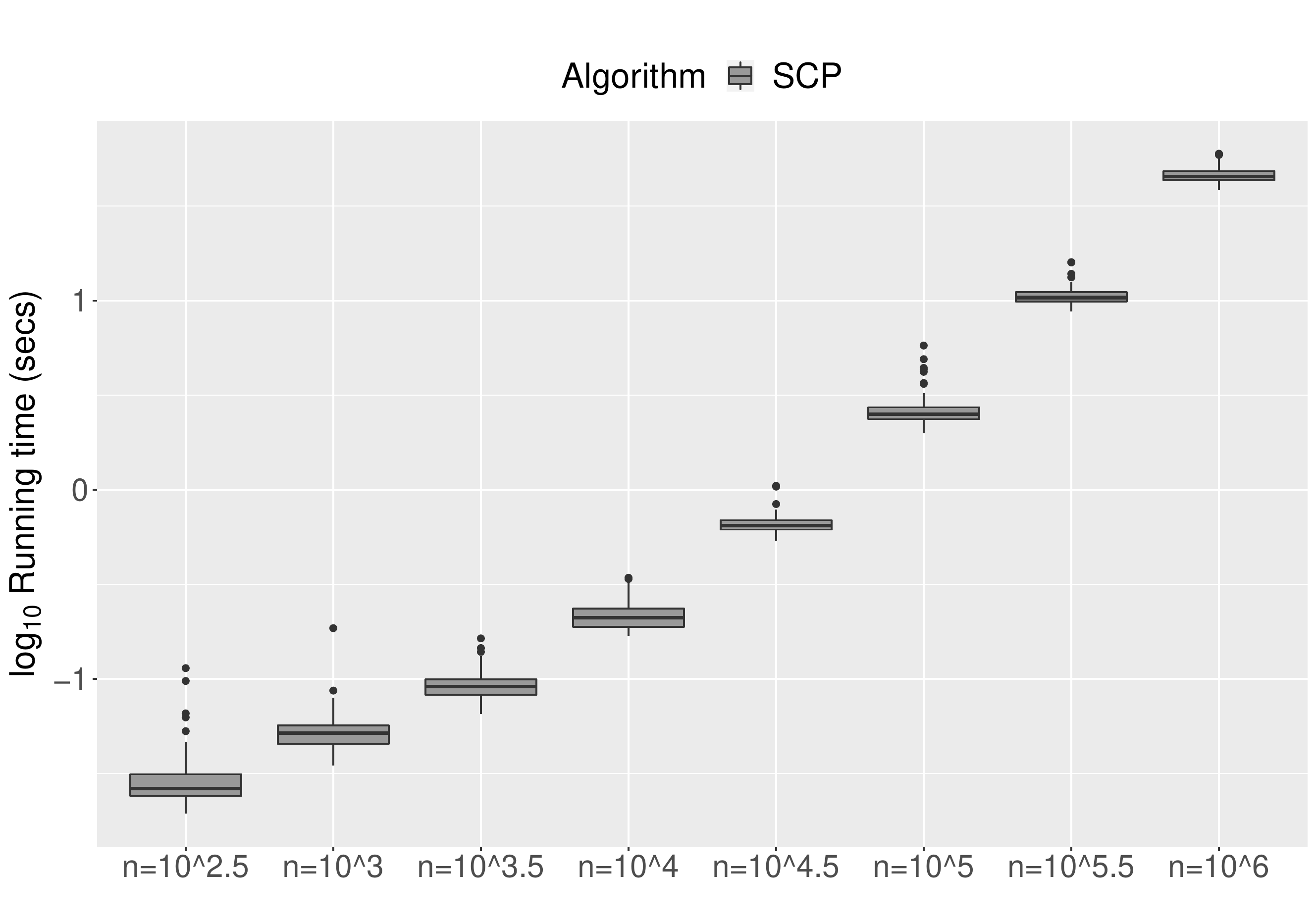}
\caption{\textcolor{black}{Computing time from \texttt{SCP} for large-scale and sparse networks under the SBM with $K=3$, $\bm\pi=(0.2, 0.3, 0.5)$, $\lambda=5$, $\beta=0.05$ and varying $n$.}}
\label{RT_SCP}
\end{figure}

\noindent
\textcolor{black}{We report the computing time for \texttt{SCP} in Setting 3 of Section 5.1. Specifically, we set $K=3$, $\bm\pi=(0.2, 0.3, 0.5)$, $\lambda=5$, $\beta=0.05$ and vary the network size $n$ from $10^{2.5}$ to $10^6$. The results from 100 data replicates are reported in Figure \ref{RT_SCP}. It is seen that it takes \texttt{SCP} less than $100$ seconds when the network has one million nodes. Specifically, this is due to the \texttt{eigs()} function in Matlab, which performs iterative solutions for eigensystems of large sparse matrices using ARPACK. We note that the computational efficiency of \texttt{eigs()} can decrease when the network density and the number of communities $K$ increase.}

\subsubsection{Comparison with \cite{gao2017achieving}}
\begin{figure}[!t]
\centering
\includegraphics[scale=0.575]{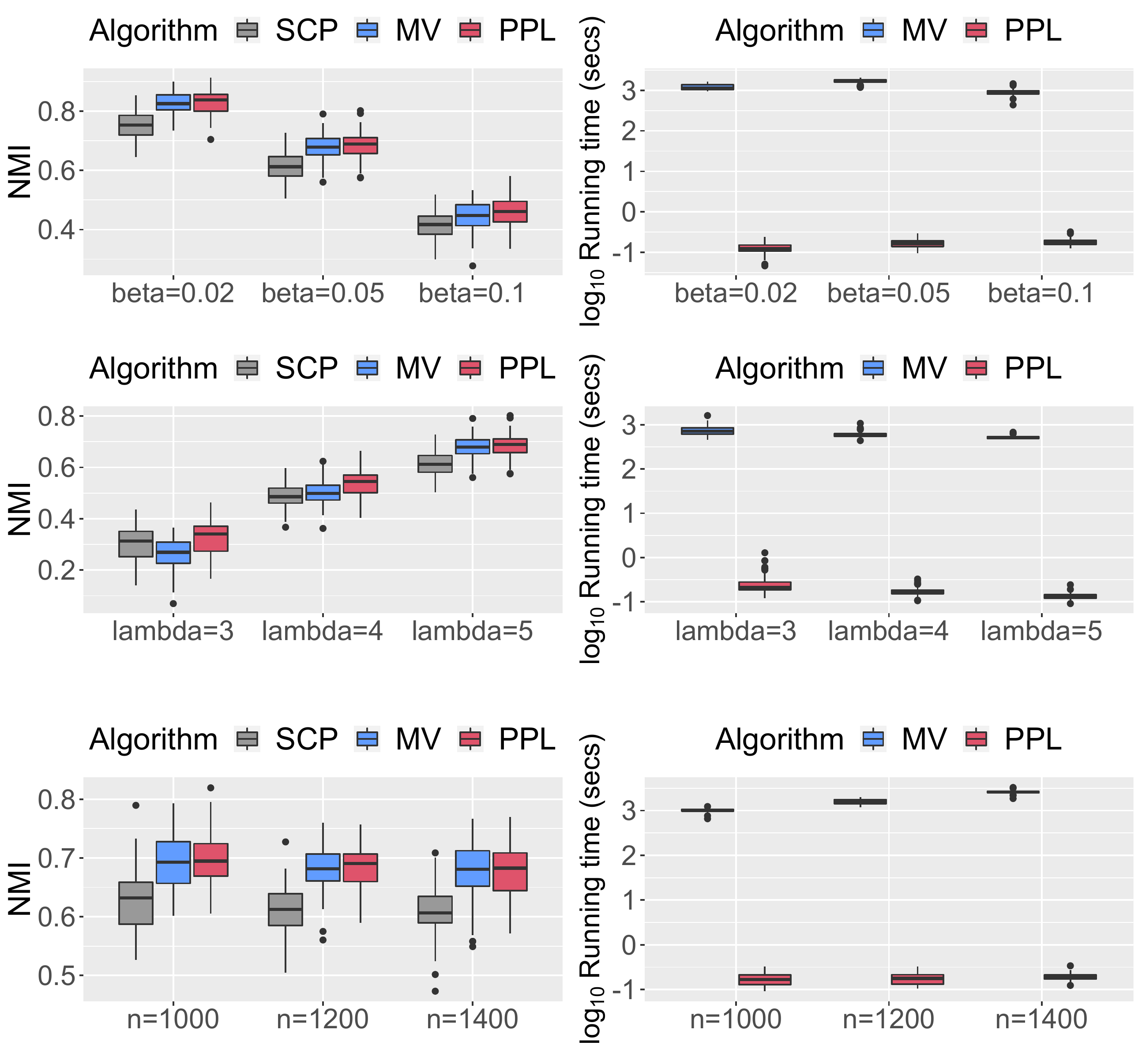}
\caption{Comparisons of the NMI and computing time from \texttt{SCP}, \texttt{MV} and \texttt{PPL} under different settings.}
\label{MV}
\end{figure}

\noindent
\textcolor{black}{
In this simulation study, we compare the performance of \texttt{SCP}, \texttt{PPL} and the majority voting method proposed in \cite{gao2017achieving} (referred to as \texttt{MV}) on networks simulated from the SBM.
Specifically, we consider the simulation Setting 3 in Section 5.1, where the parameter $\beta$ controls the ``out-in-ratio" and $\lambda$ controls the overall expected network degree. We set $K=3$ and $\bm \pi=\left(0.2, 0.3, 0.5\right)$, and we consider three scenarios, 1) varying $\beta$ while $\lambda=5$ and $n=1200$, 2) varying $\lambda$ while $\beta=0.05$ and $n=1200$, and 3) varying $n$ while $\lambda=5$ and $\beta=0.05$. Figure \ref{MV} reports the NMI from the three methods and the computing time from \texttt{PPL} and \texttt{MV}, based on 100 replications.
The running time for \texttt{PPL} does not include the initialization step, which takes no more than a few seconds.
Both \texttt{PPL} and \texttt{MV} use $\texttt{SCP}$ as the initial clustering method.
It is seen that \texttt{PPL} and \texttt{MV} have comparable clustering accuracies and they both outperform \texttt{SCP} in terms of NMI.
Moreover, \texttt{PPL} is computationally more efficient than \texttt{MV} as it needs not to repeatedly perform the leave-one-node-out spectral clustering.
}

%\bibliographystyle{asa}
%\begingroup
%\baselineskip=16.5pt
%\bibliography{ref}
%\endgroup

\end{document}